\documentclass[twocolumn,showpacs,aps,prd,amsmath,amssymb,superscriptaddress,nofootinbib,nobibnotes,floatfix]{revtex4-1}

\usepackage{graphicx}
\graphicspath{{.}{../data/files/}}
\usepackage{float}% to force figure placement 
\usepackage{bm}% bold math
\usepackage[hidelinks]{hyperref}
\usepackage{color}

\usepackage[utf8]{inputenc}
\usepackage[T1]{fontenc}
\usepackage{slashed}

\usepackage{enumerate}

\usepackage{mathalfa,amssymb,amsthm}
\usepackage{amsmath}
\usepackage[dvipsnames]{xcolor}
\usepackage[capitalise]{cleveref}
\begin{document}

%\preprint{APS/123-QED}
%=============================================================================
\title{
%Approaches for studying modifications to 
%General Relativity in the non-linear regime. A comparison
Nonlinear studies of modifications to general relativity: Comparing different approaches
   }
\author{Maxence Corman}
\email{maxence.corman@aei.mpg.de}
\affiliation{Max Planck Institute for Gravitational Physics (Albert Einstein Institute), D-14476 Potsdam, Germany}
\author{Luis Lehner}
\email{llehner@perimeterinstitute.ca}
\affiliation{%
Perimeter Institute for Theoretical Physics, Waterloo, Ontario N2L 2Y5, Canada
}%
\author{William E. East}
\email{weast@perimeterinstitute.ca}
\affiliation{%
Perimeter Institute for Theoretical Physics, Waterloo, Ontario N2L 2Y5, Canada
}%
\author{Guillaume Dideron}
\email{gdideron@uwaterloo.ca}
\affiliation{%
Perimeter Institute for Theoretical Physics, Waterloo, Ontario N2L 2Y5, Canada
}%
\affiliation{%
Department of Physics and Astronomy, University of Waterloo, Waterloo, Ontario N2L 3G1,
Canada
}%

%=============================================================================
\begin{abstract}
    Studying the dynamical, nonlinear regime of modified theories of gravity
    remains a theoretical challenge that limits our ability to test general
    relativity. Here we consider two generally applicable, but approximate
    methods for treating modifications to full general relativity that have
    been used to study binary black hole mergers and other phenomena in this
    regime, and compare solutions obtained by them to those from solving the
    full equations of motion. The first method evolves corrections to general relativity order by order in
    a perturbative expansion, while the second method introduces extra dynamical
    fields in such a way that strong hyperbolicity is recovered.
    We use shift-symmetric
    Einstein-scalar-Gauss-Bonnet gravity as a benchmark theory to illustrate
    the differences between these methods for several spacetimes of physical
    interest.  We study the formation of scalar hair about initially
    non-spinning black holes, the collision of black holes with scalar charge,
    and the inspiral and merger of binary black holes.  By directly comparing
    predictions, we assess the extent to which those from the approximate treatments
    can be meaningfully confronted with gravitational wave observations. We find
    that the order-by-order approach cannot faithfully track the solutions when
    the corrections to general relativity are non-negligible. The second 
    approach, however, can provide consistent solutions, provided the ad-hoc
    timescale over which the dynamical fields are driven to
    their target values is made short compared to the physical timescales.

\end{abstract}
%=============================================================================
\maketitle
%=============================================================================
\section{Introduction}%
Remarkably, the theory of general relativity (GR) has passed all experimental
and observational tests to date~\cite{Will:2014kxa}, including on cosmological
\cite{Clifton:2011jh} and short scales \cite{Lee:2020zjt,Kapner:2006si}, in
slow-motion, weak-curvature \cite{GRAVITY:2020gka} and strong-curvature
\cite{Kramer:2021jcw,
EHT:2019pgp,EHT:2022xqj,Volkel:2020xlc,Psaltis:2020lvx,Kocherlakota:2021dcv}
regimes, as well as in the nonlinear, high-velocity, strong-field regime
\cite{TheLIGOScientific:2016src,
Yunes:2016jcc,Baker:2017hug,Isi:2019aib,Isi:2020tac,Okounkova:2021xjv,
Abbott:2018lct,LIGOScientific:2019fpa,Abbott:2020jks,LIGOScientific:2021sio}.
The last-named has been probed, in recent years, through the gravitational wave
(GW) observations of compact object binary mergers by the LIGO, Virgo and KAGRA
detectors \cite{LIGOScientific:2014pky,VIRGO:2014yos,Somiya:2011np,Aso:2013eba}.  
In general, tests of GR
with GW observations fall into two categories: those that look for generic
types of deviations from GR, and those that target specific theories. The
former assumes the underlying gravitational signal is well described by GR (potentially
adding independent parametric deviations to the signal, whether for the inspiral, the merger
or the ringdown stages, to characterize potential deviations)
\cite{Abbott:2018lct, LIGOScientific:2019fpa,
LIGOScientific:2020tif,LIGOScientific:2021sio,Ghosh:2022xhn}.  However, in
general, a consistent modified theory will introduce deviations in a more
complicated and correlated manner, which such tests may be insensitive to.  
Even when such
searches are sensitive, there could be subtleties in trying to map constraints
on parameteric deviations to constraints on parameters in specific theories
related to the choice of priors and the actual parameters on which the
measurements are done.  Example predictions from specific theories could be
used to test, improve, and motivate new generic tests. They could also be
directly compared to the data, in principle allowing us to place the most
stringent constraints on deformations of GR
\cite{Yunes:2013dva,Berti:2015itd,Berti:2018cxi,Berti:2018vdi}.  However, in
order to make predictions of the GW signals from the
final stages of compact object mergers to compare to the observations,
solutions in beyond GR theories are required. 

There are a number of challenges associated with studying
compact object mergers in modified theories, the foremost of which is that in any theory
that modifies the principal part of the Einstein equations, the issue of a well-posed 
initial value problem must be revisited.%
\footnote{
Some theories feature non-minimally coupled fields, but the principal part of the evolution
equations for the metric is unmodified, thus naturally inheriting desirable hyperbolic properties.
Examples include the scalar-tensor theory of gravity from Ref.~\cite{Damour:1993hw} (in which neutron stars can develop 
a scalar charge) for which merging neutron stars' dynamics were presented in, e.g., Ref.~\cite{Barausse:2012da,Shibata:2013pra} ,
and Einstein-Maxwell-Dilaton gravity (in which electrically charged black holes develop a scalar charge 
\cite{Garfinkle:1990qj}), which was studied in the context of binary black hole mergers in, e.g., Ref.~\cite{Hirschmann:2017psw}.
}
In order to address this issue, several strategies have emerged in the literature for
simulating compact object mergers in modified theories:
\begin{enumerate}[i.]
\item Focus on specific modified theories where a formulation leading to well posed initial value problem is known (at least for some
    range of parameters), and \emph{solve the full evolution equations}.
\item Exploit the observation that corrections to GR ought to be subleading, and incorporate their
    effects on the dynamics in a perturbative fashion, \emph{order by order}, where corrections are evaluated with respect to
    the solution obtained at lower order, thereby maintaining the same principal structure as in GR.
\item Take the effective field theory (EFT) viewpoint that a modified theory is only valid up to some short length/timescale, and 
    \emph{``fix" the evolution equations} below this scale by introducing new dynamical fields that replace high derivative terms in the original evolution equations, allowing them to maintain the same principal structure as in GR, while including the effect of correcting terms via suitable coupling among relevant fields. 
    These new fields are given evolution equations that drive them to the values they would have in the original theory on some short timescale\footnote{A somewhat
    related approach is to introduce effective viscosity contributions at the same order in derivatives as the beyond GR contributions~\cite{deRham:2023ngf}.
    This, however, requires specific types of beyond GR theories to be feasible and, a priori, it is unclear that this same order modification might not affect
    the physics to be extracted without a specific analysis on a case-by-case basis.}.
\end{enumerate}
We expound on these different strategies below.

Applications of approach (i), {\em solving the  full evolution equations}, have mostly
focused on theories with second order equations of motion.  (Though it should
be noted that some special theories with higher than second order equations of motion can
still be well-posed and evolved nonlinearly. An example of this is quadratic gravity, which gives fourth
order equations~\cite{Noakes:1983xd,Held:2021pht,Held:2023aap,East:2023nsk}.) In this case, at sufficiently weak coupling, the
modified theory can be seen as having the principal part one would have in GR,
with some small correction. Crucially, even at weak coupling such
modifications can lead to the loss of well-posedness when using standard formulations of the equations of motion.
For example, generic examples of Horndeski theories---scalar-tensor theories
with second order equations of motion---do not lead to well posed problems  in the harmonic
formulation~\cite{Papallo:2017qvl} (as the resulting equations of motion 
are weakly, but not strongly hyperbolic). However,
general Horndeski theories have been shown to lead to a (locally) well-posed
initial value problem in a modified generalized harmonic (MGH)
formulation~\cite{Kovacs:2020pns,Kovacs:2020ywu}, so long as the length scale
associated with the beyond-GR coupling is much smaller than all the other
length scales in the problem. (At larger couplings, such theories can exhibit
the complete breakdown of hyperbolicity,
e.g.~\cite{Ripley:2019hxt,Bernard:2019fjb}.) The MGH formulation has been used
to simulate compact object mergers and compute GW signals in
example modified
theories~\cite{East:2020hgw,East:2021bqk,East:2022rqi,Corman:2022xqg}, and has
been extended to the modified CCZ4 formulation
\cite{AresteSalo:2022hua,AresteSalo:2023mmd}. Related to this are studies of
compact object mergers in Horndeski theories where the metric evolution can be
treated the same way as in GR, but the principal part of the scalar field
evolution is modified from that of the wave
equation~\cite{Figueras:2021abd,Bezares:2021dma}.  With this approach, the
dynamics of binary black holes and other compact object mergers can be studied,
and predictions for the resulting GW signals can be extracted,
without approximation (beyond that of numerical discretization error), at least
within some range of the parameter space of theories where a well-posed
formulation is known. However, there exist many classes of modified gravity
theories which almost certainly define ill-posed problems if treated as
complete theories, regardless of the value that the coupling parameter
controlling the modifications. Taking an EFT perspective, this can be seen as a
natural consequence of a specific truncation of the theory. In order to treat
such theories, a different strategy, making use of the perturbative nature of
the modifications to GR, is necessary.

Approach (ii), the {\em order-by-order approach}, treats mathematical pathologies in modified theories
by evolving separately the corrections to GR at each order (in the coupling controlling the modifications to GR) and only evaluating corrections through solutions to previous orders. Backreaction of corrections to the GR 
solution are thus only passively and perturbatively accounted for through higher order corrections to the solution, ensuring the principal part of the equations is unaltered.
For example, and focusing primarily on the two body problem, studies of dynamical Chern Simons gravity~\cite{Alexander:2009tp} and
Einstein scalar Gauss Bonnet (EsGB) gravity (discussed below)
have been pursued using
this strategy to solve for the leading order correction to the metric~\cite{Okounkova:2017yby,Okounkova:2019dfo,Okounkova:2019zjf,Okounkova:2020rqw}. 
However, the quality (or faithfulness) of
the solutions that are obtained depends not only on the corrections being small, but also on secular effects due to the accumulation of errors introduced at the perturbative order {\em staying small}. 
In the inspiral of a binary black hole, strong secular effects accumulate in the order-by-order approach 
on the radiation-reaction timescale~\cite{Okounkova:2019zjf,Okounkova:2020rqw}, making it unclear
what parts of the computed GW signal can accurately be extracted from such a computation.

An alternative approach is {\em fixing the equations} (iii), where extra fields are
introduced in a way that mocks up degrees of freedom that would have been
integrated out in an EFT truncation in such a way that they play a subleading
role in the dynamics. An ad-hoc prescription is used to drive these fields
towards satisfying specific identities on a certain chosen time/lengthscale,
while modes below this scale are effectively damped away. As a consequence of
replacing higher order terms with these extra fields, well posedness is
achieved. Correcting terms actively backreact, and are treated consistently
within the approximation in the limit of the prescription's timescale is
much smaller%
\footnote{In general, the fixing timescale cannot be made arbitrarily small, since at some point this 
may reintroduce high-frequency instabilities associated with the ill-posedness of the original problem.}
than the other physical timescales in the problem. Obtaining quality solutions this way requires that
the ad-hoc timescale be made short compared to the other physical timescales,
and that the dynamics does not lead to a significant UV cascade, so that the
solution is not sensitive to the ad-hoc scale.  For instance, an EFT which
introduces higher order modifications to GR without new
fields~\cite{Endlich:2017tqa} was studied in recent
works~\cite{Cayuso:2020lca,Lara:2021piy,
Franchini:2022ukz,Bezares:2021dma}. Assessing whether one can obtain an answer
that is insensitive to the fixing scale for the problem of interest is essential to being able to draw physical conclusions and examine
the viability of the theory under study. 
In some cases, this may pose a practical challenge, as simulating compact object mergers is computationally expensive,
and this approach, by construction, introduces an additional small scale that must
be resolved.

The goal of this study is to focus on binary black hole mergers in a benchmark
modified gravity theory where the full equations can be solved [approach (i)],
and employ such solutions to evaluate the order-by-order approach (ii) and the
fixing-the-equations approach (iii). (As well, all solutions can be compared to those
in GR for comparable physical parameters).  In Ref.~\cite{Allwright:2018rut}, a
comparison between these three approaches was carried out for a scalar field
toy-model with a known UV completion, highlighting some of the clear
shortcomings of the order-by-order strategy, while in
Ref.~\cite{Franchini:2022ukz}, spherically-symmetric scalar field collapse in
EsGB was studied using approaches (i) and (iii).  Here we study specifically
the case of binary black hole dynamics.  As binary black holes in GR do not
display a significant UV-cascade~\cite{Allwright:2018rut} and solutions remain smooth throughout, it
could be argued that corrections in a modified theory likely remain small, and
potentially all approaches could be useful\footnote{This observation follows from
the fact that the 
gravitational wave energy in multipoles above ${\ell}=2$ remains small even during coalescence.
This is especially true in the equal-mass, non-spinning case where the source 
drives an ${\ell}=2$ structure and energy in higher multipoles is smaller by at least two orders of
magnitude (e.g.~\cite{Berti:2007fi}).}. By leveraging full
solutions of binary black hole mergers, we are able to directly quantify the
errors introduced by secular effects in the order-by-order approach and by
different prescriptions for the ad-hoc dynamical fields with non-zero damping
timescales in the fixing-the-equations approach, in order to study their 
accuracy.

The particular benchmark theory we will focus on is shift-symmetric EsGB gravity.
This theory introduces deviations from GR at small curvature length
scales and yields solutions that are distinct from those in GR.
Indeed, vacuum black holes evolve to develop stable 
scalar clouds 
\cite{Kanti:1997br,Yunes:2011we,Sotiriou:2013qea,Sotiriou:2014pfa,Antoniou:2017acq,
Papageorgiou:2022umj},
and hence can differ qualitatively from GR in the strong field regime, 
while still passing weak field tests.
EsGB is thus a promising theory to perform theory-specific tests of GR 
in the strong field, dynamical regime.
From a theoretical point of view, shift-symmetric EsGB gravity can be 
motivated as the leading order
scalar-tensor theory of gravity whose resulting equations of motion
are invariant under a shift in the scalar field. 
As a result, significant recent work has gone into modeling compact object mergers in
EsGB gravity both
in post-Newtonian 
\cite{Yagi:2011xp,Sennett:2016klh,Julie:2019sab,Shiralilou:2020gah,Shiralilou:2021mfl,
vanGemeren:2023rhh,Julie:2022qux,Julie:2023ncq} 
and numerical relativity using either approach (i) 
\cite{East:2020hgw,East:2020hgw,Ripley:2022cdh,East:2022rqi,Corman:2022xqg,AresteSalo:2022hua,AresteSalo:2023mmd}
or (ii) \cite{Witek:2018dmd,Silva:2020omi,Okounkova:2020rqw,Elley:2022ept}.
From a practical standpoint, we will study this theory because 
we can make use of recently developed 
computational methods~\cite{East:2020hgw} to solve the equations of motion for black hole 
spacetimes without approximation, 
using the MGH formulation. Another practical reason is that one can use 
an EFT argument to motivate this theory 
\cite{Weinberg:2008hq} which, as mentioned, underlies the use of the order-by-order 
and fixing the equation approaches. 

In this paper, we describe our methods for numerically solving the equations of 
EsGB gravity using the MGH formulation, order-by-order, and fixing-the-equations approaches,
all within the same infrastructure, and apply these to study the dynamics of black hole scalar hair
formation and black hole mergers. One of our main results is to highlight the inability
of the order-by-order approach to faithfully track the solution in the regime
where the corrections to the GR waveform become non-negligible.
We also confirm that the fixing equations approach can provide consistent solutions
when the ad-hoc timescales are made sufficiently small, and
motivate how to choose the associated parameters in this method. Interestingly,
we find here (as in Ref.~\cite{Corman:2022xqg}) that the inspiral-merger-ring down waveforms primarily differ from those
in GR by their phase evolution, without significantly deviating in amplitude, 
elevating options for encoding beyond GR waveforms by restricting 
to changes in phase (see e.g.~\cite{PhysRevD.100.104036,PhysRevD.89.082001,PhysRevD.107.044020,Dideron:2022tap}).

The remainder of this work is organized as follows. 
In Sec.~\ref{sec:esgb}, we review shift-symmetric EsGB gravity.
In Sec.~\ref{sec:method}, we describe the three different approaches we consider
to evolve this theory, including the equations of motion we numerically solve for. 
We discuss their numerical implementation as well as the 
numerical methods for analyzing the
results in Sec.~\ref{sec:numerics}.  
In Sec.~\ref{sec:results}, we present and discuss our results for several physically interesting spacetimes,
including the dynamical formation of scalar hair around non-spinning black holes
in axisymmetry and head-on
and quasi-circular binary black hole mergers.  
We conclude
in Sec.~\ref{sec:conclusion}. We give details on the numerical implementation of
the evolution equations in Appendix~\ref{app:implementation},
and discuss the numerical accuracy of our simulations in
Appendix~\ref{app:convergence}.
In Appendix~\ref{app:fixingconstraints}, we describe how constraints behave 
in the fixing-the-equations approach, while in Appendices~\ref{app:anharm_osc} and \ref{app:fixingmodel}, 
we present toy models to illustrate potential pitfalls 
in the order-by-order and fixing-the-equations approaches, respectively. 
We use geometric units:
$G=c=1$, a metric sign convention of $-+++$, lower case Latin letters to index
spacetime indices, and lower case Greek letters to index spatial indices.

%=============================================================================
\section{Einstein scalar Gauss Bonnet gravity\label{sec:esgb}}%
The benchmark modified gravity theory that we consider in this study is shift-symmetric EsGB gravity,
which has an action given by
\begin{align}
\label{eq:esgb_action}
    S
    =&
    \frac{1}{16\pi}\int d^4x\sqrt{-g}
   \left(
        R
    -   \left(\nabla\phi\right)^2
    +   2\lambda\phi\mathcal{G}
    \right)
    ,
\end{align}
where $g$ is the determinant of the spacetime metric and
$\mathcal{G}$ is the Gauss-Bonnet scalar given by
\begin{align}
    \mathcal{G}
    \equiv
    R^2 - 4R_{ab}R^{ab} + R_{abcd}R^{abcd}
    .
\end{align}
Here, $\lambda$ is a constant coupling parameter that, in
geometric units, has dimensions of length squared.
As the Gauss-Bonnet scalar $\mathcal{G}$ is a total derivative
in four dimensions, the action of
EsGB gravity is preserved
up to total derivatives under constant shifts in the scalar field:
$\phi\to\phi+\textrm{constant}$.
Schwarzschild and Kerr black holes are not stationary solutions
to this theory: starting with vacuum initial data, black holes 
dynamically develop stable scalar clouds (scalar hair).
The stationary state is a black hole with a nonzero scalar charge $Q_{\rm SF}$,
so long as the coupling normalized by the black hole mass $M$,
$\lambda/M^2$, is sufficiently small so that the theory does not breakdown
\cite{Sotiriou:2013qea,Sotiriou:2014pfa,Ripley:2019aqj,East:2020hgw}.
In particular, studies have found that
the regularity of black hole solutions and hyperbolicity of the corresponding equations of motion sets
$\lambda/M^2\lesssim 0.23$ for non-spinning black holes
\cite{Sotiriou:2014pfa,Ripley:2019aqj}.
Neutron stars, in contrast to black holes, do not have a scalar charge in EsGB gravity
\cite{Yagi:2011xp,Julie:2019sab}.
For black holes, at large radius the scalar field falls off like 
$\phi = Q_{\rm{SF}}/r + \mathcal{O}\left(1/r^2\right)$,
and therefore one expects 
black hole binaries to emit scalar radiation,
thus increasing the inspiral rate of binaries
\cite{Yagi:2015oca,Yagi:2011xp}.
The most stringent observational bounds on the theory
come from comparing several black hole-neutron star
GW signals to post-Newtonian results for EsGB and give a constraint of
(after restoring dimensions) $\sqrt{\lambda}\lesssim 2.5$ km 
\cite{Lyu:2022gdr}. 

The covariant equations of motion for shift-symmetric EsGB gravity are
\begin{align}
\label{eq:eom_esgb_scalar}
   \Box\phi
   +
   \lambda\mathcal{G}
   &=
   0
   ,\\
\label{eq:eom_esgb_tensor}
   R_{ab}
   -
   \frac{1}{2}g_{ab}R
   -
   \nabla_a\phi\nabla_b\phi
   +
   \frac{1}{2}\left(\nabla\phi\right)^2g_{ab}
   + \nonumber \\
   2\lambda
    \delta^{efcd}_{ijg(a}g_{b)d}R^{ij}{}_{ef}
   \nabla^g\nabla_c\phi
   &=
   0
   ,
\end{align}
where $\delta^{abcd}_{efgh}$ is the generalized Kronecker delta tensor.
In the next section, we describe the three approaches we follow to study relevant
systems governed by these equations of motion.

%=============================================================================
\section{Formalism\label{sec:method}}%
%=============================================================================
\subsection{Modified generalized harmonic formulation\label{sec:mgh}}%
We study the full, nonperturbative, shift-symmetric
EsGB equations in the MGH formulation~\cite{Kovacs:2020pns,Kovacs:2020ywu}
using the implementation and methods of Ref.~\cite{East:2020hgw}.
In this formulation, one introduces two auxiliary
Lorenzian metrics $\tilde{g}^{mn}$ and $\hat{g}^{mn}$
that obey certain causality conditions.
Adopting a gauge in the MGH formulation amounts to choosing the
functional form of the auxiliary metrics,
as well as the functional form of the source function $H^c$.
There is
a large degree of freedom in choosing the auxiliary metrics as functions of the physical metric
$g_{ab}$, but 
as in Ref.~\cite{East:2020hgw} we choose
 $\tilde{g}^{ab} = g^{ab} - \tilde{A} n^{a}n^{b}$ and
$ \hat{g}^{ab} = g^{ab} - \hat{A} n^{a}n^{b}$,
where $n^{a}$ is the (timelike) unit vector orthogonal to the spacelike
hypersurfaces we evolve on, and $\tilde{A}$ and $\hat{A}$ are constants
set to 0.2 and 0.4, respectively.
%=============================================================================
\subsection{Order by order approach\label{sec:ora}}%
We first review the order-by-order approach\footnote{This approach is often referred
to as the \emph{order reduction approach} in the literature, and we will use the abbreviation
ORA in some places.}, and refer the reader
to
Refs.~\cite{Okounkova:2017yby,Okounkova:2018abo,Okounkova:2018pql,
Witek:2018dmd,Okounkova:2019dfo,
Okounkova:2019zep,Okounkova:2019zjf,Okounkova:2020rqw,Okounkova:2021xjv,Okounkova:2022grv}
for more details
on the theory, implementation in generalized harmonic coordinates, and application
to computing GWs.

The method outlined here is applicable to any
beyond-GR theory with a continuous limit to GR, but we will restrict ourselves
to shift-symmetric ESGB
gravity. For such theories, the approach is as follows.
We perturb the theory about GR in powers of the small coupling parameter. We collect
equations of motion at each order in the coupling, creating a tower of equations, with
each level acquiring the same principal part as the background GR system.
That way, well-posedness
of the initial value problem in GR carries over to this framework\footnote{Assuming
suitable initial and boundary data are provided for each order in the expansion.}, even
though the original theory may not have a well-posed
initial value formulation.

More concretely, we introduce a dimensionless order-counting parameter $\epsilon$
which keeps
track of powers of $\lambda$ ($\epsilon$ can be set to one later on),
and expand the metric and scalar field as power series in $\epsilon$
\begin{subequations}
   \label{eq:perturbative_expansion_fields}
   \begin{align}
      g_{ab}
           &= g_{ab}^{(0)} +
      \sum_{k=1}^{\infty}\epsilon^k h_{ab}^{(k)}
      ,\\
      \phi
      &=
      \sum_{k=0}^{\infty}\epsilon^k\phi^{(k)}
      .
   \end{align}
\end{subequations}
We emphasize that this is not a weak field expansion. We always raise and lower indices
with the background metric $g_{ab}^{(0)}$, e.g.,
\begin{equation}
        h^{(k) ab} = g^{(0)ac}g^{(0)bd} h_{cd}^{(k)}.
\end{equation}
This expansion is now inserted into the field equations,
which are likewise expanded in powers
of $\epsilon$, and collected at orders homogeneous in $\epsilon^k$.
This results in a tower of
equations that need to be solved at progressively increasing order.

At zeroth order in $\epsilon$ we obtain,
\begin{subequations}
\begin{align}\label{eq:eom_ora_0}
        G_{ab}^{(0)} \left[g_{ab}^{(0)}\right] &= 8 \pi T_{ab}^{(0)}
        ,\\
        \Box^{(0)}\phi^{(0)} &= 0,
\end{align}
\end{subequations}
where $G_{ab}^{(0)} \left[g_{ab}^{(0)}\right]$ is the Einstein tensor of the background
metric $g^{(0)}$, $\Box^{(0)}$ is the associated d'Alembert operator, $T_{ab}^{(0)}$
is the stress-energy tensor of $\phi^{(0)}$ with derivatives constructed
from $g_{ab}^{(0)}$.
Since the zeroth order scalar field has no source, we can take $\phi^{(0)}=0$.
This implies that at zeroth order, the system will simply be
\begin{align}\label{eq:order_0}
        G_{ab}^{(0)} \left[g_{ab}^{(0)}\right] = 0
        ,
\end{align}
i.e., the GR Einstein field equations with solution
\begin{align}
        \left(g_{ab}^{(0)},\phi^{(0)}\right) = \left(g_{ab}^{\rm{GR}},0\right).
\end{align}

Continuing to linear order in $\epsilon$, we find the system
\begin{subequations}
\begin{align}
        G_{ab}^{(1)} \left[g_{ab}^{(0)},h_{ab}^{(1)}\right] &= 8 \pi T_{ab}^{(1)}
        \label{eq:ee_1} \\
        &-
        2 \lambda \delta^{efcd}_{ijg(a}g^{(0)}_{b)d}R^{(0)ij}{}_{ef}\nabla^{(0)g}\nabla^{(0)}_c
        \phi^{(0)}
        ,\nonumber \\
        \Box^{(0)}\phi^{(1)} +\Box^{(1)}\phi^{(0)}  &= -\lambda \mathcal{G}^{(0)}.
        \label{eq:phi_1}
\end{align}
\end{subequations}
Note the terms modifying GR in these equations appear strictly as source terms.

Here, $G_{ab}^{(1)} \left[g_{ab}^{(0)},h_{ab}^{(1)}\right]$ is the linearized Einstein
operator, built with the covariant derivative $\nabla^{(0)}$ compatible with $g^{(0)}$,
acting on the metric deformation $h^{(1)}$.
The d'Alembert operator receives a correction $\Box^{(1)}$
that depends on the metric deformation $h^{(1)}$.
$\mathcal{G}^{(0)}$ is the Gauss-Bonnet invariant
evaluated on the background spacetime metric $g^{(0)}$. Finally, $T_{ab}^{(1)}$ is the
first-order perturbation to the stress-energy tensor. It is straightforward
to see that, when evaluated on the $\mathcal{O}(\epsilon^0)$ solution, the right-hand side
of Eq.~\eqref{eq:ee_1} vanishes\footnote{Since $T_{ab}^{(1)}$
is quadratic in $\phi$, $T_{ab}^{(1)}$ has pieces both linear and quadratic
in $\phi^{(0)}$
(the quadratic in $\phi^{(0)}$ pieces are linear in $h^{(1)}$). Therefore, both
the Gauss-Bonnet term in Einstein equations and $T^{(1)}$ are built from pieces
linear and quadratic in $\phi^{(0)}$. We just found that $\phi^{(0)}=0$, such that
when evaluated on the $\mathcal{O}(\epsilon^0)$ solution these both vanish.}.
Therefore, at $\mathcal{O}(\epsilon^1)$ in perturbation theory
evaluated on the background solution,
we have the system
\begin{align}\label{eq:order_1}
        G_{ab}^{(1)} \left[g_{ab}^{(0)},h_{ab}^{(1)}\right] = 0, \ \ \
        \Box^{(0)}\phi^{(1)}  = -\lambda \mathcal{G}^{(0)}.
\end{align}
Since at first order in $\epsilon$ the metric itself is not deformed,
we can set $h^{(1)}_{ab}=0$.
The scalar field on the other hand is sourced by the curvature
of the background spacetime, hence develops a nontrivial scalar profile.
To summarize, the solution at $\mathcal{O}(\epsilon)$ is
\begin{equation}
        \left(g_{ab}^{(1)},\phi^{(1)}\right) = \left(0,\phi^{(1)} \right).
\end{equation}
So far, we have just derived the equations in the test-field, or {\em decoupling limit}.
We now consider the equations to order $\epsilon^2$, which is the lowest order
where the metric perturbation is
sourced by $\phi^{(1)}$.
Hence, any backreaction on the system's dynamics begins to enter at second order
in the small coupling parameter. The field equations at $\mathcal{O}(\epsilon^2)$ read
(after taking account of $\phi^{(0)}=0$ and $h^{(1)}=0$)
\begin{subequations}
\begin{align}\label{eq:order_2}
        G_{ab}^{(0)} \left[h_{ab}^{(2)}\right] &= 8 \pi T_{ab}^{(2)} \\
        & -
        2 \lambda \delta^{efcd}_{ijg(a}g^{(0)}_{b)d}R^{(0)ij}{}_{ef}\nabla^{(0)g}\nabla^{(0)}_c
        \phi^{(1)}
        ,\nonumber \\
        \Box^{(0)}\phi^{(2)}  &= 0,
\end{align}
\end{subequations}
where $G_{ab}^{(0)} \left[h_{ab}^{(2)}\right]$ is the second order Einstein operator,
the Gauss-Bonnet term in Einstein equations is formed from the background metric
and non-vanishing
first order scalar field $\phi^{(1)}$ (hence does not contribute to principal part), while
$T_{ab}^{(2)}$ is the first nonvanishing contribution to the stress-energy tensor
\begin{align}
T^{(2)}_{ab} = \frac{1}{8 \pi}
        \left[ \nabla^{(0)}_a\phi^{(1)}\nabla^{(0)}_b\phi^{(1)} -\frac{1}{2} g^{(0)}_{ab} \nabla^{(0)}_c
        \phi^{(1)} \nabla^{(0)c} \phi^{(1)}
         \right].
\end{align}
Therefore, we can consistently set $\phi^{(2)}=0$
and compute the first nonvanishing corrections
to the metric components by solving the Einstein equations at second order,
with the linear-order
scalar field and background solution for the metric as an input.
The details of the numerical implementation of these equations can be found in
Appendix~\ref{app:implementation}. We next describe some of the features and caveats
one inherits as a result of solving the equations of motion perturbatively.

\subsubsection{Scaling of solutions}\label{sec:scaling_solns}
Since the coordinates have dimensions of length,
the covariant derivative $\nabla$ can be made dimensionless by factorizing
out a factor of $M^{-1}$ where $M$ is the total ADM mass.
Similarly the Riemann tensor may be made dimensionless by
pulling out a factor of $M^{-2}$.
If we apply this reasoning to the first order equation for the
scalar field (\ref{eq:phi_1}) we find
\begin{align}
        M^{-2} \Box^{(0)}\phi^{(1)}  = -\lambda M^{-4} \mathcal{G}^{(0)},
\end{align}
so that we may define a dimensionless scalar field $\Phi$ as
\begin{align}\label{eq:scaling_phi}
        \phi^{(1)}  = \frac{\lambda}{M^2} \Phi.
\end{align}
Similarly, even though $[h^{(k)}]=[g]=[L]^0$ is dimensionless, it depends on $\lambda/M^2$,
such that one may define a scale free metric deformation $\Upsilon$ such that
\begin{align}\label{eq:scaling_metric}
        h^{(2)}_{ab}  = \frac{\lambda^2}{M^4} \Upsilon_{ab}.
\end{align}
In practice, this means that once we computed a solution for $h^{(2)}$ and $\phi^{(1)}$
at a given coupling, we can scale them accordingly
to find a solution at a different coupling
provided we remain within the regime of validity of perturbation theory, which we describe
in the next section.

\subsubsection{Regime of validity}\label{sec:regime_validity}
Since this approach is perturbative, it will break down at some point.
There are at least two possible ways this
approach could break down. First, at some instant in time the
correction from the perturbative series could fail to be
small compared to the background solution. Second, even
if the instantaneous corrections are small, over longer times there could
be a secular drift between the
perturbative solution and true solution such that at some point 
the perturbative scheme is no longer valid.
Practically speaking, the salient question is whether there is some regime
where the corrections to a GR waveform, say from a binary black hole, are large
enough to have observational relevance, but still small enough for this approach
to accurately compute them.

\subsubsection*{Instantaneous regime of validity}\label{sec:inst_regime_validity}
A necessary condition for the validity of the perturbative scheme is that the expansions for 
the metric and scalar field
\eqref{eq:perturbative_expansion_fields} are convergent.
One can check for violations of this by comparing
successive terms in the series expansion.
Since we are only solving for the first order scalar field,
we cannot assess the convergence of $\phi$.
However, we can compare the magnitudes of $g_{ab}^{(0)}$
and $h_{ab}^{(2)}$.  Roughly, the series will converge provided
\begin{align}\label{eq:inst_val}
        || h_{ab}^{(2)}|| \lesssim ||g_{ab}^{(0)}||,
\end{align}
where $|| \cdot ||$ is an $L^2$ norm.
As discussed in the previous section, the magnitude of the metric
perturbation depends on the coupling constant via~\eqref{eq:scaling_metric}.
This means we can translate~\eqref{eq:inst_val} into a rough condition on the maximum value
allowed by $\lambda/M^2$ at any given time,
\begin{align}\label{eq:l_max}
        \left| \frac{\lambda}{M^2}\right|_{\rm{max}} \sim C
        \left( \frac{||g_{ab}^{(0)}||}{||\Upsilon_{ab}||} \right)_{\rm{min}}^{1/2}
        ,
\end{align}
where $C \lesssim 1$ is some factor, and the norm on right hand side is
evaluated pointwise on the computational domain.  As one approaches the merger
and nonlinearities become important, we expect increasingly tighter upper
bounds on the value of $\lambda$ compared to early times. We stress this is a
na\"ive estimate, as corrections may be small at any given time, but their
effects accumulate to a significant degree\footnote{A painful illustration,
Argentine daily inflation in 2023 was a ``small'' $0.2\%$ but accumulated to
$211\%$ in the year.}.

\subsubsection*{Secular regime of validity}\label{sec:sec_regime_validity}
In the order-by-order approach, the binary black hole's inspiral rate is
governed by the GR background. The GR background sources a scalar field, which
in turn sources the second order metric perturbation, but this perturbation
does not backreact on the background.  This means that the black holes'
trajectories are purely determined by the GR solution. When solving the full
equations, however, the black holes will inspiral faster than in GR due to the
energy loss to the scalar field.  Since we do not capture this correction to
the rate of inspiral, the second order solution is contaminated by secular
effects.  This is a generic feature of solving a system of equations
perturbatively~\cite{logan2013applied}, and in Appendix~\ref{app:anharm_osc} we
illustrate this using an anharmonic oscillator as a toy model.  This was
observed in earlier applications of this method to binary black hole inspirals
in dynamical Chern-Simons and EsGB gravity
\cite{Okounkova:2019zjf,Okounkova:2020rqw}.  In those works, the secular
breakdown of perturbation theory was observed in the amplitude of the second
order correction to the waveform by ramping on the coupling of the theory at
various starting times during the inspiral for the same background system.
Doing so, it was found that the amplitude of the correction to the metric at
merger was larger for simulation with earlier start times, and this secular
growth as a function of start times roughly followed a quadratic dependence 
(see, e.g. Fig. 4 of Ref.~\cite{Okounkova:2020rqw}). In an attempt to mitigate
secular effects from the inspiral, the coupling of the theory was thus turned
on as close to the merger as possible---thus ignoring beyond GR effects prior
to that point.
We emphasize that this set-up does not correctly capture a binary system in
EsGB gravity that started at an infinite time in the past.  In this study, we
apply the same procedure, but unlike in previous studies, we compare the
resulting waveform to solutions obtained by solving the full EsGB equations.
This gives us a way to quantify the error introduced by this scheme.

%=============================================================================
\subsection{Fixing equations approach\label{sec:fixing}}%
The \emph{fixing-the-equations approach} \cite{Cayuso:2017iqc} 
(also \cite{Allwright:2018rut,Cayuso:2020lca}), is
inspired by EFT arguments and related to strategies developed to fix issues in 
dissipative relativistic hydrodynamics \cite{Israel:1976a,Israel:1976tn}. The key strategy is to
adopt a prescription to dynamically control the high frequency behavior of a given theory, 
motivated by respecting the separation of scales already assumed in obtaining the original equations
in the first place. This is achieved by modifying, in a 
somewhat ad hoc way, the higher order contributions 
to the equations of motion 
at {\em short wavelengths}  
through the introduction of auxiliary massive fields. The auxiliary fields are evolved according to
driver equations that force them to approach the long wavelength behavior dictated by the original theory, 
while effectively dissipating shorter wavelength components.
As a consequence, modes with wavelengths above a certain scale fully backreact dynamically respecting the original
system of equations. If the system's dynamics does not display a significant UV cascade, the solution
is largely insensitive to the details of the driver prescription (e.g. Ref.~\cite{geroch}). 
Otherwise, it does\footnote{This sensitivity highlights that the theory under study,
in the particular case being consider, evolves outside the regime of applicability of the EFT.}, thereby providing a powerful
way to discern whether the scenario studied satisfies the assumptions made in designing the beyond GR
theory under study. 
This strategy has been successfully implemented in different models and 
spacetimes of physical interest 
\cite{Bezares:2021yek,Bezares:2021dma,Franchini:2022ukz,Lara:2021piy,Cayuso:2023aht}. In those
works, the particular structure of the equations being studied, or the
assumption of special symmetries, allowed for the implementation of this approach with a small number of fields.
In this work, we make no such assumptions and apply this approach in a generic manner  
by introducing {\em eleven} massive fields, denoted by $\mathbf{P}$. We replace the 
equations of motion~\cref{eq:eom_esgb_scalar,eq:eom_esgb_tensor} with the following fixed
system of equations,
\begin{align}
\label{eq:eom_esgb_scalar_fix}
	\Box\phi
	+
	\Pi^{(\phi)}
	&=
	0
   ,\\
\label{eq:eom_esgb_tensor_fix}
	R_{ab}
   -
	\frac{1}{2}g_{ab} R 
	-
   &\nabla_a\phi\nabla_b\phi
	+
   \frac{1}{2}\left(\nabla\phi\right)^2g_{ab} 
	+ \Pi^{(g)}_{ab}
	\\
    &= F_{ab}
	, \\
\label{eq:fixing_eqn}
	\sigma
        g^{ab}
	\partial_a \partial_b \mathbf{P}
	&=
	\partial_{0} \mathbf{P}
	+
	\kappa \left( \mathbf{P} -\mathbf{S} \right),		
\end{align}
where $F_{ab}$ is some function of the MGH constraint that vanishes
when the constraint is satisfied,
 $\mathbf{P}= \left(\Pi^{(\phi)},\Pi^{(g)}_{ab}\right)$ encompasses the 
metric and scalar auxiliary fields, and
$\kappa$ and $\sigma$ are constants that determine the timescale over which 
the auxiliary fields
are driven to their corresponding desired values $\mathbf{S}$ (see Appendix~\ref{app:fixing} for the
derivation of this term).
These parameters 
also determine the length scale below which
short wavelength features are effectively damped away. 

Associated with these parameters we define two timescales
\begin{equation}
	t_1 \equiv 2\sigma, \ \ \ t_2 \equiv \sqrt{\frac{\sigma}{\kappa}}
    \ , 
\end{equation}
which combine to give the slowest\footnote{There is also a faster timescale that 
can be relevant for numerical applications, as it has to be suitably resolved when employing
an explicit time integrator.} timescale for
the variable $\mathbf{P}$ to decay to its target solution $\mathbf{S}$ 
, given by $T^{-1} = -t_1^{-1} + \sqrt{-t_2^{-2} + t_1^{-2}}$.
We stress here that there are many ways one could drive the 
fixing variables $\mathbf{P}$ to 
their true solution $\mathbf{S}$. From an EFT point of view, subleading terms to
the equations of EsGB would come with one higher power of curvature, thus two additional
derivatives of the gravitational field. Thus, we find a wave
operator prescription to be natural in light of this observation and the nature of the Einstein equations. Nevertheless,
this is just a matter of choice if a strong UV cascade does not take place.
 
We note that a non-trivial departure from a stationary solution satisfying $\mathbf{P}= \mathbf{S}$ 
can be obtained for a non-vanishing value 
of $\sigma$ (due to contributions from $\sigma g^{\mu \nu} \partial_\mu \partial_\nu \mathbf{P}$). 
However, by making this parameter sufficiently small, one can minimize the 
difference\footnote{Alternative choices can be made so that this is not an issue,
e.g., by adding to the source a suitable contribution of $\partial_\mu \partial_\nu \mathbf{S}$, considering
a different type of driver equation, etc. We do not concern ourselves with such options in this work.} between 
$\mathbf{P}$ and $\mathbf{S}$. We refer the reader to Appendix~\ref{app:fixing},
where we provide further details on the numerical implementation of the fixed system of
equations and explain how different choices of $\kappa$ and $\sigma$
will affect the solution.

Naturally, the solution to this system of equations will depend on the value of
the ad hoc timescales, though if these are made sufficiently small compared to
the other physical scales, and there is no significant cascade towards
the UV, we expect this dependency to be subleading.
We will thus check how our solutions depend on the fixing parameters and their impact on
physical observables and contrast it with the full solution obtained though the MGH formulation
 of shift-symmetric EsGB gravity.
We note that the equations of motion for EsGB gravity can only be stably 
evolved in time for
weakly-coupled solutions \cite{Ripley:2019irj,Kovacs:2020pns,Kovacs:2020ywu,
East:2020hgw,Ripley:2022cdh}. That is, when the Gauss-Bonnet
corrections to the spacetime geometry remain sufficiently small compared to the 
smallest curvature
length scale in the solution. 
We will also use the fixing-the-equations approach 
to briefly explore regimes of the parameter space 
where the MGH formulation breaks down.

%=============================================================================
\section{Numerical implementation\label{sec:numerics}}%
We implement and evolve the order-by-order approach and fixed system of equations
with the same numerical infrastructure as the one we use to evolve the full EsGB equations of motion 
\cite{East:2020hgw,East:2011aa}. In Appendix~\ref{app:implementation}, 
we derive the explicit forms of the equations we solve in each
formalism.
The gauge we use
is the modified (by the auxiliary metric) version of the
damped harmonic gauge (which includes the special case of $H^c = 0$) 
\cite{Lindblom:2009tu,Choptuik:2009ww}
which has been found to
work well for a large number of highly dynamical
spacetimes. We sometimes also fix $H^c$ to be constant in time for
cases where we wish to maintain Kerr-Schild like coordinates.
We provide details on the numerical resolution and convergence of solution in
Appendix~\ref{app:convergence}.

For initial data, we do not implement a method to solve for the Hamiltonian and 
momentum constraint equations for general $\phi$, but instead choose
$\phi=\partial_t \phi = 0$, in which case the constraint equations of EsGB are the
same as in GR (see Ref.~\cite{East:2020hgw})\footnote{
See Ref.~\cite{Brady:2023dgu} for a methods for 
solving the constraint equations
for EsGB gravity with non-trivial $\phi$ using a modified version of the
Conformal-Transverse-Traceless method.}.
We thus consider single or binary vacuum GR black hole
solutions. The latter are
constructed either using the conformal thin sandwich method described in
Ref.~\cite{East:2012zn} when performing head-on collisions, or using the black hole
puncture method with the \texttt{TwoPunctures}
\cite{Ansorg:2004ds,Paschalidis:2013oya,code_repo_tp} 
code when performing quasi-circular inspirals.
For the scalar field, we slowly ramp on the coupling of the theory as described
in Appendix B of Ref.~\cite{Okounkova:2019zjf} over a period of time (typically $\sim 100M$,
with $M$  the total ADM mass of the spacetime)
shorter than the time to merge.
We first focus on the scalarization of isolated non-spinning black holes for different
coupling parameters $\lambda$. We study couplings where
one expects the approximate treatments to do well, but
also consider parameters approaching, and in
one case exceeding, the maximum values where the hyperbolicity of the theory 
breaks down, i.e. $\lambda/M^2 \gtrsim 0.23$. 
We then study equal mass head-on collisions with an initial 
separation of $75M$.
Here again, we consider a range of coupling parameters, and determine if and when
the approximate methods break down.
We finish by using our methods to evolve quasi-circular inspirals for a system
with parameters consistent with GW150914 \cite{LIGOScientific:2016vlm}.
While there is a range of mass and spin parameters consistent with GW150914,
we choose to use the same mass parameters as the one used in 
Fig.1 of the GW150914 detection
paper \cite{LIGOScientific:2016aoc} and set the spins to zero.
The system has initial masses $m_1= 0.5497M$ and $m_2 = 0.4502M$, 
giving a mass ratio of $q=1.221$. 
We choose an initial separation of $9.8M$ such that, in GR, the black holes merge
after $\sim 6$ orbits (roughly $1600 \,m_2$). Here we consider couplings of 
$\lambda/m_2^2=\{ 0,0.02,0.04,0.1 \}$.
See, e.g., Eq. (14) in Ref.~\cite{Corman:2022xqg} for a comparison of our coupling convention to other works.

We use many of the same diagnostics introduced in Refs.~\cite{East:2020hgw,Corman:2022xqg},
which we briefly review here. To determine the 
scalar and gravitational radiation,
we compute the Newman-Penrose scalar $\Psi_4$ and scalar field $\phi$
on finite-radius coordinate spheres (up to $r = 100M$) and
decompose these quantities into spin $-2$ and $0$ 
weighted spherical harmonics. We do not concern ourselves with potentially subtle
effects due to slightly different gauges/physical extraction radii at this 
time~\cite{Lehner:2007ip}. Instead, we rely
on the fact that at such distances, their effect, as well as corrections to GR, would
be significantly smaller than the physical waveforms measured. 

During the evolution, we track any apparent horizons (AH) present at a given time,
and measure their corresponding areas and angular momentum $J_{\rm BH}$.
From this, we compute a black hole mass $M_{\rm BH}$
via the Christodoulou formula \cite{Christodoulou:1970wf}.
We measure the average value of the
scalar field on the black hole apparent horizons and compute the scalar charge
from the asymptotic value of the scalar field at a large radius $r = 100M$:
$\phi = Q/r + \mathcal{O}\left(1/r^2\right)$, where our initial conditions
fix $\phi \rightarrow 0$ at spatial infinity.
When evolving the spacetime using the order-by-order approach, we also
compute an approximation of the instantaneous regime of validity of the method by evaluating
\cref{eq:l_max}.
Additionally, in the order-by-order approach $\Psi_4$ can also be expanded
about the GR solution as,
\begin{equation}
   \label{eq:perturbative_psi4}
      \Psi_{4}
           = \Psi_4^{(0)} +
      \sum_{k=1}^{\infty}\epsilon^k \Psi_{4}^{(k)}
      .
\end{equation}
Substituting the expanded metric \cref{eq:perturbative_expansion_fields} into expression
for $\Psi_4$ (see Appendix C of Ref.~\cite{Okounkova:2019dfo}), one finds that the leading-order
correction to gravitational radiation will be linear in the leading-order correction
to the spacetime metric. In practice, this means we can scale out the $\lambda$
dependence and write,
\begin{equation}
   \label{eq:psi4_dimless}
	\Psi_{4}^{(2)} =  \left(\frac{\lambda}{M^2}\right)^2 \Delta \Psi_4
    ,
\end{equation}
such that a given solution can be scaled to 
predict the value of $\Psi_4$ for any coupling, provided it is within the
regime of validity of the perturbative expansion.
We often find it useful to quote the relative errors between the solution we
obtain solving the full system of equations and the order-by-order or fixing
system. We denote these quantities by $e$, e.g. for the scalar charge we have
\begin{equation}
e_Q \equiv \frac{Q_{\rm Full} - Q_{\rm ORA/Fix}}{Q_{\rm Full}}.
\end{equation}

%=============================================================================
\section{Results\label{sec:results}}%
We provide a detailed comparison of the three different approaches used to
study extensions of GR in the strong-field regime below, but summarize our main
findings first. We considered three systems: the dynamical formation of scalar
hair around non-spinning black holes in axisymmetry, head-on equal mass black
hole collisions, and quasi-circular binary black hole mergers.  In general, we
find that the order-by-order approach is accurate for sufficiently small
couplings, but breaks down as soon as the deviations become large enough to be
of observational relevance.  As expected, the order-by-order solution is
subject to secular effects in the amplitude of the waveform in quasi-circular
inspirals and we conclude that this approach cannot be used in the way it has
been implemented to place useful constraints on beyond GR theories.  The
fixing-the-equations approach, however, is able to track the scalarization of
the black hole as well as solutions for head-on collisions, even at large
couplings, provided we choose the fixing parameters judiciously.  When applying
the fixing-the-equations approach to the inspiral and merger of binary black
holes, we were unable to obtain faithful solutions, but we infer that one would
be able to do this using shorter fixing timescales, together with higher
resolution, than we consider here. We next study each system in detail.

%=============================================================================
\subsection{Black hole scalarization and saturation}\label{sec:single_bh}
We first present results for the case of a single, non-spinning, stationary black hole 
in axisymmetry. We set initial data and evolve the spacetime in Kerr-Schild
coordinates using the three methods outlined in Sec.~\ref{sec:method}.
Convergence results can be found in Appendix~\ref{app:convergence}.
Our main conclusion in this section is that for small enough couplings, the order-by-order 
approach accurately reproduces the formation of scalar hairy black hole solutions,
but nonlinear effects become important for couplings below the threshold for which
perturbation theory breaks down. The fixing-the-equations approach, however, is (after
extrapolating the fixing parameters) able to track the scalarization of the
black hole even at large couplings when the fixing parameters are chosen 
such that their associated length scale
is at or below the black hole size.

\begin{figure*}
    \includegraphics[width=0.99\columnwidth,draft=false]{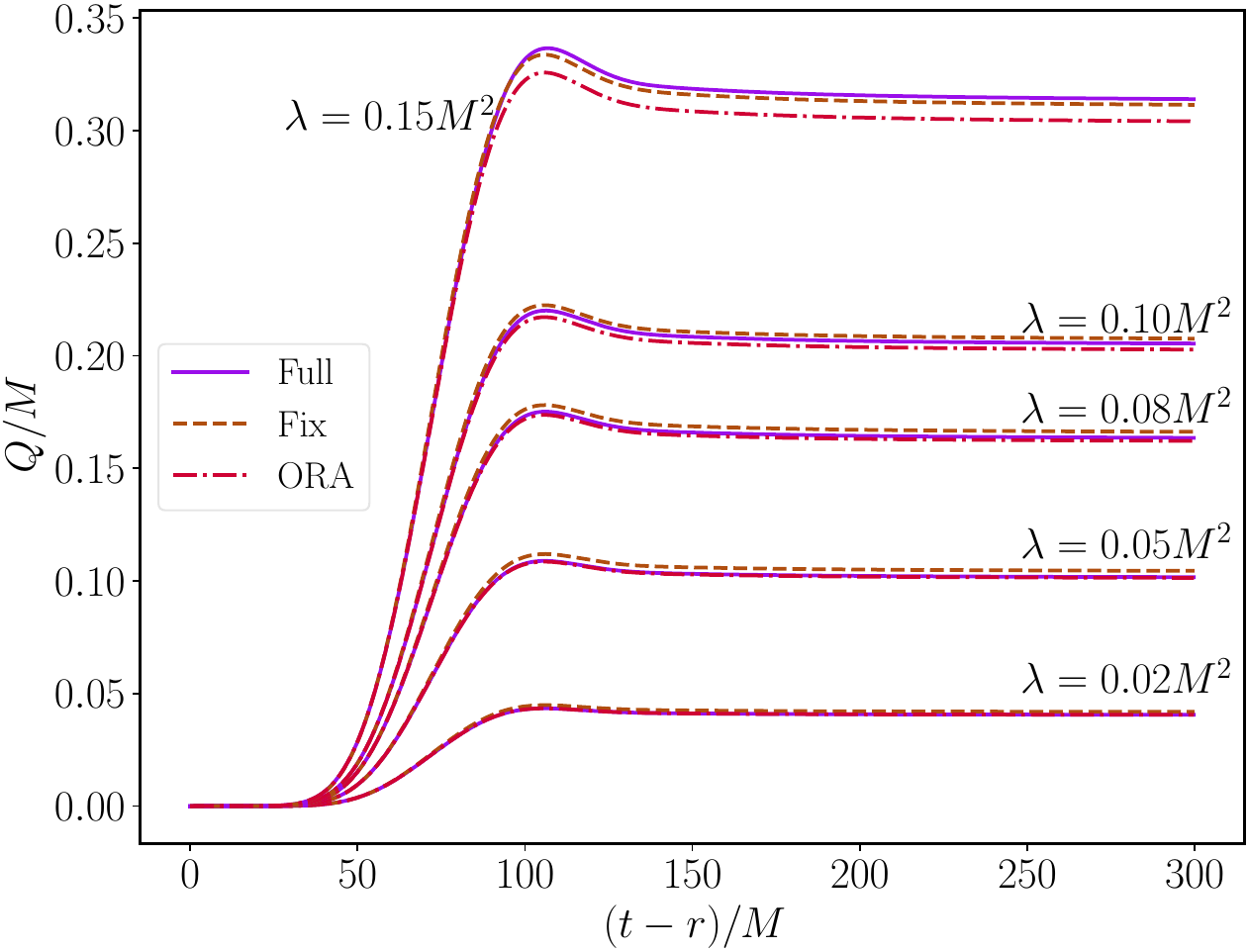}
      \includegraphics[width=0.99\columnwidth,draft=false]{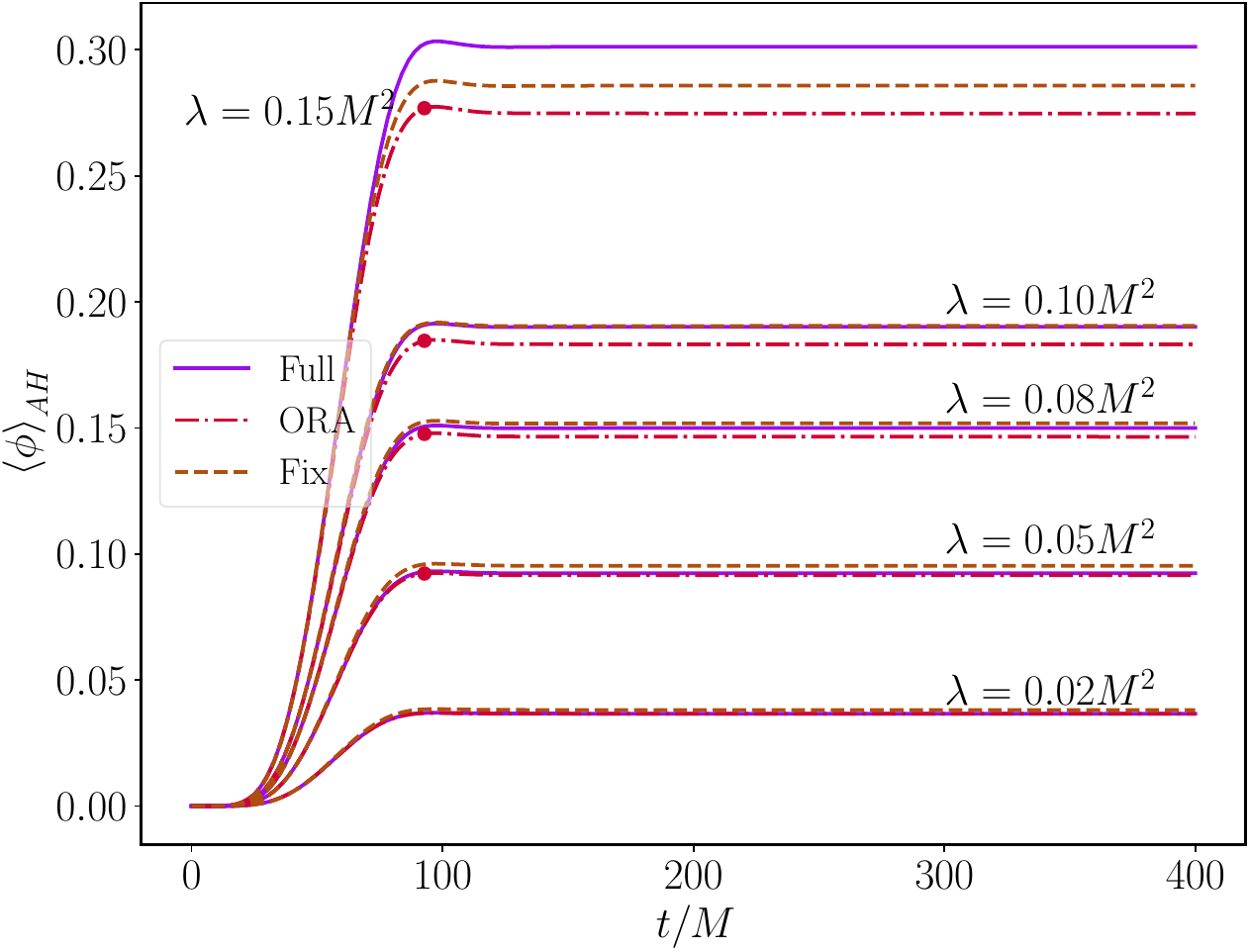}
	\caption{Left: Scalar charge measured at a large distance of $r =100M$ as 
	a function of look-back time, $(t-r)/M$, 
	for different couplings and the three different
	approaches. 
	The solid purple curves correspond to the full solutions, the dashed brown
	curves to the solutions obtained by solving the fixed system of 
	equations, and the curves labeled `ORA' to the solutions solved through
	an order-by-order approach.
	Right:  Average value of the scalar field on the black hole horizon as a function
	of time $t/M$.
\label{fig:bh_all_couplings}
}
\end{figure*}

To illustrate these conclusions, in Fig.~\ref{fig:bh_all_couplings} we plot the scalar field value averaged over the
black hole horizon $\left<\phi\right>_{AH}$ as a function of time (right) 
along with the scalar charge $Q/M$
(left), as a function of retarded time $(t-r)/M$,
for different values of coupling parameter 
$\lambda/M^2=\{0.02,0.05,0.08,0.1,0.15\}$ and the three different methods.

In the order-by-order approach, the scalar field  
scales linearly with the coupling [see \cref{eq:scaling_phi}]. Comparing these curves
to the values of the scalar field when solving the full equations of motions, we find
that this approximation is valid for small couplings with relative errors
of 
$\left( e_Q, e_{\left<\phi\right>_{\rm AH}} \right)= \left( 0.05\%, 0.14\% \right)$ 
for a coupling of 
$\lambda = 0.02 M^2$ when computed at late times. However the validity of approximation
quickly worsens with errors of 
$\left( e_Q, e_{\left<\phi\right>{\rm AH}} \right) = \left( 0.8\%,2.3\% \right)$
and $\left( 3.1\%,8.8\% \right)$
for couplings of $\lambda = 0.08 M^2$ and $0.15$ respectively. 
Note that the relative error is larger for
the value of the scalar field on the apparent horizon, as expected, since this
is where the deviation from GR is the largest. 
Recall from Sec.~\ref{sec:inst_regime_validity} that the order-by-order scheme
requires the modifications to the spacetime to be a convergent perturbative series 
around GR, which can be turned into a constraint on the coupling of the theory
of the form given by~\cref{eq:l_max}.  
Evaluating ~\cref{eq:l_max} on each numerical slice, using the $g_{xx}$ metric component,
we find 
that $\lambda/M^2 \lesssim 0.9 C$, where the minimum 
comes from the strong-field region outside the apparent horizon of the black hole
when the scalar field is growing the fastest, roughly before it settles down to
its lower stationary value (indicated by a red dot in Fig.~\ref{fig:bh_all_couplings}).

\begin{figure*}
    \includegraphics[width=0.99\columnwidth,draft=false]{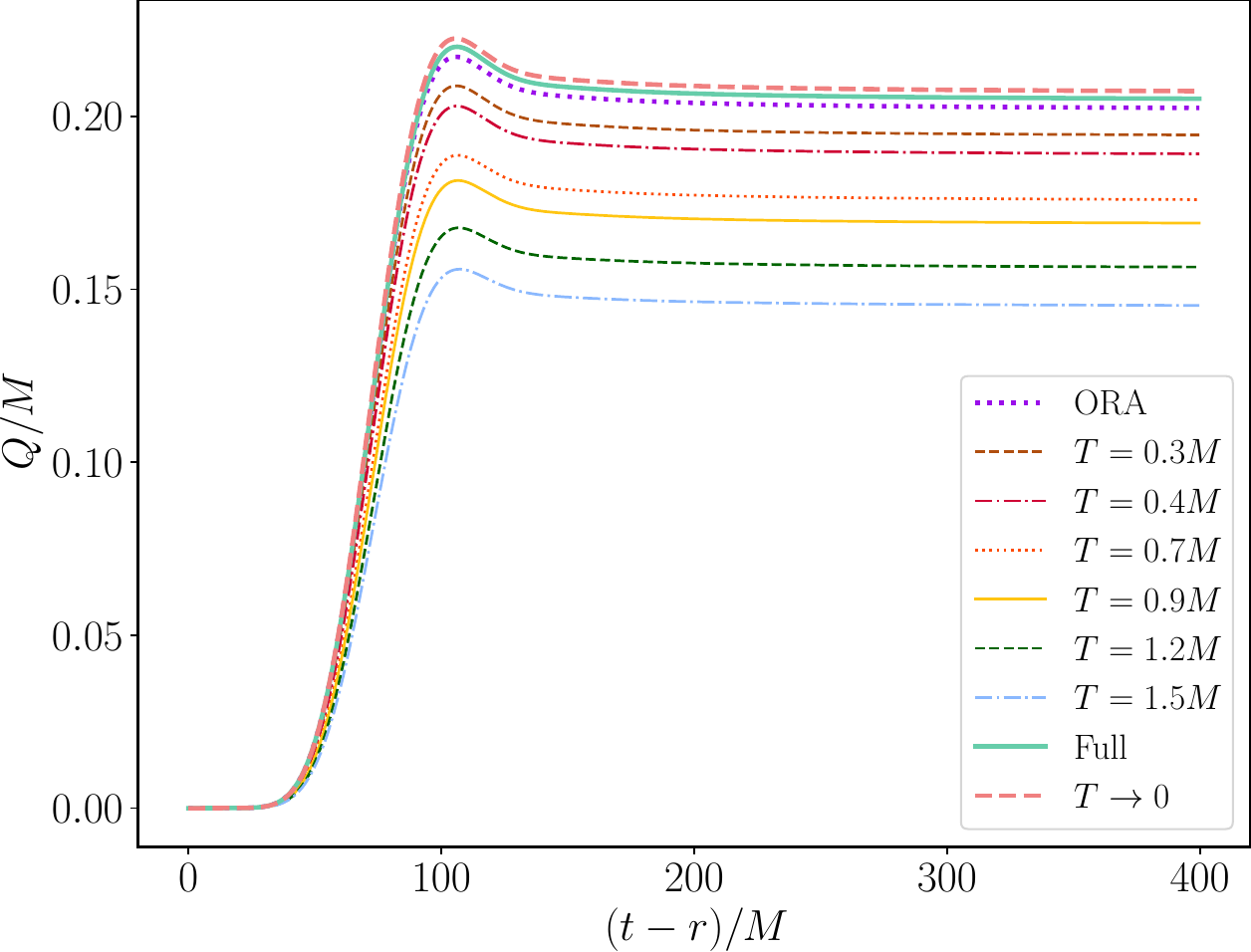}
      \includegraphics[width=0.99\columnwidth,draft=false]{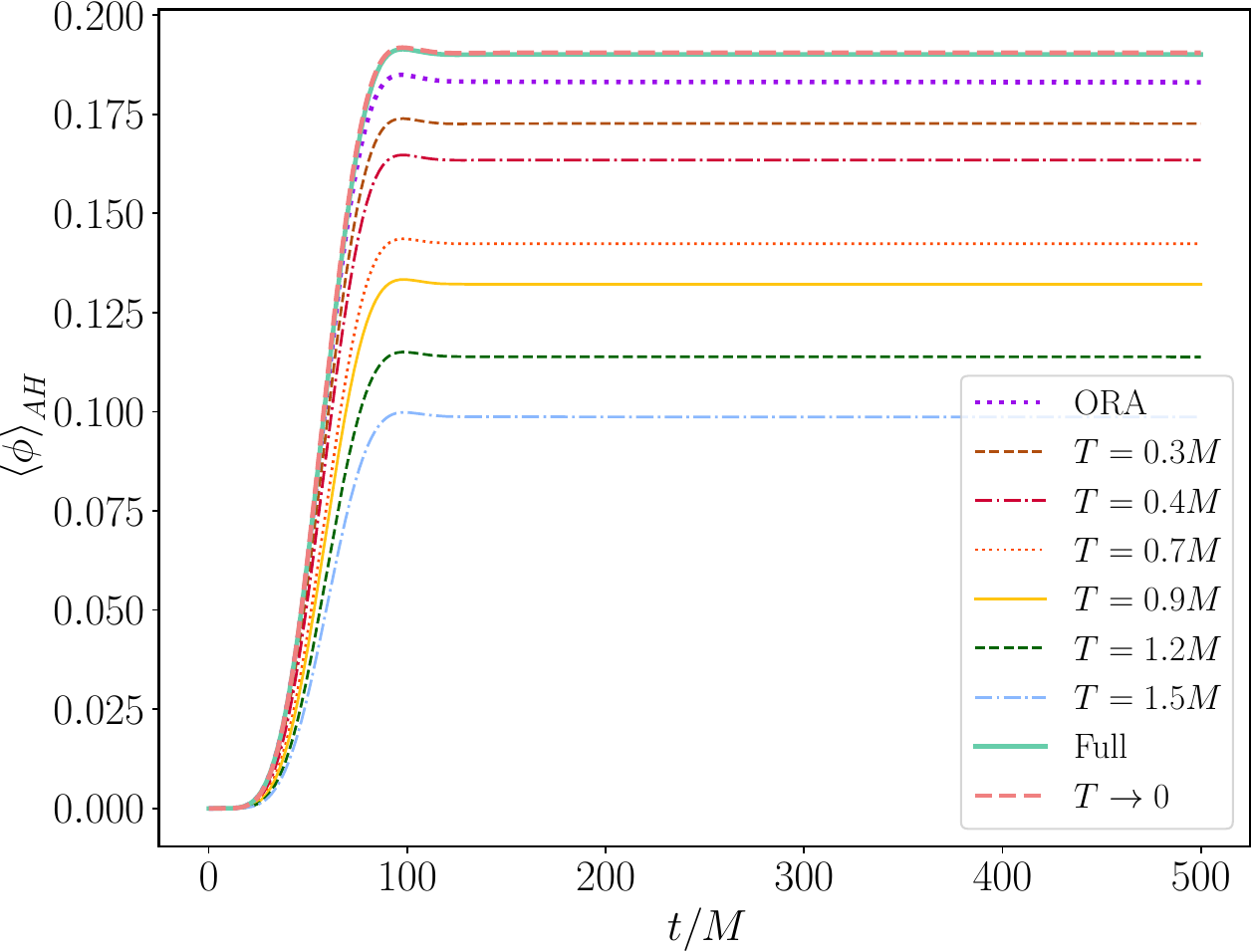}
	\caption{Left: Scalar charge measured at a large distance of $r =100M$ as
	a function of look-back time, $(t-r)/M$, for a coupling of $\lambda=0.1M^2$ and
	for the set of fixing parameters used to extrapolate the fixed solution
	to $T \rightarrow 0$, where 
	$T$ refers to 
	the slowest decay rate of the solution to the simplified driver equation.
	We also show the full solution and order-by-order solution for comparison.
        Right:  The equivalent of the left panel, but for 
	average value of the scalar field on the black hole horizon.
\label{fig:bh_l0p1}
}
\end{figure*}

Figure~\ref{fig:bh_all_couplings} also shows the corresponding 
values of the scalar charge and 
scalar field averaged on the black hole horizon when evolving the fixed system
of equations. These results were obtained
with different 
$t_1/M=\{0.2,0.3,0.5,0.6,0.8,1.0\}$ and varying $t_2$ so that $t_1^2=0.9t_2^2$.
We then fit the
scalar field and scalar charge values. This is
shown in Fig.~\ref{fig:bh_l0p1} for a coupling of
$\lambda/M^2=0.1$. For this fit, we
assume the solutions take the form
$F=F_0+F_1 T$ where $T$ is the slowest decay rate of the solution to
the simplified driver equation with a given set of fixing parameters~\cref{eq:toy_fix_eqn}
\footnote{A similar fit was performed in Ref.~\cite{Cayuso:2023aht}.}.

The values we compare to the full solutions are obtained by extrapolating the
fits to $T \rightarrow 0$ and are shown in Fig.~\ref{fig:bh_all_couplings}.
We find that the disagreement with the full solution is
$\left( e_Q, e_{\left<\phi\right>_{\rm AH}} \right)= \left( 3.2\%,4\% \right)$
for a coupling of $\lambda = 0.02 M^2$, and decreases to values of
$\left( 0.2\%,1\% \right)$ for a coupling of $\lambda=0.1M^2$, yet then increases
for larger couplings with an error of $\left( 0.8\%,5\% \right)$ for 
a coupling value of $0.15$.
We attribute the larger errors at weak couplings to extrapolation rather than numerical
errors. For instance, for a coupling of $\lambda=0.1M^2$ the extrapolation error on the
scalar charge (obtained
by comparing the solution assuming a linear and quadratic fit) is $0.6\%$, whereas
the truncation error is $0.02\%$.

This is also confirmed by the observation
that for a coupling of $0.05$, the error in the extrapolated solution halves
when solving the system for smaller values of
$t_1/M=\{0.16,0.18,0.2,0.3,0.5,0.6\}$ 
and assuming a quadratic fit in $T$, $F=F_0+F_1 T + F_2 T^2$.

%=============================================================================
\subsection{Beyond hyperbolicity break-down}\label{sec:single_bh}
Here, we consider a case where $\lambda/M^2 = 0.25$. We recall that the regularity of 
black hole solutions and the hyperbolicity of the theory requires 
$\lambda/M^2 \leq 0.23$ for non-spinning black holes 
\cite{Ripley:2019aqj,Sotiriou:2013qea}. 
Thus, the full system can not be evolved in
time however small the time step of interest, as the problem is ill-posed.
The order-by-order approach, with its passive accounting for the role of correcting
terms, will be unaffected by this breakdown. At best, the impact may be surmised by accounting
for sufficiently high orders in the perturbation scheme and, potentially, resumming their effects.

The fixing-the-equations approach, on the other hand, can regularize the problem
and allow for stable evolutions, provided
the ad-hoc scales are {\em larger} than the scales that trigger the breakdown of hyperbolicity, or 
allow for evolutions until some time where an instability takes over, if the scales
are smaller. 
As shown with the simple model described in Appendix~\ref{app:fixingmodel}, fixing a
potentially elliptic equation results in a dispersion relation that captures exponentially
growing modes at long wavelengths, but controls them at shorter ones, leading to
plane wave modes at high frequencies. The scale at which the modes are controlled is
governed by the damping parameters.
As expected, we also find here that the solution depends on the ad-hoc parameters, as
seen in Ref.~\cite{Franchini:2022ukz}. While this feature can be seen as artificial, it allows
for exploring the behavior of a given system (at least locally), and potentially provides
a solution which could be regarded as approximating that of a suitable UV completion of the theory.

\begin{figure}
	\includegraphics[width=0.99\columnwidth,draft=false]{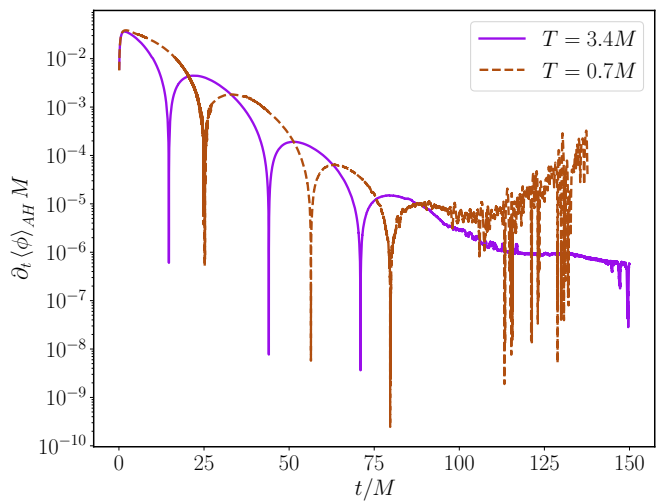}
	\caption{Behavior of the time derivative of the
	scalar field integrated on the horizon of an isolated black hole for cases with  
    a coupling value of $\lambda/M^2=0.25$ and $T=0.7$ and $3.4M$ in the fixing-the-equations approach.
    For the short time scale case,
	exponentially growing modes associated with the breakdown of hyperbolicity in the original problem
    are still allowed above the frequency cutoff, driving an instability that cause the simulation to fail.
	On the other hand, for the long time scale case, such modes are ``aggressively'' controlled to
	the point of removing the unstable behavior. 
\label{fig:too_high_single}
}
\end{figure}

As an illustration, we show the value of the time derivative of 
scalar field integrated over 
the apparent horizon, normalized by its area, for the 
case $\lambda/M^2=0.25$ in Fig.~\ref{fig:too_high_single}. 
We consider two cases with the following fixing parameters: 
 (i) $T=0.7M$ 
with $\{t_1,t_2\}=\{0.1, 0.2\}M$ and (ii) $T=3.4M$ with $\{t_1,t_2\}=\{1,1.4\}M$.
For case (i), the solution is followed until $t\approx 140M$, at which time the code crashes due
to an unstable evolution. This is a result of the low frequency behavior describing
exponentially growing modes which are not being effectively controlled by the fixing terms. Such growing
modes result in the apparent horizon eventually being lost, at which point the calculation cannot be continued.
On the other hand, for case (ii), a well behaved
evolution is obtained, running past $1000M$ with no signs of instability. Here, the damping parameters
introduce a maximum frequency cut-off and both stability and convergence are achieved. Of course,
in this case the chosen parameters define a scale larger than that of the black hole. Such a choice essentially
amounts to forcing an UV completion, and the behavior that is obtained is then highly dependent on the fixing
schema and parameters. The salient point of this exercise is to highlight how the fixing-the-equations approach captures
complex behavior of corrections to GR consistently at long wavelengths and provides a way to examine their
phenomenology under controlled circumstances. 

%=============================================================================
\subsection{Head on collisions of scalarized black holes}\label{sec:head_on}
We next study non-spinning, binary black hole mergers in EsGB gravity, restricting to the axisymmetric
case of a head-on collision first. This allows us to explore the impact of different
parameters, including a range of coupling values, as well as fixing
parameters, when applying the corresponding approach. 
Since we are mainly interested in comparing the three approaches to evolving the
system, we only consider equal mass collisions for simplicity. We label the cases considered
in terms of the dimensionless coupling $\lambda/m^2$, where $m$ is the 
mass of either black hole in the system. Our results in this section are consistent
with those observed in single black hole spacetimes, namely the order-by-order approach
does well at small couplings but ceases to be a good approximation even for couplings
below which the perturbative approximation breaks down as estimated from the ``instantaneous regime of validity.''
On the other hand, the 
fixing-the-equations approach does well even for moderately large couplings.
As discussed in Sec.~\ref{sec:numerics}, we start with initial data where
$\phi$ and $\partial_t \phi$ are identically zero, but slowly ramp on the scalar field
over a period of $100M$. We choose the initial separation of the black holes to
be $75M$, and set their initial velocities to the values corresponding to the binary
being marginally bound at infinite separation.
The black holes merge around $315M$, which is sufficiently 
long for the individual black holes to develop their scalar hair well prior to merger.

As a rough indication of the instantaneous regime of validity of the
order-by-order approach that would be inferred from the size of the correction to the metric, 
we note that when evaluating~\cref{eq:l_max} for the head-on collision presented in this study,
we find that $\lambda/m^2 \lesssim 0.2 C$, where the minimum value comes at merger when the
spacetime is most highly perturbed.

In addition to the instantaneous regime of validity, our results will also
have a secular regime of validity for the order reduced approach. This also arises from the perturbative
nature of the approach and was discussed in Sec.~\ref{sec:sec_regime_validity}.
In Fig.~\ref{fig:bhbh_axi_ora}, we show the leading order EsGB correction to the
gravitational waveform $\Delta \Psi_{4,20}$ (after scaling out the $\lambda$ dependence,
see \cref{eq:psi4_dimless})
for a simulation with the same GR background, but different start times for turning
on the EsGB coupling (hence different times over which the secular growth can
accumulate). We find that the longer the inspiral
length $L_{\rm insp}= t_{\rm peak} - t_{\rm start} - t_{\rm ramp}$, the larger the amplitude
of correction to the gravitational waveform at merger. (Here
 $t_{\rm peak}$
is the time at which the GR solution for $r M \Psi_{4,20}^{(0)}$ peaks, 
$t_{\rm start}$ is the time at which we start turning on the EsGB coupling,
and $t_{\rm ramp}$ is the time over which the coupling is ramped on.)
However, for the particular system we study, 
we find that 
secular effects are small in comparison to the correction to the GR waveform at any given
coupling we consider.
We attribute this to the fact that, in axisymmetry, for the coupling values 
and the initial separation we
consider, there is not much time for
the full solution to undergo significant dephasing compared to the GR solution.
This will no longer be true in the case of binary black hole inspirals,
which we discuss in Sec.~\ref{sec:inspirals}.

\begin{figure}
      \includegraphics[width=\columnwidth,draft=false]{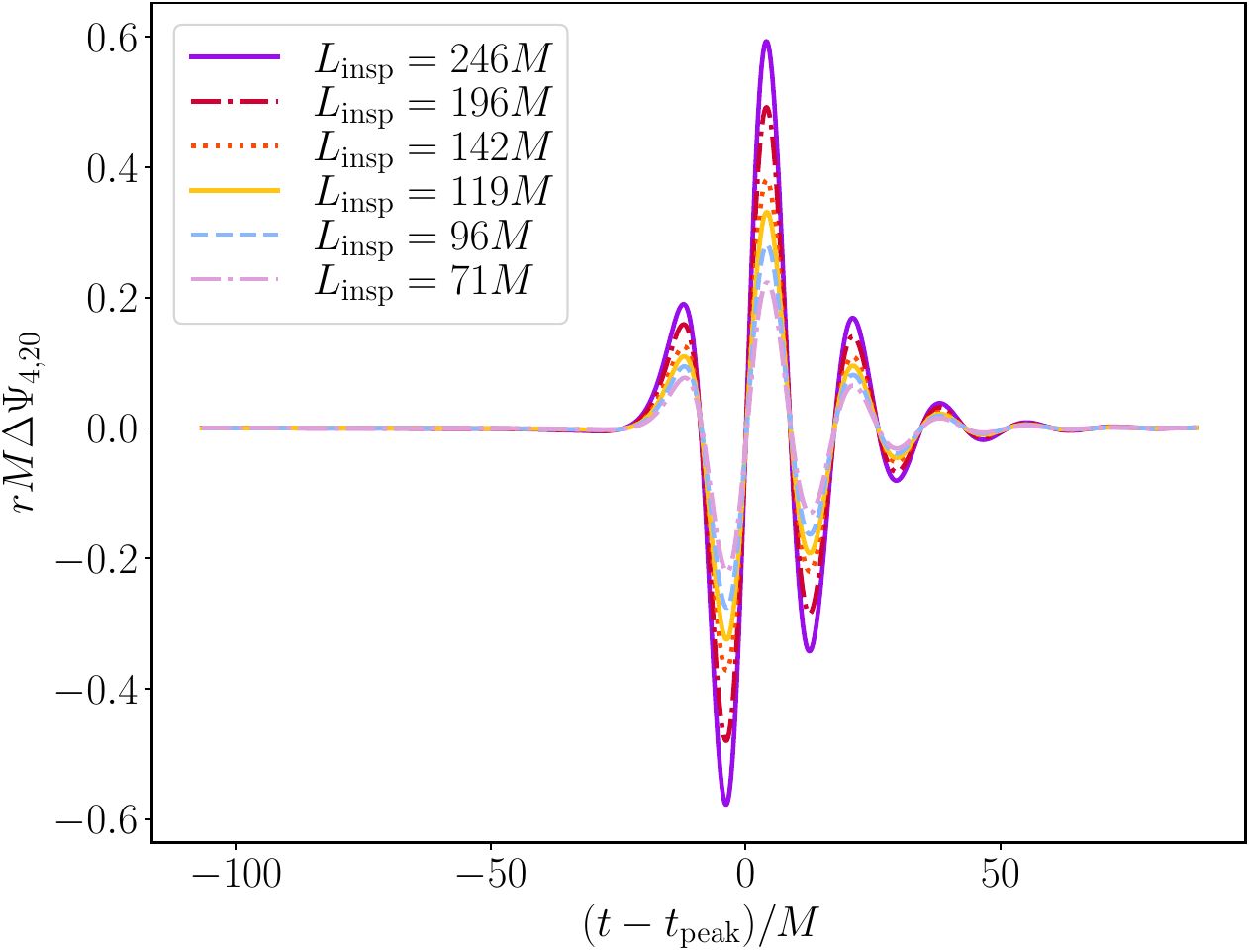}
	\caption{Leading-order correction to gravitational waveform as a 
	function of time measured with respect to 
	time at which the GR solution for $r M \Psi_{4,20}^{(0)}$ peaks,$t_{\rm peak}$.
	Each curve corresponds to a simulation
	with a different start time for turning on the coupling of theory as discussed
	in Sec.~\ref{sec:sec_regime_validity}.
	This shows secular growth
	in the leading-order gravitational waveform as a function
	of the inspiral length of waveform 
	$L_{\rm insp}= t_{\rm peak} - t_{\rm start} - t_{\rm ramp}$. 
\label{fig:bhbh_axi_ora}
}
\end{figure}

In the remainder of the figures in this section,
we therefore show waveforms with an inspiral length of $L_{\rm insp} = 246M$, which 
corresponds to turning on the coupling at $t_{\rm start} = 0$.
Figure~\ref{fig:bhbh_axi_ora} shows the gravitational waveforms for coupling values of
$\lambda = 0.04 m^2$ (left) and $0.15$ (right) as well as GR (solid grey lines)
for the three different approaches. 
The waveforms in the fixing-the-equations approach were obtained using the following combination of
fixing parameters: $\{t_1,t_2\}=\{0.4,0.6\}m$.
We find that for a coupling value of $\lambda = 0.04 m^2$, all three approaches agree 
relatively well. The amplitude of the full solution peaks $0.17M$
before the GR solution, corresponding to a difference of $0.05\%$, 
and the fixed and 
order-by-order solution agree within $15\%$ and $4\%$ with the full solution.
The peak 
amplitude of the full solution increases by $0.24\%$ compared to GR while 
the peak amplitude of the fixed and order-by-order solution increase by $0.32\%$
and $0.39\%$ respectively compared to GR.

For a relatively large
coupling of $\lambda = 0.15 m^2$,
we find that the amplitude of the full solution peaks $2.4M$ or $0.7\%$
before the GR solution and the fixed and
order-by-order solution agree within $24\%$ and $12\%$ with the full solution.
The peak of the amplitude of the full solution increases by $3.7\%$ compared to GR
and the fixed and order-by-order solution increase by $4.1\%$ and $40.6\%$ respectively.
The very large error in the amplitude of $\Psi_{4,20}$ in the order-by-order approach
is not surprising given we found that the perturbative approximation breaks down for
coupling values larger than $\sim 0.04$, yet it is interesting to note that the error
in time of peak remains small. This is consistent with the intuition we gained from 
studying the anharmonic oscillator in Appendix~\ref{app:anharm_osc}.

\begin{figure*}
    \includegraphics[width=\columnwidth,draft=false]{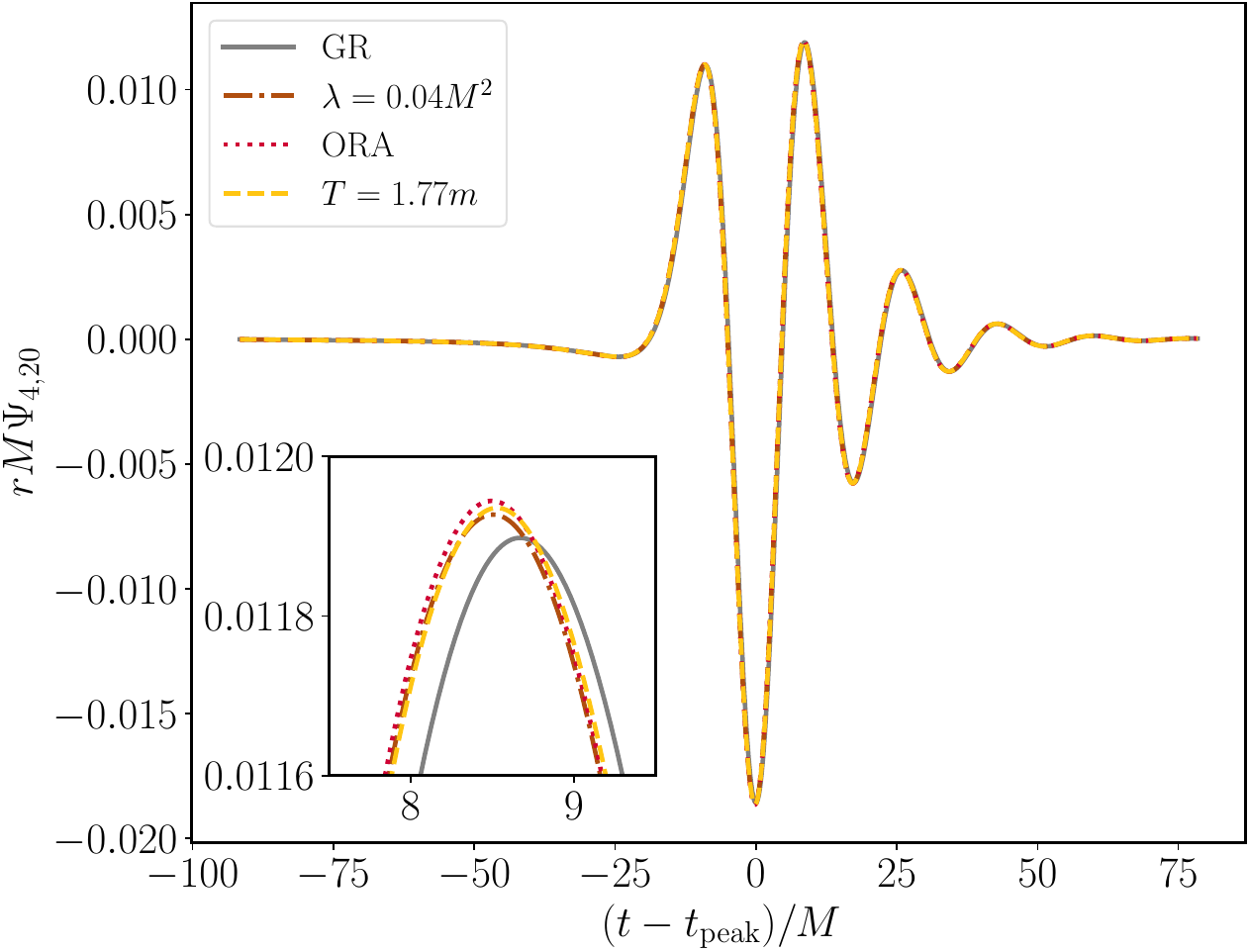}
      \includegraphics[width=\columnwidth,draft=false]{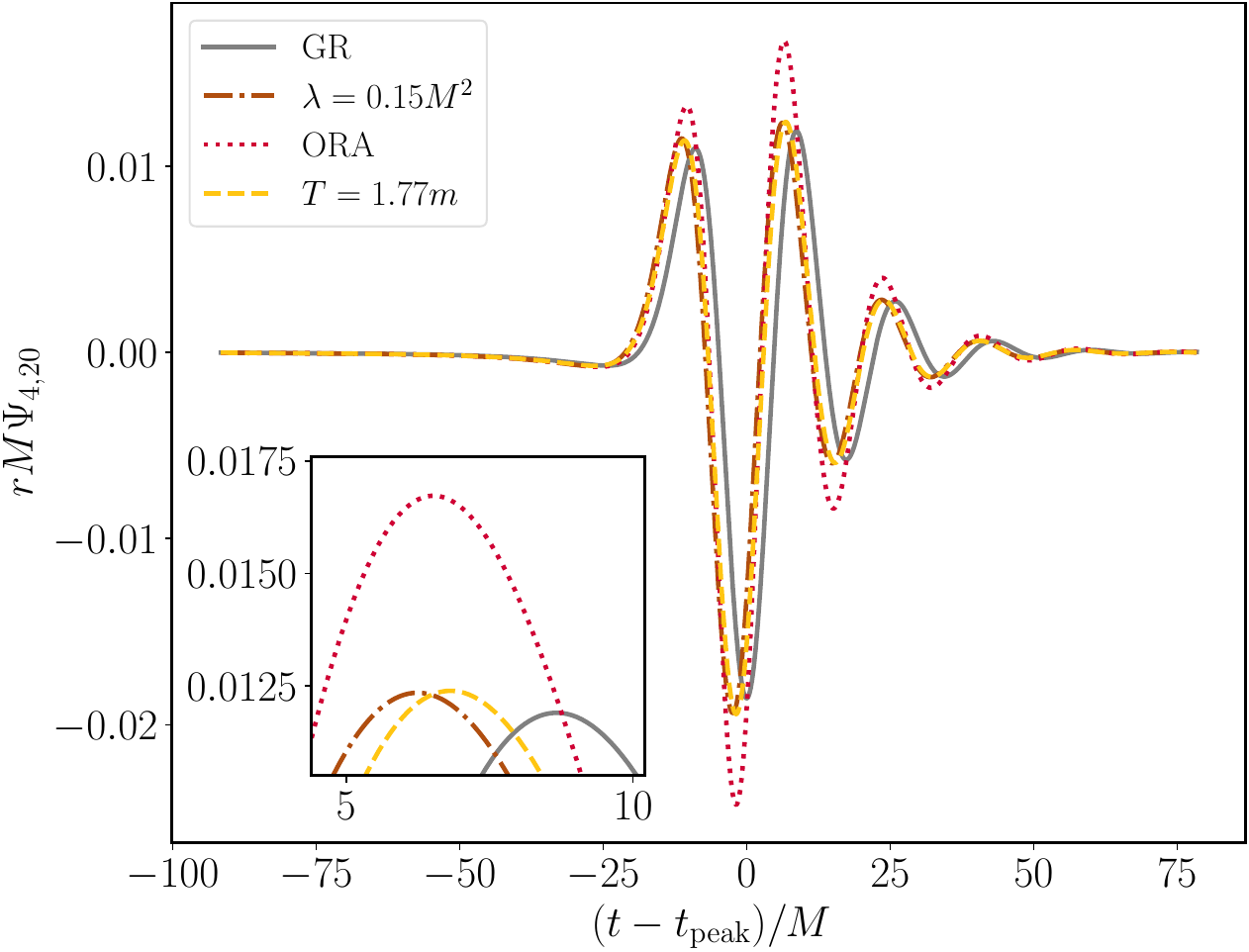}
	\caption{Gravitational radiation from the head-on collision of an equal mass binary
	with $\lambda=0.04m^2$ (left) and $0.15$ (right)
	in the three approaches. The waveforms in the fixing-the-equations approach were obtained
    with fixing parameter values of $\{t_1,t_2\}=\{0.4,0.6\}m$.
	All waveforms were obtained by turning on the coupling of theory at zero
	coordinate time. The value $\lambda_{\rm max}/m^2=0.038$ corresponds to the
	largest allowed value for the perturbation scheme to be valid when assuming
	$C=0.2$.
\label{fig:bhbh_axi_ora}
}
\end{figure*}

\begin{figure}
    \includegraphics[width=\columnwidth,draft=false]{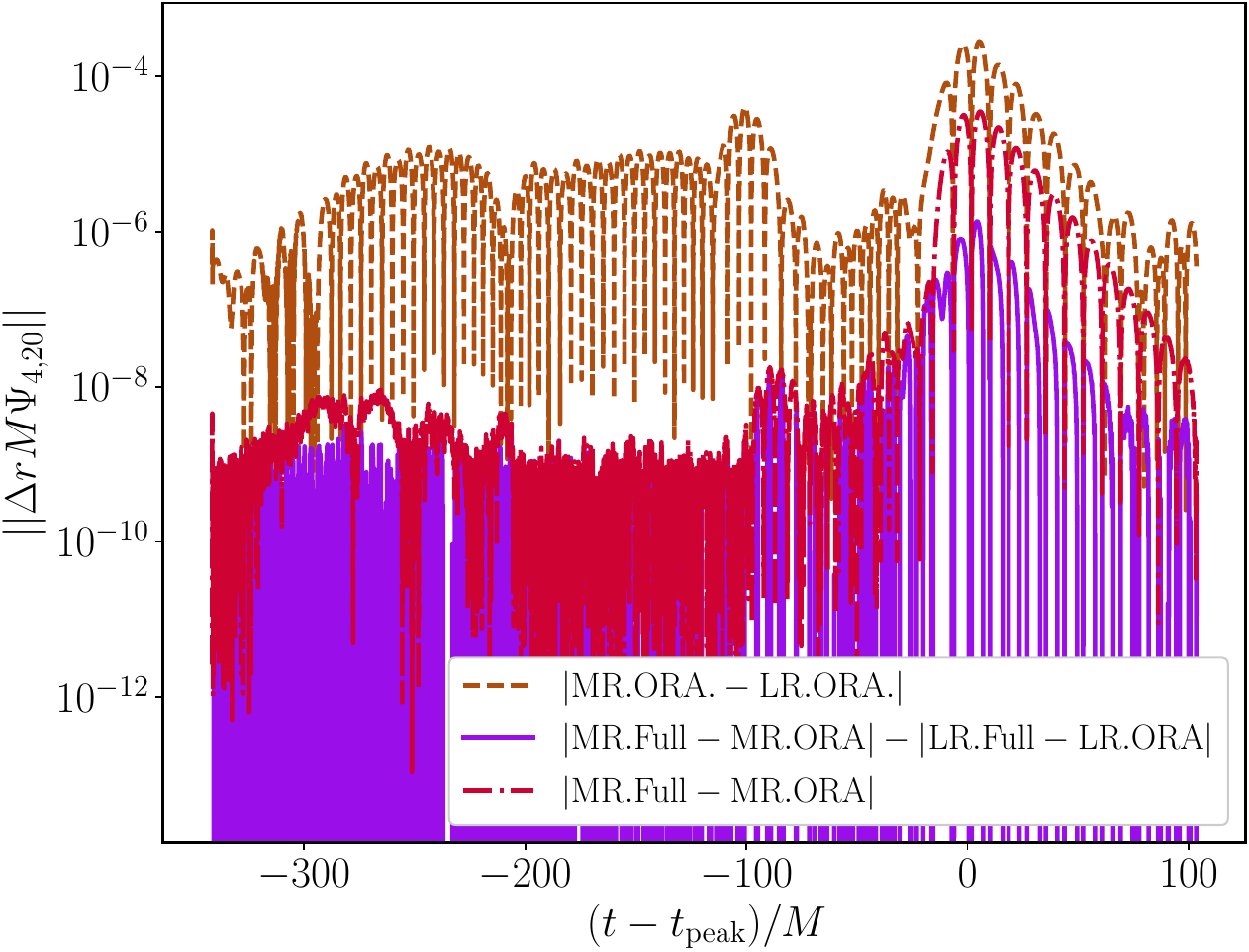}
    \caption{We show the difference between the low and medium resolution (respectively labelled ``LR" and ``MR") of the 
	amplitude of the gravitational waveform for non-spinning equal mass head-on
	collision with coupling of $\lambda=0.04m^2$
	when solving system using the order-by-order approach (dashed brown
	curve). This is larger than the difference between the order-by-order and full
	solution at the medium resolution (red dash-dotted curve). However
	we also show the difference of the difference between the
	order-by-order and full GW amplitude at low and medium resolutions 
	(solid purple curve). This
	provides evidence that the truncation error roughly cancels between the order
	by order and full runs.
\label{fig:bhbh_axi_cnv_ora}
}
\end{figure}

Restricting to the $\lambda=0.04m^2$ run for
the order-by-order approach, we compare the difference in the amplitude of the waveform
with the fully nonlinear solution to the numerical errors in the simulation.
In Fig.~\ref{fig:bhbh_axi_cnv_ora}, we show that, although the numerical 
error in the GW amplitude (dashed curve)
is larger than the difference between the full and order-by-order solution
(dash-dotted curve), the latter is larger than the difference between the full
and order-by-order solutions at two resolutions. This arises because the
dominant truncation error in our simulations does not depend strongly on the
approach used or the coupling and thus partially cancels out when looking at
the difference in the GW amplitude between the full and order-by-order solution
using the same resolution.

\begin{figure*}
     \includegraphics[width=0.99\columnwidth,draft=false]{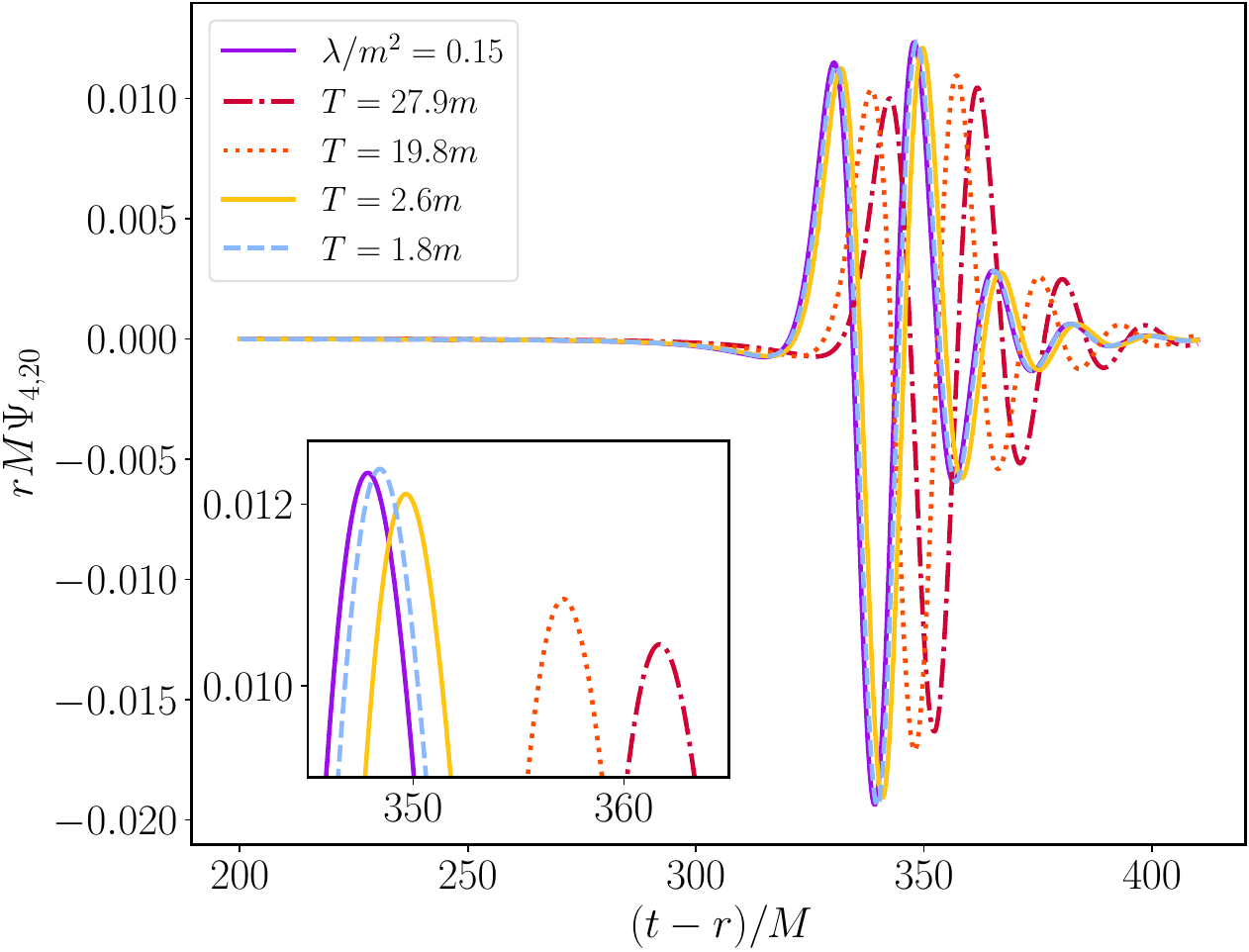}
     \includegraphics[width=0.99\columnwidth,draft=false]{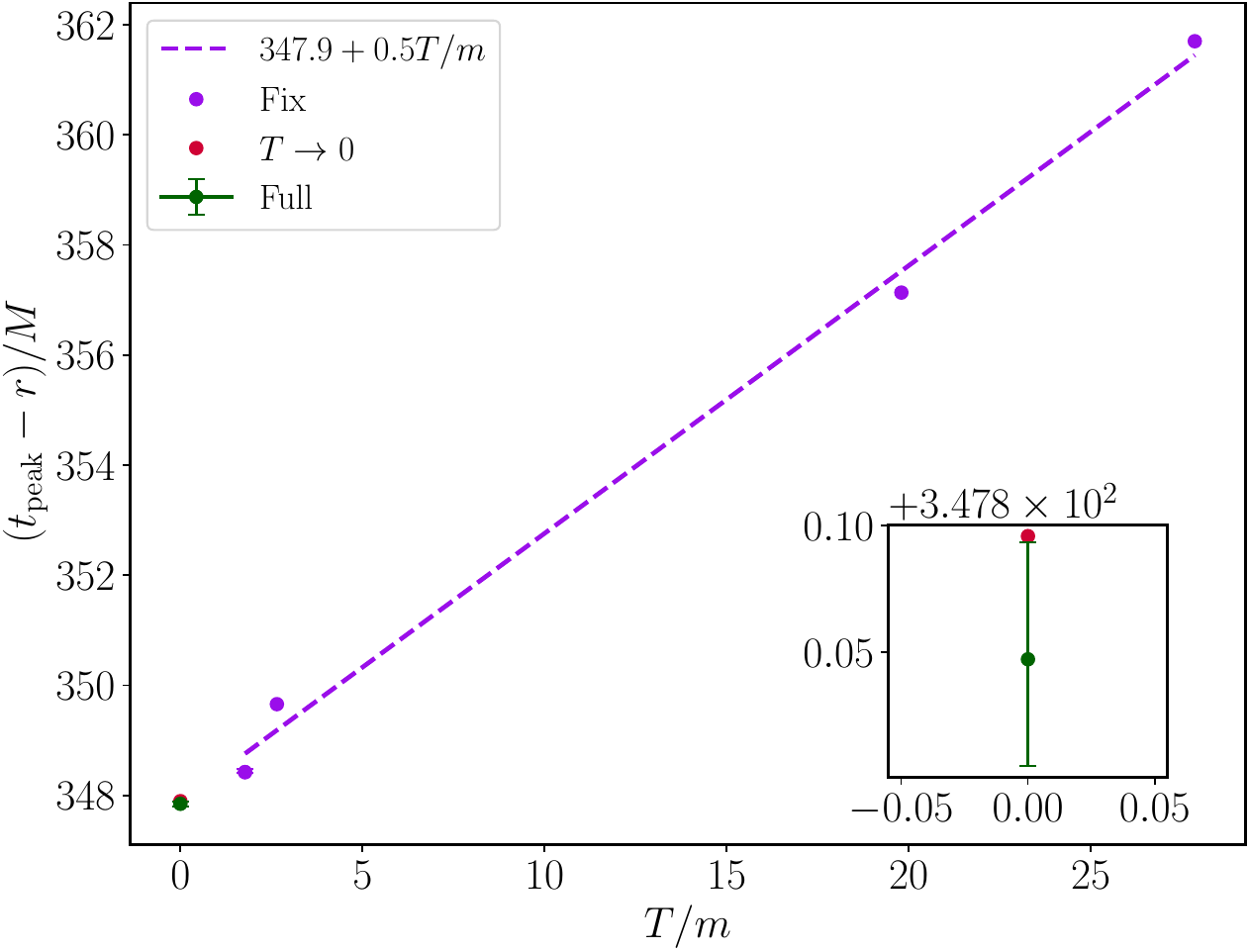}
	\caption{Left: Gravitational waveforms for a coupling value of $\lambda=0.15m^2$ 
	when solving the full system of equations (solid purple curve) and for four
	different combinations of fixing parameters with different decay rates $T$. Right:
	Time at which amplitude of waveforms peak and corresponding linear fit. The 
	error bars represent estimates for the truncation error of full solution.
\label{fig:bhbh_axi_fix}
}
\end{figure*}

\begin{figure}
     \includegraphics[width=\columnwidth,draft=false]{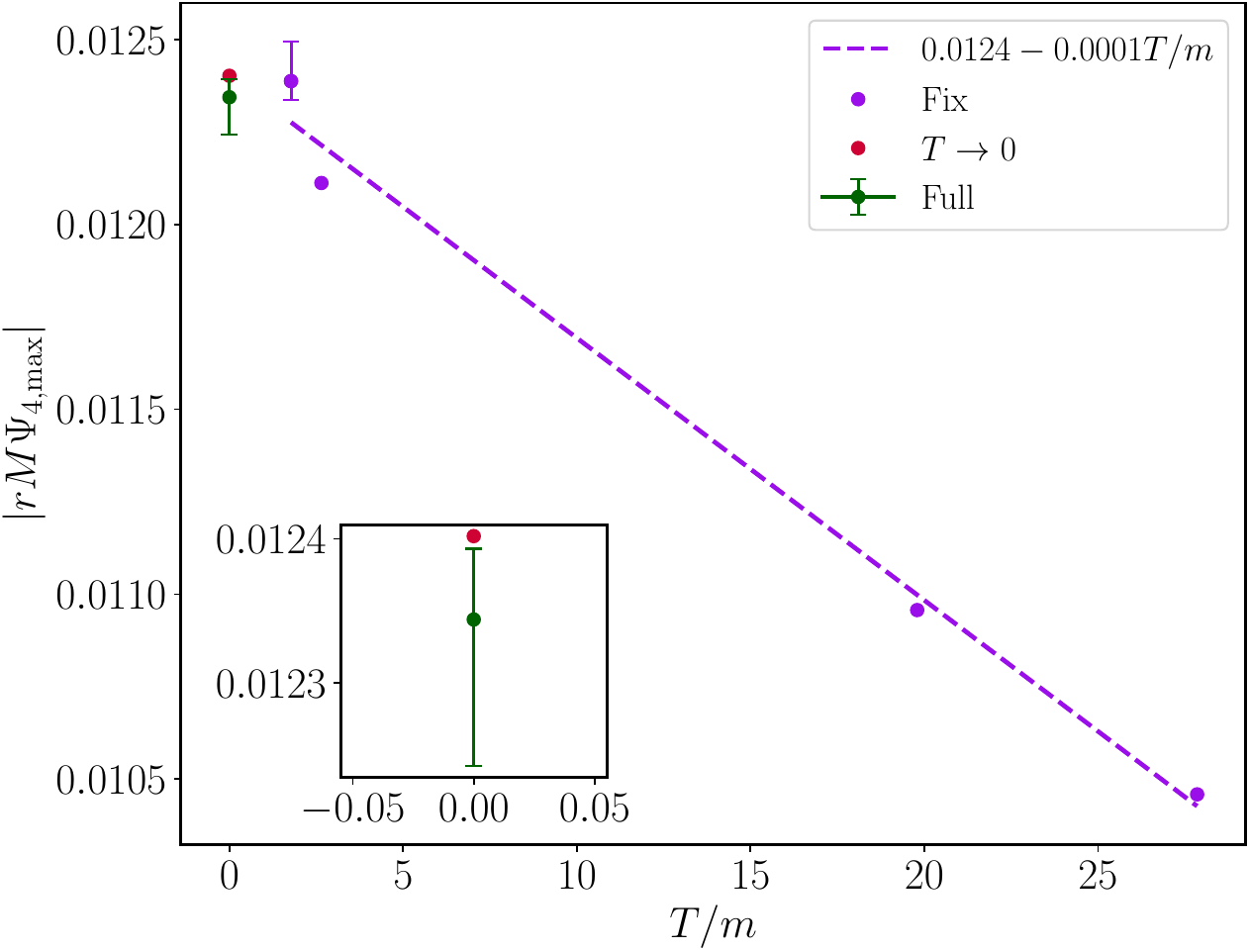}
	\caption{Peak amplitude of gravitational signal 
	for a coupling value of $\lambda=0.15m^2$
        when solving the fixed system of equations for four
        different combinations of fixing parameters with different decay rates $T$
	and corresponding linear fit.
\label{fig:bhbh_amp_fix}
}
\end{figure}

The waveforms obtained with the fixing-the-equations approach in Fig.~\ref{fig:bhbh_axi_ora} correspond
to a single combination of fixing parameters. However, in a similar fashion
to the single black hole case, one can evolve the system for a combination of parameters and 
extrapolate how the solution would behave as $T \rightarrow 0$.
The left panel of Fig.~\ref{fig:bhbh_axi_fix} shows the gravitational waveforms 
when evolving the
fixed system of equation assuming a coupling value of $\lambda=0.15m^2$ and
using the following parameters 
for $t_1=\{0.4,0.3\}m$ while keeping $t_2=0.6m$ fixed, 
as well as letting $t_1=\{0.4,0.3\}m$ but keeping $t_2=2m$ fixed.
In other words, the waveforms have a
slowest decay rate of the solution to
the simplified driver equation~\cref{eq:toy_fix_eqn} of
$T = \{1.8,2.6,19.8,27.9\}m$. Fitting the amplitude and the time at peak amplitude
for a behavior $F=F_0+F_1 T$, as shown Figs.~\ref{fig:bhbh_axi_fix} (right panel) and
\ref{fig:bhbh_amp_fix}, we
then extract what the values would be as $T \rightarrow 0$, obtaining a relative error
of $\left(e_{t_{\rm peak}},e_{|\Psi_{4}|} \right)= (0.014,0.47) \%$.

%=============================================================================
\subsection{Quasi-circular inspirals of scalarized black holes}\label{sec:inspirals}
\begin{figure*}
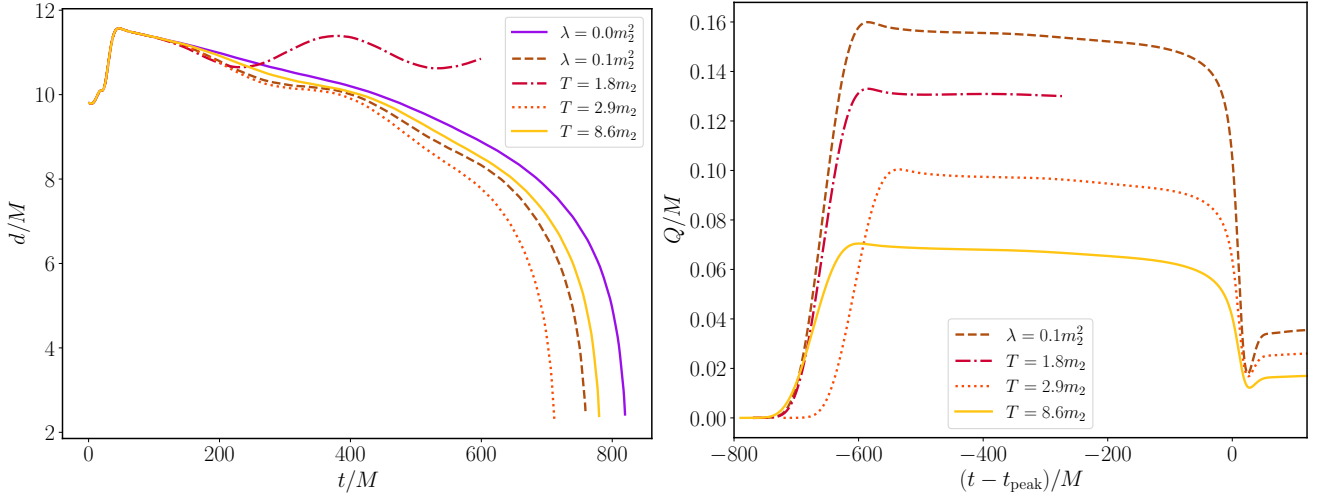

	\includegraphics[width=0.99\columnwidth,draft=false]{{trajectories}.pdf}
	\includegraphics[width=0.99\columnwidth,draft=false]{{Qsf}.pdf}
	\caption{(Left) 
	Coordinate separation between black holes of quasi-circular inspiral
	of GW150914 like binary for a coupling of $\lambda=0.1 m_2^2$ when solving
	the full equations and fixed system of equations with different fixing
	parameters. We also show the coordinate separation of the black holes in GR
	for comparison. (Right) Scalar charge measured at $100M$ as a function of time with
	respect to time when amplitude of
	gravitational waveform is maximum, $t_{\rm peak}$,
	for the solutions shown in left panel, with the exception of the 
	fixed solution with slowest decay rate of $T=1.8m_2$.
\label{fig:coor_sep}
}
\end{figure*}

%=============================================================================
Finally, we consider the quasi-circular inspiral and merger of binary black holes.
Here we focus on one GW150914-like system consisting
of two non-spinning black holes with masses $m_2=0.4502 M$ and $m_1=0.5497 M$, giving
a mass ratio of 1.22. We consider an initial
separation of $9.8 M$ corresponding to a system that undergoes $\sim 6$ orbital periods before merging in GR.
We solve the full equations for coupling values of $\lambda/m_2^2=\{0,0.02,0.04,0.1 \}$
and turn on the coupling at $t_{\rm start} = 0$ but over a period of $200M$.
For the largest coupling, we also solve the system using the fixing equations approach.
We first tried fixing parameters 
that were found to give an accurate representation of the full
solution when studying head-on collisions, 
$\{t_1,t_2\}=\{0.75,0.9\}m_2$
but found that the black holes
stopped inspiralling into each other shortly after the black holes acquired their 
individual charges (see Fig.~\ref{fig:coor_sep}). 
We attribute this behavior to insufficient numerical resolution to: (i) resolve the 
small values of $t_1$ and $t_2$ (which
we recall, determine the final value of the scalar charge that the black holes acquire) and
(ii) resolve the faster timescale in the decay defined by the fixed equations 
which is $\approx T/4$.
Though the grid spacing on the finest adaptive mesh refinement level is $dx = 0.016M$,
for points at a distance of $\approx10M$ from both black holes, the minimum linear grid spacing
becomes comparable to $t_1/5$, so this scale is barely resolved\footnote{For comparison,
this is roughly $3$ times coarser than the resolution employed in the axisymmetric cases
studied.}.
We thus instead consider two sets of fixing parameters with longer fixing timescales than 
the abovementioned run, namely 
$\{t_1,t_2\}=\{2.2,2.3\}m_2$ and $\{t_1,t_2\}=\{5.6,6.0\}m_2$.
When performing the order-by-order approach,
as explained in Sec.~\ref{sec:sec_regime_validity}, because of secular growth during
the inspiral, we turn on the coupling of the theory as close to merger as possible, by
first evolving the binary black hole system in GR, and then ramping on the source terms for
the scalar field and metric perturbation at some later time $t_{\rm start}$.

\begin{figure}
	\includegraphics[width=\columnwidth,draft=false]{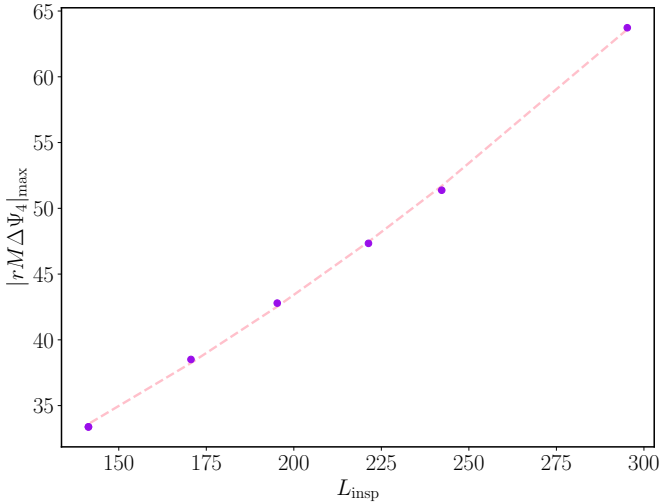}
	\caption{Peak amplitude of the EsGB correction to the gravitational waveform 
	as a function of inspiral length for quasi-circular binary inspiral in the order-by-order approach. 
\label{fig:bhbh_3d_sec}
}
\end{figure}

We performed six simulations with different start times for
turning on the EsGB correction in the order-by-order approach. In Fig.~\ref{fig:bhbh_3d_sec},
we show the peak amplitude of the gravitational waveform as a function of inspiral length
$L_{\rm insp}= t_{\rm peak} - t_{\rm start} - t_{\rm ramp}$. We find that the secular
growth, as reflected in the peak amplitude, roughly
behaves quadratically with the inspiral length
(see Fig.~7 of Ref.~\cite{Okounkova:2019zjf} and Fig.~4 of Ref.~\cite{Okounkova:2020rqw}).
In a similar fashion to Refs.~\cite{Okounkova:2019zjf,Okounkova:2020rqw}, which considered
evolutions starting 
close to merger time in the hope of reducing the amount of secular growth, 
we choose the simulation with the closest start time to merger (unless stated otherwise), 
corresponding to an inspiral length of
$\sim 140 M$, as our representative waveform for the order by
order approach. 
Of course, this is an artificial set-up which ignores the difference in dynamics prior to such
time with respect to GR. 
We only adopt this approach here to compare with such works.

\begin{figure*}
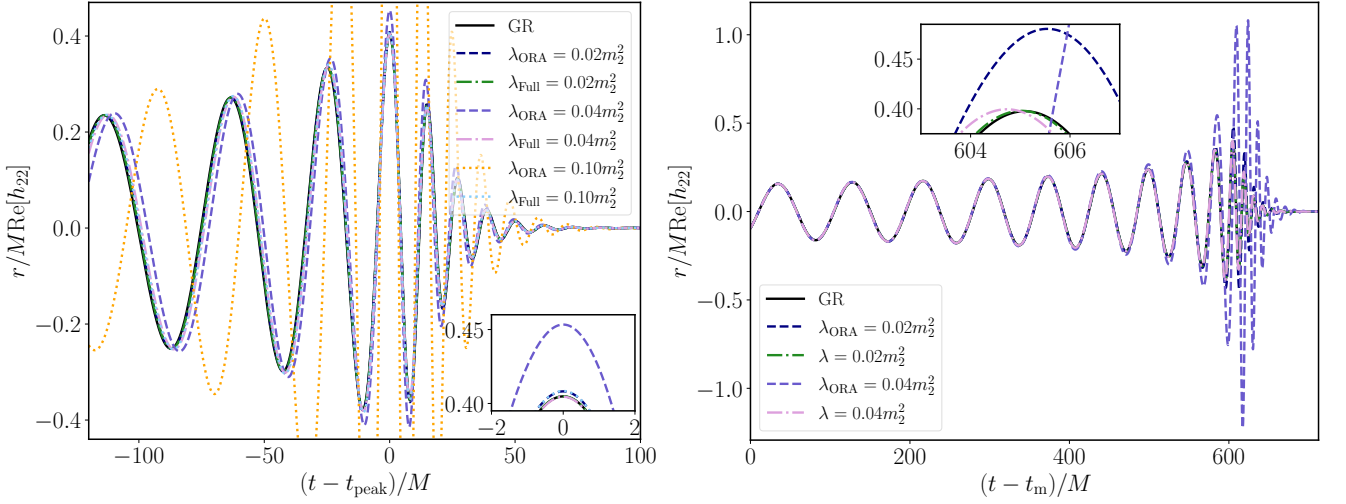

	\includegraphics[width=\columnwidth,draft=false]{{strain_3d_merger_ora}.pdf}
	\includegraphics[width=\columnwidth,draft=false]{{h_plus_ora}.pdf}
	\caption{
        Gravitational wave radiation in the order-by-order approach and when
        solving the full equations for coupling values of $\lambda =
        \{0.02,0.04,0.1\} m_2^2$ as well as GR ($\lambda =0$).  We show the
        real part of the $\ell = m = 2$ spherical harmonic of the GW strain
        $h$.  Left: The waveforms are aligned in phase and time at peak
        amplitude and the inspiral length of the ORA waveform is $\sim 140 M$.
        Right: The waveforms are aligned in phase and time at a frequency of
        $\omega_m M = 0.064$ and the inspiral length of the ORA waveform is
	$\sim 600 M$.  The $\lambda = 0.1 m_2^2$ case is not
        included for easier visualization of the better behaved solutions. 
\label{fig:bhbh_3d_ora}
}
\end{figure*}

As mentioned in Sec.~\ref{sec:regime_validity}, besides the secular breakdown,
there is also an instantaneous regime of validity. For the waveform with 
inspiral length of $\sim 140 M$, evaluating~\cref{eq:l_max} gives 
$\lambda/m_2^2 \lesssim 0.13 C$ with the minimum occurring at merger, where the strong-field effects are
the largest.
In the left panel of Fig.~\ref{fig:bhbh_3d_ora}, 
we compare the real part of the $\ell=m=2$ harmonic of the GW  
strain\footnote{The strain $h\equiv h_{+} + i h_{\times}$ 
is related to $\Psi_4$ through $\Psi_4 = \ddot{h}$. 
We numerically integrate
$\Psi_4$ using fixed frequency integration \cite{Reisswig:2010di}.} $h_{22}$ of
our representative order-by-order waveform to the gravitational
wave strain one obtains when solving the full system of equations 
for coupling values of $\lambda = \{0.02,0.04,0.1\} m_2^2$. 
We align the waveforms in time and phase at the peak of the complex amplitude of the waveform.
In the right panel of Fig.~\ref{fig:bhbh_3d_ora}, we again compare the real
part of the $\ell=m=2$ harmonic of the GW
strain one obtains when solving the full system to the order-by-order waveform
for coupling values of $\lambda = \{0.02,0.04\} m_2^2$ (where we omitted 0.1 for clarity),
but we consider order-by-order solution obtained when turning on the scalar field at
$t_{\rm start}=0$ to illustrate the effect of a longer inspiral. We align the waveforms in time
and phase at some fiducial frequency $\omega_m M = 0.064$ early on in the inspiral.
Note that the coupling values are already approaching the regime where estimates 
based on the size of the metric contribution [e.g.~\cref{eq:l_max}] suggest 
the perturbative scheme in the order-by-order approach may be breaking down;
Though, on the other hand, we find that the deviations from GR for values below
these values are very small, and hence not very useful to assess
the validity of the order-by-order approach.
To that end, we also show the waveform assuming GR for comparison.
We find that both the order-by-order and full waveform have a phase shift relative
to the GR solution, but the accumulation of secular effects in the waveform obtained
from turning on scalar field at $t_{\rm start}=0$ leads to the EsGB waveform inspiraling
slower than the GR waveform, which is in disagreement
with the expectation that the EsGB waveform should inspiral
faster due to the additional energy loss to the scalar field. 
We verify this statement 
by confirming that the order-by-order solution does inspiral faster than
GR when using the representative waveform which has a shorter inspiral length as in left
panel of Fig.~\ref{fig:bhbh_3d_ora}.
However, we find that
even when using the shortest inspiral length possible,
the order-by-order approach overestimates the amplitude shift and
underestimates the dephasing relative to GR. This is consistent with the
spurious increase in amplitude of oscillations we found in
Appendix~\ref{app:anharm_osc} when solving the anharmonic oscillator order by
order over timescales that are sufficiently long to obtain noticeable
dephasing.

\begin{figure*}
	\includegraphics[width=\columnwidth,draft=false]{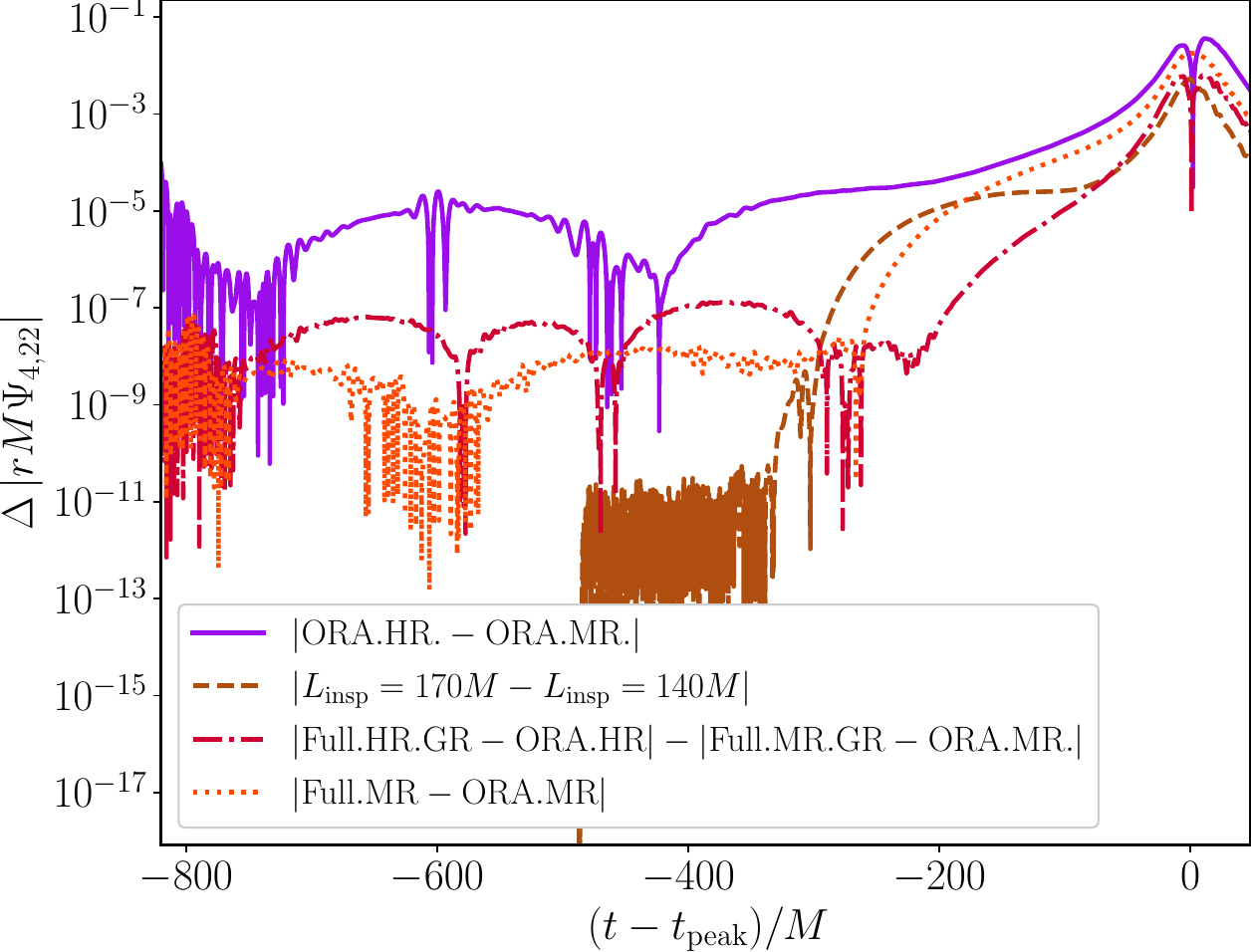}
	\includegraphics[width=\columnwidth,draft=false]{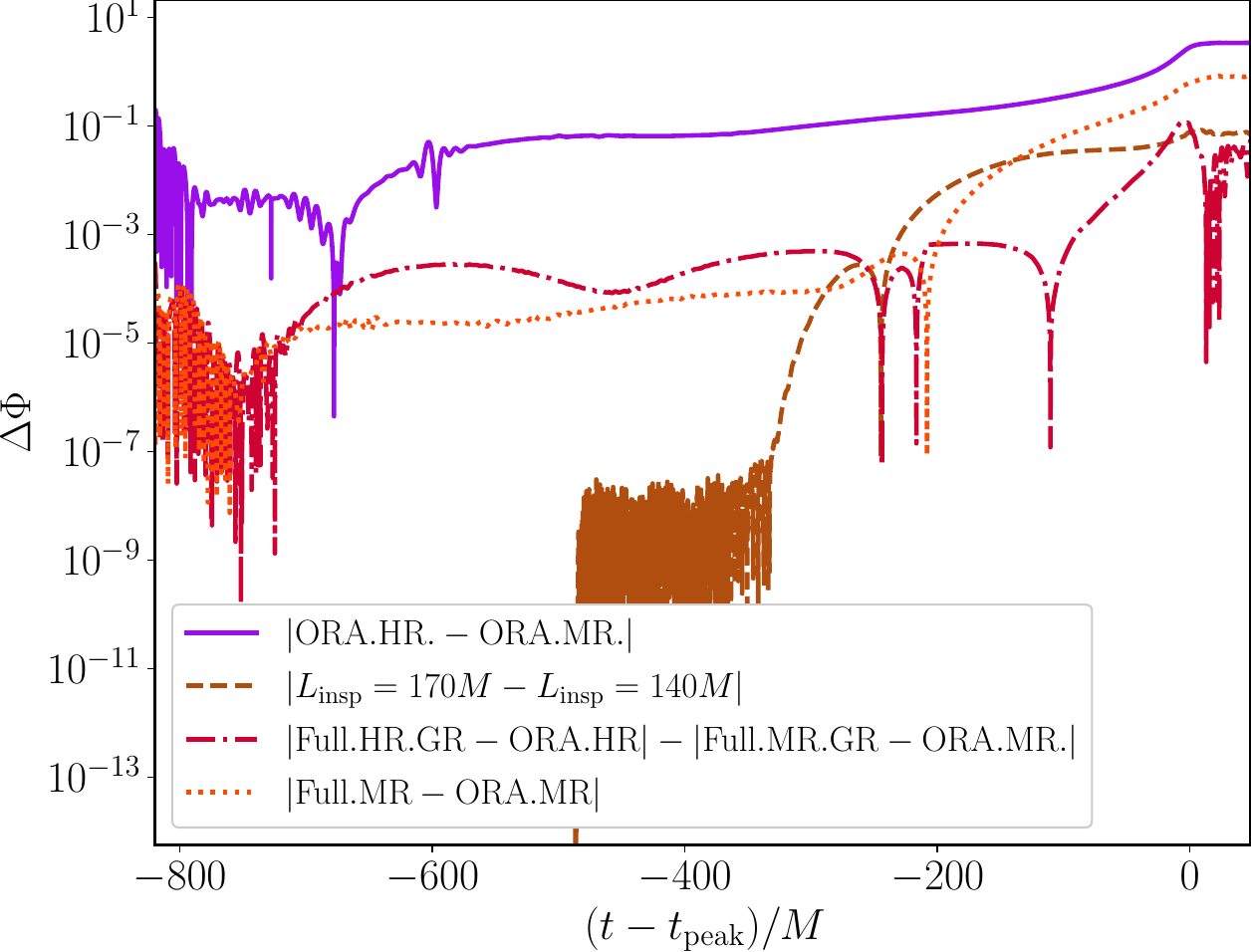}
	\caption{Difference between high and medium resolutions of the
	amplitude (left) and phase (right) of the gravitational waveform for the GW150914-like binary 
    with coupling $\lambda = 0.04 m_2^2$ in the order-by-order approach
	(solid curve). We also show the difference in amplitude and phase when turning
	on the coupling at two different times in order-by-order approach (dashed curve), 
	as well as the difference between solving the full equations and the order-by-order 
    equations at medium resolution (dotted curve) and the difference
	between the full and order-by-order solutions at high and medium resolution
	(dash-dotted). This provides evidence that the truncation error cancels between
	the order-by-order and full runs.
\label{fig:bhbh_3d_errors}
}
\end{figure*}

In Fig.~\ref{fig:bhbh_3d_errors}, we quantify how the difference between the order-by-order and 
full solution for a coupling value of
$\lambda = 0.04 m_2^2$ compares to the truncation error of our simulations.
To do so, we split the $\ell=m=2$ harmonic of $\Psi_4$ into a real phase $\Phi$
and amplitude $|\Psi_{4,22}|$ so that 
$\Psi_{4,22}(r,t) = |\Psi_{4,22}| e^{-i \Phi(r,t)} $ and consider
the differences in each quantity separately.
Figure~\ref{fig:bhbh_3d_errors} shows that, just as for the head-on collision,
the difference in amplitude $r M \Delta |\Psi_{4,22}|$ 
(left) and phase $\Delta \Phi$ (right) of the gravitational
waveform between the high and medium resolutions of the order-by-order solution (solid
curve) is larger
than the difference between the full and order-by-order solution at medium resolution
(dashed curve).
However, because we use the same numerical resolution for carrying out the order-by-order
and full solution and find that the dominant truncation error in our simulations
does not depend strongly on the approach used or coupling, the truncation error
will partially cancel out when calculating the
difference in the phase and amplitude between the order-by-order and full solution
using the same resolution. 
This is illustrated by the dash-dotted curve in Fig.~\ref{fig:bhbh_3d_errors}, where
we estimate the truncation error in the difference between the order-by-order
and full solution by
comparing the order-by-order and full solutions at two different
resolutions (dash-dotted curve), finding it to be smaller than the difference
obtained at the medium resolution. 
We also present the difference in amplitude and phase
of waveforms when turning on the coupling at slightly different times which
is smaller than the difference between the full and order-by-order solution.
These findings imply that one can not take at face value
bounds obtained on EsGB gravity with the order-by-order approach \cite{Okounkova:2020rqw}.

\begin{figure*}
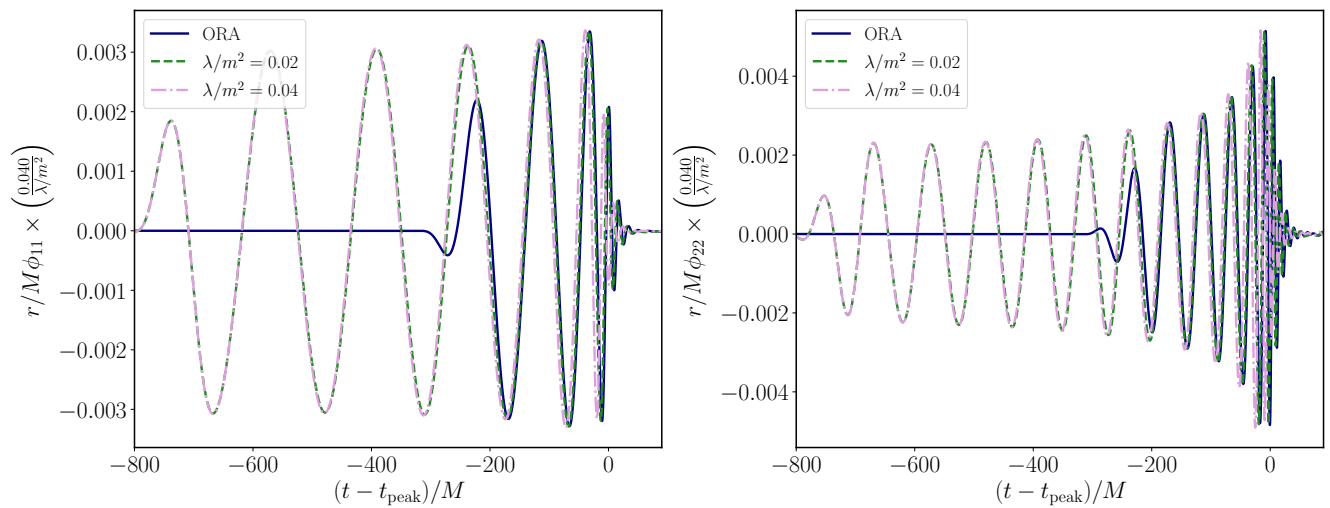

	\includegraphics[width=\columnwidth,draft=false]{{phi_ll1}.pdf}
	\includegraphics[width=\columnwidth,draft=false]{{phi_ll2}.pdf}

	\caption{Scalar waveforms as a function of look-back time measured from the time at which
	the gravitational waveform in GR peaks, rescaled by extraction radius $r = 100M$ and
	scaling out the decoupling behavior for the full solutions shown in 
    Fig.~\ref{fig:bhbh_3d_ora}, as well as the order-by-order solution. We show the
	$\ell=m=2$ and $\ell=m=1$ spherical harmonic components.
\label{fig:bhbh_3d_sf}
}
\end{figure*}

\begin{figure*}
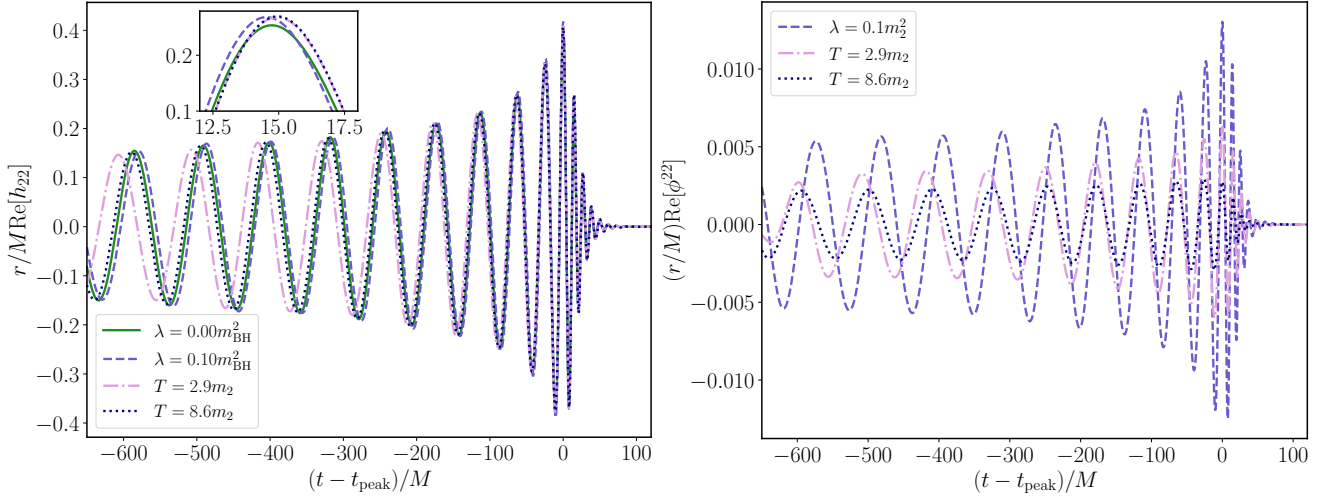

	\includegraphics[width=0.99\columnwidth,draft=false]{{strain_3d_merger}.pdf}
	\includegraphics[width=0.99\columnwidth,draft=false]{{phi_ll2_merger}.pdf}
	\caption{Gravitational (left) and scalar (right) waveforms for the GW150914-like
	binary of coupling $\lambda=0.1m_2^2$ when solving the full
	and fixed
	system of equations with two sets of fixing parameters. The GR waveform is
	shown for comparison. We align the waveform in phase and time at the
	peak of the amplitude of gravitational waveform.
\label{fig:bhbh_3d_fix_merger}
}
\end{figure*}

\begin{figure*}
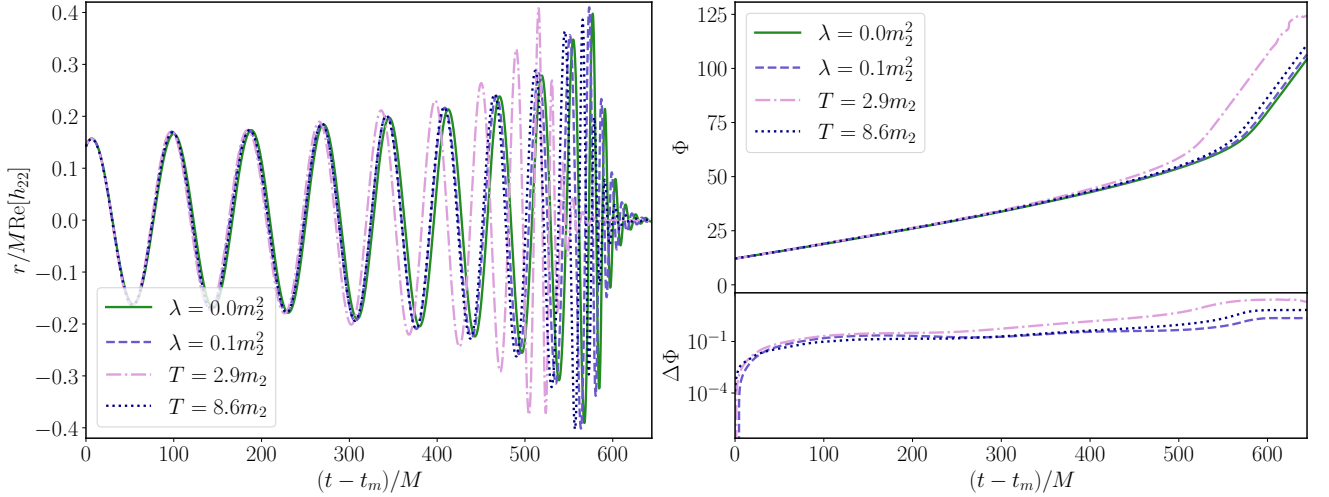

	\includegraphics[width=0.99\columnwidth,draft=false]{{h_plus}.pdf}
	\includegraphics[width=0.99\columnwidth,draft=false]{{Phi_h_subplt}.pdf}
	\caption{Left: Gravitational waveforms for the GW150914-like  
        binary with coupling $\lambda=0.1m_2^2$ when solving the full and fixed 
        system of equations with two sets of fixing parameters. All the waveforms
	are aligned in time and phase at some fiducial frequency chosen to be
	$\omega_m M = 0.065$. 
	Right: The GW
	phase of the aligned waveforms $\Phi$. 
	The bottom plot shows the phase difference between the EsGB and GR waveforms. 
	Time is measured
	with respect to the time where the waveforms are aligned in frequency and phase. 
\label{fig:bhbh_3d_align_freq}
}
\end{figure*}

\begin{figure}
	\includegraphics[width=0.99\columnwidth,draft=false]{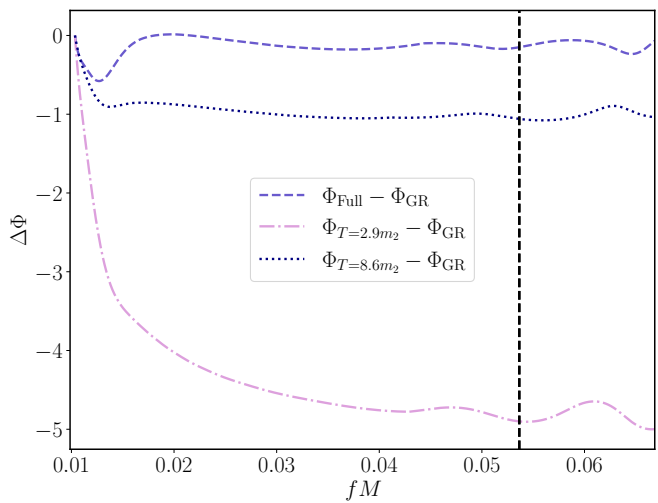}
	\caption{Left: The dephasing compared to GR as a function
	of GW frequency $f M$
	for the GW150914-like
        binary with coupling $\lambda=0.1m_2^2$ when solving the full and fixed
        system of equations. The dashed vertical line represents the frequency at which 
	the waveform peaks in GR.
\label{fig:bhbh_3d_fix_delta_phi}
}
\end{figure}

We finish this section by presenting the results of evolving the GW150914-like binary
assuming a coupling of $\lambda = 0.1 m_2^2$, 
when solving the full and fixed system of equation.
In Fig.~\ref{fig:bhbh_3d_fix_merger}, we show the gravitational (left) and scalar (right)
waveforms aligned in time and phase at the peak of the amplitude of the 
gravitational waveforms. 
We show results for two sets of fixing parameters chosen to have 
longer timescales than the ones that gave us the most accurate results in the head-on collisions
to ensure that all relevant timescales are properly resolved.
As shown in the right panel of Fig.~\ref{fig:coor_sep}, we find that, 
in agreement with the results above, the smaller the slowest decay timescale is, 
the larger---and hence, the closer to the full solution---the scalar charge is,
which in turn leads to a faster the inspiral
(see, for instance, the phase of the GW in the right panel of 
Fig.~\ref{fig:bhbh_3d_align_freq}).
However, unlike for the head-on collisions, we find 
that, despite the scalar charge
and scalar waveforms being noticeably smaller in amplitude than the full solution,
the two fixed solutions merge earlier than the full solution.
The careful reader will have noticed that from the
trajectories of the black holes shown in Fig.~\ref{fig:coor_sep} one would
conclude that the fixing solution with the largest decaying timescale, namely $T=8.6m_2$,
merges later than the full solution. However, after aligning the waveforms
at merger, as in Fig.~\ref{fig:bhbh_3d_fix_merger}, or alternatively
at some fiducial frequency, as in Fig.~\ref{fig:bhbh_3d_align_freq},
we conclude that
the fixing solution lags the full solution, although by a very little amount.
The right panel of Fig.~\ref{fig:bhbh_3d_align_freq} shows the phase of each solution
in the top panel, and the difference between the phase in GR and the full or fixed solution
after alignment. This confirms the statement that the shorter the decay rate of the
fixed solution, the faster the inspiral and the larger the error compared with the full
solution, which is shown as a function of GW frequency 
in Fig.~\ref{fig:bhbh_3d_fix_delta_phi}.

The reason for this requires further investigation. This is likely due 
to a lack of numerical resolution, but another possible cause is inconsistent initial data. It
is possible that as we slowly turn on the scalar field, the fixed solution moves to
a different orbit---e.g., a slightly more eccentric one---compared with the full solution. 
Another source of error for the inspiral case that is worth pointing out 
is that the initial velocities of the individual black holes
in the inspiral case are roughly $3\times$ larger than in the head-on collisions
where we set the initial velocities to the values corresponding to the
binary being marginally bound at infinite separation. 
However, remember that the fixing parameters $\{t_1,t_2\}$
determine the timescale over which the auxiliary fields are driven to their corresponding
true value, and thus if the timescale over which the true solution changes decreases, 
then one would have to decrease the fixing parameters accordingly to keep the error
in the fixed solution unchanged. We verified that the error in the amplitude and time
of merger in the head-on collision for a fixed set of fixing parameters increases with
the initial speed of the individual black holes.
The study of all these effects is beyond the scope of this paper.

\section{Discussion and Conclusion}\label{sec:conclusion}
In this work, we have performed the first systematic and fully self-consistent comparison
of the three different numerical approaches commonly used in the literature to study
extensions of GR in the strong-field, highly dynamical regime.
The three approaches we compared are: (i) solving the full evolution equations, (ii)
solving the equations order by order, and (iii) solving a fixed system of evolution
equations.
By leveraging our ability to solve the full equations of motions in a particular theory,
we were able to quantify the errors introduced by secular effects in the order-by-order
approach and by different prescriptions for the ad-hoc dynamical fields with finite
damping timescales in the fixing-the-equations approach.
We focused on shift-symmetric EsGB gravity as a benchmark, and considered the dynamical
formation of scalar hair around non-spinning black holes in axisymmetry, 
head-on black hole collisions, and quasi-circular binary black hole mergers.

For single black hole spacetimes, we found that,
for small enough couplings, the order-by-order approach accurately 
reproduces the formation of scalar hairy black hole solutions, but that
nonlinear effects become important for couplings below the threshold
for which perturbation theory breaks down. The fixing-the-equations approach, however,
is able to track the scalarization of the black hole even at large couplings,
provided we choose a good set of fixing parameters. 
We guided and justified our choices for a good combination of fixing parameters 
by analytically
solving for a simpler fixing equation that retained most of the features of the full
solution. This toy equation illustrated that the decaying solution is 
subject to two different timescales. We define good parameters as ones that will
lead to both solutions decaying at roughly the same rate while avoiding oscillatory
and singular behavior. Another requirement is that the decay rate of 
both solutions be
smaller than all the relevant timescales in system under study. Of course, in a numerical
implementation, care must be exercised to ensure that the resulting timescales are not too small
as to become poorly resolved.
However, even better solutions can be obtained through extrapolation
to vanishing timescales over which the auxiliary fields are driven to their 
corresponding true value.
Finally, for our particular choice of driver equation,
making the fixing timescales sufficiently short
was important to avoid spurious spatial
dependence in the stationary solution. Of course, alternative driver choices could be devised, e.g., by adding
a suitable contribution to source term or considering a different order of driver equation. However,
these would introduce only mild effects as we see no significant energy cascade to the UV. Thus, differences
introduced by the choice of driver are not essential, as argued in Refs.~\cite{geroch,Cayuso:2023aht} and
we do not concern ourselves with other options for this work. 

We next studied the head-on collision of an equal-mass
binary black hole in axisymmetry, considering a range of coupling values. Our results were entirely
consistent with the single black hole spacetimes. We found that
the order-by-order approach does
well at small couplings, but ceases to yield a good approximation even for 
couplings below which the perturbative approximation breaks down. The fixing 
approach, on the other hand, does well even for moderately large couplings provided
the fixing parameters are chosen wisely. In the order-by-order approach,
we found that the dominant source of error arises in the amplitude of the gravitational waveform, rather
than the merger time.

We finished our study by considering the astrophysically relevant inspiral and merger of
binary black holes without any continuous symmetries. Here, we focused on a single system
with parameters similar to the GW150914 event. As expected, the order-by-order 
solution is subject to secular effects in the amplitude of the waveform.
We found that these errors are small only when the coupling is small enough
that the deviations from GR in the waveform---which occur primarily as a dephasing---are negligible. 
We conclude that one cannot use this approach
in the way it has been implemented to place useful constraints on EsGB gravity. 
As this issue is generic, it will likely prevent one from obtaining faithful
solutions using this approach in other beyond GR theories.

In this study, for computational expedience, we only applied the
fixing-the-equations approach to a moderately large coupling quasicircular
merger. We adopted approximate initial data and accounted for the scalarized
solution by a transition to the corresponding coupling over an extended time
frame. With this choice, employing fixing parameters with a short time for the
damping parameters gave an unphysical behavior, which we believe is due to a
lack of resolution for the timescales involved. With the same resolution, we
evolved the system for two other sets of parameters with somewhat longer
timescales, which gave better results, but we were unable to see a trend one
could extrapolate. Based on the results for head-on collisions, we expect that
one will be able to obtain accurate solutions in this approach using
sufficiently short fixing timescales with sufficiently high resolution, though
as a practical matter, the higher resolution requirements are an impediment.
Constructing initial data and a different combination of fixing parameters may
also help improve the accuracy of the solution. This observation was supported
by additional studies of head-on collisions starting at increasingly larger
speeds.

We note that recent work \cite{Lara:2024rwa} applied the fixing-the-equations approach
to study spontaneous scalarization in EsGB gravity taking the decoupling limit (i.e.
considering a test field) and
proposed a new driver equation that improved the accuracy of their
stationary solutions. Though it is unclear whether such choice was a consequence of
implementing the approach under a dual-frame strategy.

The question remains whether one could modify the order-by-order approach in a
way that would give useful results, e.g., by ``renormalizing" in some
way~\cite{GalvezGhersi:2021sxs}. One option may be to renormalize the
correction the GW signal, as opposed to working with the full
spacetime solution.  Given that the main source of error seems to be in the
GW amplitude, while the main effect of the modified gravity (in
EsGB, as well as other examples like the higher derivative gravity theory
considered with the fixing-the-equations approach in
Ref.~\cite{Cayuso:2023aht}) seems to be in the phase evolution,
one could imagine devising a way to interpret the correction derived in the order-by-order
approach as solely one to the frequency evolution.

We have only simulated the dynamics of one of the simplest EsGB gravity
theories that gives scalar hairy black holes. Although we believe our
comparison should also apply to other kinds of sGB couplings---and the general
conclusions should be broadly applicable to modified gravity theories---it would be
interesting to perform the comparison we did on coupling functions that allow
for the effect of (de)scalarization, which will change the second order
equations in the order-by-order approach, and also might affect the timescales
we consider in the fixing-the-equations approach. Similarly including spin can
significantly impact the dynamics of binaries. Finally, our results argue for
applying the fixing-the-equations approach to other classes of theories which
cannot be formulated in a well-posed manner. We note that in those theories,
unlike the one studied here (at sufficiently small coupling), it may not be
possible to make the fixing timescales arbitrarily small, and instead one may
need to establish that there is a regime where these timescales are
sufficiently small so that the results are insensitive to them, but not too
small as to reintroduce the instabilities associated with ill-posedness.

%=============================================================================
%\acknowledgments
\section{Acknowledgments}
We thank Miguel Bezares, Ramiro Cayuso, Guillermo Lara, Peter James Nee, 
Harald Pfeiffer, Justin Ripley, 
and Thomas Sotiriou
for discussions.
W.E. and L.L. acknowledge support from the Natural Sciences and Engineering Research 
Council of Canada through a Discovery
Grant.
L.L. also thanks financial support via the Carlo Fidani Rainer
Weiss Chair at Perimeter Institute and CIFAR. 
W.E. is also supported by an Ontario Ministry of Colleges and Universities Early Researcher Award.
This research was
supported in part by Perimeter Institute for Theoretical Physics. Research at
Perimeter Institute is supported in part by the Government of Canada through
the Department of Innovation, Science and Economic Development and by the
Province of Ontario through the Ministry of Colleges and Universities. This
research was enabled in part by support provided by SciNet (www.scinethpc.ca)
and the Digital Research Alliance of Canada (alliancecan.ca).  Calculations
were performed on the Symmetry cluster at Perimeter Institute, the Niagara
cluster at the University of Toronto, and the Urania cluster at the Max Planck
Institute for Gravitational Physics.

%=============================================================================
%\bibliographystyle{apsrev4-1.bst}
\bibliography{./main}
%=============================================================================

\appendix
%=============================================================================
%=============================================================================
\section{Implementation of evolution equations for shift-symmetric EsGB gravity}
\label{app:implementation}
Here we give some details on the numerical implementation of the order-by-order
and fixing-the-equations approach in the numerical scheme we use to evolve the
full EsGB equations. To do so, we first write down the MGH equations of motion 
in the form we numerically evolve, namely
\begin{align}
\label{eq:time_evo_system}
   \begin{pmatrix}
      \mathcal{A}_{ab}{}^{ef}
   &  \mathcal{B}_{ab}
   \\ \mathcal{C}^{ef}
   &  \mathcal{D}
   \end{pmatrix}
   \partial_0^2
   \begin{pmatrix}
      g_{ef}
   \\ \phi
   \end{pmatrix}
   =  
   \begin{pmatrix}
      \mathcal{S}_{ab}^{(g)}
   \\ \mathcal{S}^{(\phi)}
   \end{pmatrix}
   ,
\end{align}
where the explicit forms of the components are derived in Appendix A of 
Ref.~\cite{East:2020hgw}. We note that this includes constraint damping
terms~\cite{Gundlach_2005} designed to damp away violations of the MGH constraint
\begin{align}\label{eq:mgh_constraint}
    C^a \equiv H^a - \tilde{g}^{bc} \nabla_b \nabla_c x^a  =0
    .
\end{align}
In all three approaches, we set the constraint damping timescale to
be $\sim M$. 
Using this hyperbolic system of equations, we directly evolve the
22 variables (after accounting for the symmetries of the metric components)
$\{g_{ab},\partial_t g_{ab},\phi,\partial_t \phi\}$ through a method-of-lines algorithm.

The goal now is to rewrite
the order-by-order \cref{eq:order_0,eq:order_1,eq:order_2}
and fixed \cref{eq:eom_esgb_scalar_fix,eq:eom_esgb_tensor_fix,eq:fixing_eqn} system of equations 
in the same form as \cref{eq:time_evo_system}, such that we can re-use the already existing
numerical scheme.

\subsection{Order by order approach}
\label{app:oro}
Considering the order-by-order approach first, recall that 
to evolve the first order metric perturbation, we need to evolve
three systems of equations simultaneously: one for the GR background $g^{(0)}_{ab}$
\cref{eq:order_0},
one for the first order scalar field $\phi^{(1)}$ sourced by background curvature \cref{eq:order_1},
and one for the metric perturbation $h^{(2)}_{ab}$ \cref{eq:order_2} sourced by background metric
and $\phi^{(1)}$. 
Let us rewrite these equations in the same form as \cref{eq:time_evo_system}. We do this order by order.
At zeroth order in $\epsilon$, we obtained \cref{eq:eom_ora_0}, i.e.
a pure GR Einstein field equation. This is then equivalent to
the system we solve for \cref{eq:time_evo_system} provided we set all the terms 
introducing modifications from GR to zero,
\begin{align}
\label{eq:time_evo_system_0}
   \begin{pmatrix}
      \mathcal{A}_{ab}{}^{ef}
   &  0
   \\ 0
   &  \mathcal{D}
   \end{pmatrix}
   \partial_0^2
   \begin{pmatrix}
           g^{(0)}_{ef}
           \\ \phi^{(0)}
   \end{pmatrix}
   =
   \begin{pmatrix}
      \mathcal{S}_{ab}^{(g)}
      \\ \mathcal{S}^{(\phi)}
   \end{pmatrix}
   ,
\end{align}
all evaluated using $(g_{ab},\phi)= (g_{ab}^{(0)},\phi^{(0)})$ and setting $\lambda = 0$.
Since the scalar field equation has no source, we also take $\phi^{(0)}=0$
which simplifies the
system even further. 
In fact we no longer have to solve the scalar field equation.
We will use this to simultaneously solve for the background metric 
and first order scalar field.

At first order in $\epsilon$, we find that the equations of motion for the first order perturbation to the metric
and scalar field are given by \cref{eq:order_1}, i.e., the Einstein equations do not have a source
term and the scalar field is sourced by the curvature of the background spacetime only. Rewriting
this system in the form we are numerically implementing gives,
\begin{align}
\label{eq:time_evo_system_1}
   \begin{pmatrix}
      \mathcal{A}_{ab}{}^{ef}
   &  0
   \\ 0
   &  \mathcal{D}
   \end{pmatrix}
   \partial_0^2
   \begin{pmatrix}
           h^{(1)}_{ef}
           \\ \phi^{(1)}
   \end{pmatrix}
   =
   \begin{pmatrix}
      \mathcal{S}_{ab}^{(g)}
   \\ \mathcal{S}^{(\phi)}
   \end{pmatrix}
   ,
\end{align}
where all the functions involving the metric in the scalar sector
are evaluated using the zeroth order GR solution
$g^{(0)}$. The metric sector does not depend on the first order scalar field or in other words
it is not sourced. This means that we can solve for both the zeroth order metric and first order
scalar field by solving the system
\begin{align}
\label{eq:time_evo_system_0p1}
   \begin{pmatrix}
      \mathcal{A}_{ab}{}^{ef}
   &  0
   \\ \mathcal{C}^{ef}
   &  \mathcal{D}
   \end{pmatrix}
   \partial_0^2
   \begin{pmatrix}
           g^{(0)}_{ef}
           \\ \phi^{(1)}
   \end{pmatrix}
   =
   \begin{pmatrix}
      \mathcal{S}_{ab}^{(g)}
      \\ \mathcal{S}^{(\phi)}
   \end{pmatrix}
   .
\end{align}
Instead of solving for the second order perturbation directly, we choose to solve for the full
metric, but making the perturbation sufficiently small that
any $\mathcal{O}(\epsilon^3)$ term can be neglected. The system we then solve for is,
\begin{align}
\label{eq:time_evo_system_ora}
   \begin{pmatrix}
      \mathcal{A}_{ab}{}^{ef}
   &  \mathcal{B}_{ab}
   \\ 0
   &  1
   \end{pmatrix}
   \partial_0^2
   \begin{pmatrix}
      g_{ef}
\\ \phi^{(1)}
   \end{pmatrix}
   =
   \begin{pmatrix}
      \mathcal{S}_{ab}^{(g)}
\\ \partial_0^2\phi^{(1)}
   \end{pmatrix}
   ,
\end{align}
where
\begin{subequations}
\begin{align}
   \mathcal{A}_{ab}{}^{ef}
   \equiv&
    A_{ab}{}^{00ef}
   ,\\
   \mathcal{B}_{ab}
   \equiv&
    B_{ab}{}^{00}
   ,\\
   \mathcal{S}_{ab}^{(g)}
   \equiv&
   -  2A_{ab}{}^{(0\alpha) ef}\partial_0\partial_{\alpha}g_{ef}
   -   A_{ab}{}^{\alpha\beta ef}\partial_{\alpha}\partial_{\beta}g_{ef}\\
   &-  2B_{ab}{}^{(0\alpha)}\partial_0\partial_{\alpha}\phi
   -   B_{ab}{}^{\alpha\beta}\partial_{\alpha}\partial_{\beta}\phi
   -   F_{ab}^{(g)} \nonumber
\end{align}
\end{subequations}
where $B_{ab}{}^{cd}$ is evaluated using the background metric as any contribution coming from using
the full metric would give a fourth order correction.
Moving to the $A_{ab}{}^{cdef}$ term, the two contributions are the following (see also Eqs. (A8) 
and (A23) in \cite{East:2020hgw}),
\onecolumngrid
\begin{align}
    A_{ab}{}^{cdef}
    =&
    \delta_a^e\delta_b^fg^{cd}
-       \left(
            \delta_a^f\delta_b^dg^{ce}
    -   \delta_a^f\hat{g}_b{}^d\tilde{g}^{ce}
    \right)
-       \left(
            \delta_a^c\delta_b^fg^{de}
    -   \delta_b^f\hat{g}_a{}^c\tilde{g}^{de}
    \right)
+       \left(
            \delta_a^c\delta_b^dg^{ef}
    -   \delta_{(a}^d\hat{g}^c_{b)}\tilde{g}^{ef}
    \right)
    \nonumber\\&
-       \left(
            g^{ce}\hat{g}_{ab}
    +   \hat{g}^{ce}g_{ab}
    -   \frac{1}{2}g_{ab}g^{ce}\hat{g}
    \right)\tilde{g}^{df}
+       \frac{1}{2}\left(
            g^{cd}\hat{g}_{ab}
    +   g_{ab}\hat{g}^{cd}
    -   \frac{1}{2}g_{ab}g^{cd}\hat{g}
    \right)\tilde{g}^{ef}
\end{align}
and
\begin{align}
   A_{ab}{}^{cdef}
   =&
   \lambda g^{jd}g^{ie}g^{gl} \Delta^{cf\gamma}_{ijgab}
      \left[
         \partial_l\partial_{\gamma}\phi
      -  \Gamma^m_{l\gamma}\partial_m\phi
      \right]   
  .
\end{align}
\twocolumngrid
The first one is the same as if one were to evolve GR in the MGH formulation, while the
second one includes the terms arising from introducing modification to GR.
We use the full metric for the first term, but again choose the perturbation to be small enough
that any higher order terms are negligible.
However, for the same reasons as for $B_{ab}{}^{cd}$, we evaluate the second term using the background metric
and first order scalar field. Note that, since we are using the background metric in both $B_{ab}{}^{cd}$
and the non-GR contribution to $\mathcal{A}_{ab}{}^{ef}$, these terms can be included in the right-hand side
of the equation such that one recovers the same principal part as in GR as
required. 
Applying similar reasoning, one can also simplify some of the contributions to the source terms $F^{(g)}_{ab}$.
We test our implementation of the order-by-order approach by verifying that the appropriate scaling relations
\cref{eq:scaling_phi,eq:scaling_metric} are satisfied.

\subsection{Fixing equations approach}
\label{app:fixing}
Let us now consider the fixing-the-equations approach, outlined in Sec.~\ref{sec:fixing},
and rewrite the fixed system of equations 
\cref{eq:eom_esgb_scalar_fix,eq:eom_esgb_tensor_fix,eq:fixing_eqn} 
in the same form as \cref{eq:time_evo_system}.
To do so, we move all the EsGB contribution to the right-hand side of
\cref{eq:time_evo_system} where the source terms are,
\onecolumngrid
\begin{align}
\label{eq:time_evo_system_fix_0}
   \begin{pmatrix}
	   \mathcal{A}_{ab}{}^{{(\rm GR)} ef}
   &  0
   \\ 0
   &  \mathcal{D}
   \end{pmatrix}
   \partial_0^2
   \begin{pmatrix}
      g_{ef}
   \\ \phi
   \end{pmatrix}
   =  
   \begin{pmatrix}
	   \mathcal{S}_{ab}^{{(\rm GR)} (g)} + \mathcal{S}_{ab}^{{\rm (GB)}(g)} - \mathcal{B}_{ab}\partial_0^2 \phi - \mathcal{A}_{ab}{}^{{(\rm GB)} ef}\partial_0^2g_{ef}
	   \\ \mathcal{S}^{{\rm (GR)}(\phi)}+\mathcal{S}^{{\rm (GB)}(\phi)} - \mathcal{C}^{ef}\partial_0^2g_{ef}
   \end{pmatrix}
   .
\end{align}
\twocolumngrid
Note that, by doing so, we have recovered the same
principal part as in GR. Note that we also split the purely GR and additional EsGB contributions 
in the source terms. Introducing the source terms 
$\mathbf{S}=\left(\tilde{S}^{(g)}_{ab},\tilde{S}^{(\phi)}\right)$, 
which include all contributions from EsGB,
one has
\begin{align}
\label{eq:time_evo_system_fix_1}
   \begin{pmatrix}
	   \mathcal{A}_{ab}{}^{{(\rm GR)} ef}
   &  0
   \\ 0
   &  \mathcal{D}
   \end{pmatrix}
   \partial_0^2
   \begin{pmatrix}
      g_{ef}
   \\ \phi
   \end{pmatrix}
   =  
   \begin{pmatrix}
	   \mathcal{S}_{ab}^{{(\rm GR)} (g)} 
	   \\ \mathcal{S}^{{\rm (GR)}(\phi)}
   \end{pmatrix}
	-
   \begin{pmatrix}
	   \tilde{S}^{(g)}_{ab}
	   \\ \tilde{S}^{(\phi)}
   \end{pmatrix}
   .
\end{align}
Introducing eleven massive fields $\mathbf{P}= \left(\Pi^{(\phi)},\Pi^{(g)}_{ab}\right)$,
we write the system as,
\begin{align}
\label{eq:time_evo_system_fix_2}
   \begin{pmatrix}
	   \mathcal{A}_{ab}{}^{{(\rm GR)} ef}
   &  0
   \\ 0
   &  \mathcal{D}
   \end{pmatrix}
   \partial_0^2
   \begin{pmatrix}
      g_{ef}
   \\ \phi
   \end{pmatrix}
   =  
   \begin{pmatrix}
	   \mathcal{S}_{ab}^{{(\rm GR)} (g)}
	   \\ \mathcal{S}^{{\rm (GR)}(\phi)}
   \end{pmatrix}
	-
   \begin{pmatrix}
	   \Pi^{(g)}_{ab}
	   \\ \Pi^{(\phi)}
   \end{pmatrix}
   .
\end{align}
When $\Pi = \tilde{S}$, the equations are `unfixed.' To fix the equations, we choose the
driver equation \cref{eq:fixing_eqn},
\begin{align}
	\sigma
        g^{ab}
	\partial_a\partial_b \mathbf{P}
	=
	&\partial_{0} \mathbf{P}
	+
	\kappa \left( \mathbf{P} -\mathbf{S} \right) .
\end{align}
Note that the source terms depend on second derivatives in time of the metric and 
scalar field $\mathbf{S}=\mathbf{S}\left[\ldots \partial_{0}^{2} g, \partial_{0}^{2} \phi \ldots\right]$. We compute those by applying order reduction, i.e. we substitute
the zeroth order solutions to the second time derivatives, obtained by solving for
\cref{eq:time_evo_system_fix_1} with $\mathbf{S}=0$. This implies that the 
solutions to the fixed system 
\cref{eq:eom_esgb_scalar_fix,eq:eom_esgb_tensor_fix,eq:fixing_eqn}
are valid to first order in the coupling of the theory.

Finally, in order to justify and guide our choices for the fixing parameters 
$\{\sigma,\kappa\}$, we analytically solve for a simpler fixing equation
(where we set any spatial dependence to zero),
albeit with most of the features of the true fixed equation,
\begin{equation}\label{eq:toy_fix_eqn}
    \sigma \frac{d^2P}{dt^2} + \frac{dP}{dt} + \kappa (P - 1) = 0 
.
\end{equation}
The general solution to this equation, subject to $P(0)=dP/dt(0)=0$, is
\begin{align}\label{dampingsoln}
P(t)= -e^{a_+ t} \frac{(-\sqrt{-4 \kappa \sigma + 1} +4 \kappa \sigma - 1)}{2(4\kappa\sigma-1)} \\
- e^{a_- t} \frac{(\sqrt{-4 \kappa \sigma + 1} +4 \kappa \sigma - 1)}{2(4\kappa\sigma-1)} + 1 \nonumber
,
\end{align}
where
\begin{equation}
a_+ = \frac{-1+\sqrt{1-4\kappa\sigma}}{2\sigma} \, \,\, , \, \, \, 
a_- = -\frac{1+\sqrt{1-4\kappa\sigma}}{2\sigma}
.
\end{equation}

The solutions to this system will decay, but depending on the values chosen for the
fixing parameters, the solutions might undergo an oscillatory decay. Large oscillations
will lead to stages where the system overshoots the original one, a feature that we consider it best to avoid.
To keep both solutions decaying at roughly the same rate, and the frequency of oscillations small,
one could choose $4\kappa \sigma \simeq 1$. In that case, the decay rate is 
$T \approx 2\sigma$, though care should be exercised to avoid singular behavior indicated by
the solution, Eq.~\eqref{dampingsoln}. The timescale should also be taken to be smaller than the smallest relevant 
scales in the problem being studied, in order to approach the original system; however, if 
the timescale $T$ is made too short, it may lead to stiff behavior in the numerical evolution, requiring higher resolution.
Finally we note that, depending on the driver equation adopted, a small $\sigma$ is
desirable to avoid spurious spatial dependence. 

%=============================================================================
\section{Numerical convergence and consistency tests}
\label{app:convergence}
Here we present convergence results and consistency tests for the implementation of
the order-by-order and fixing-the-equations approach in the modified 
generalized harmonic code \cite{East:2020hgw} that we use. 

\begin{figure*}
     \includegraphics[width=\columnwidth,draft=false]{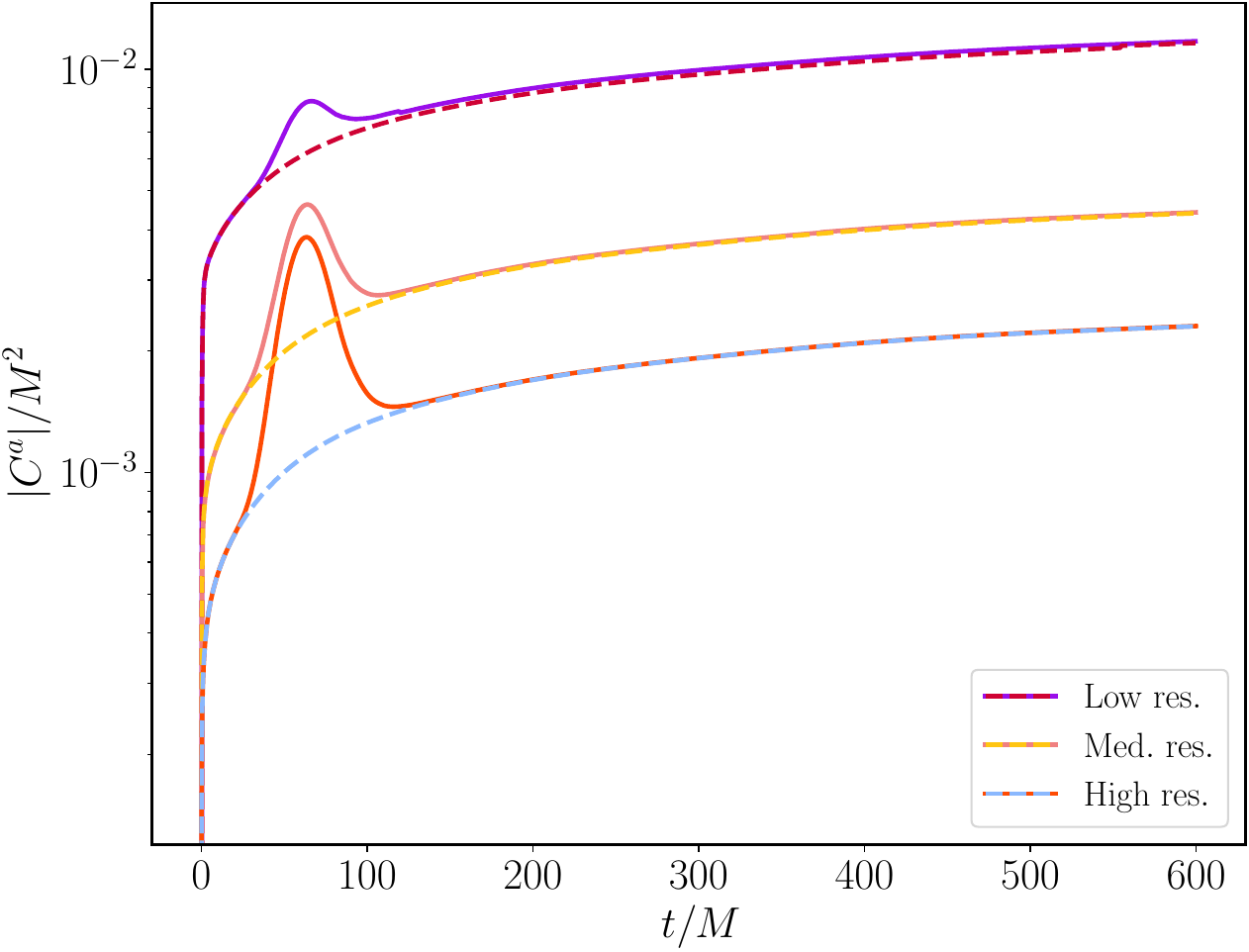}
     \includegraphics[width=\columnwidth,draft=false]{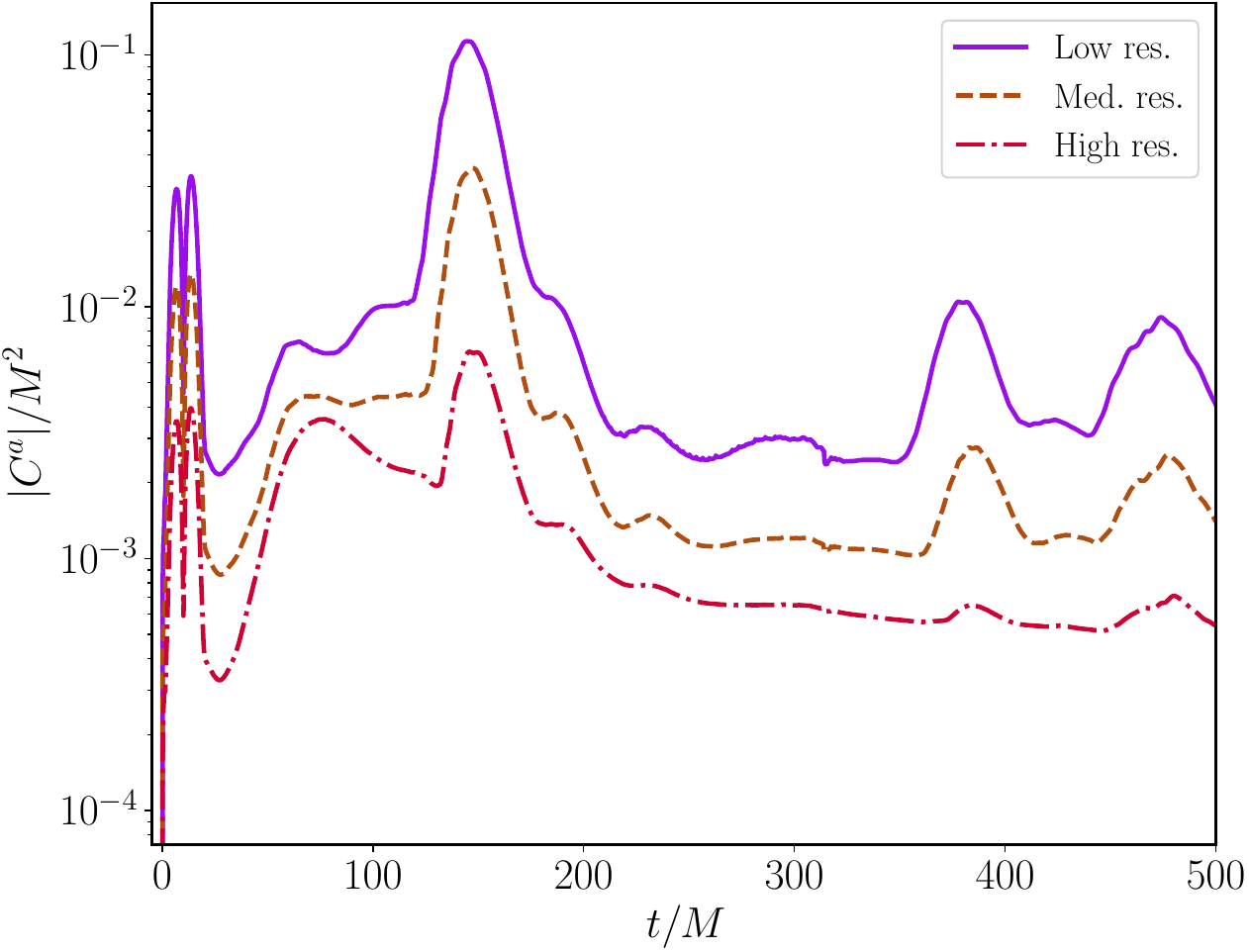}
	\caption{Integrated norm of the constraint violation $C_a$ as a function of
	coordinate time (in units of the total mass) for the scalarization of a single
	black hole spacetime with a coupling value of $\lambda = 0.1 M^2$ (left and 
	dashed curves) and
	for 
	an equal-mass head-on collision with $\lambda=0.1M^2$ (right) at three resolutions.
	The medium and high resolutions have $1.5$ and $2 \times$ the resolution of the
	low resolution simulation. The solid lines in the left panel are the corresponding
	constraint violations in GR.
\label{fig:cnst_axi}
}
\end{figure*}

In the left panel of Fig.~\ref{fig:cnst_axi}, we show the integrated MGH constraint
violation $|C_a|$ as a function of coordinate time for a single black hole spacetime
with $\lambda=0.1 M^2$ (dashed curves) as well as GR (solid curves)
and several resolutions, demonstrating that this quantity is converging to zero
with third-order, as expected.
Here, the lowest resolution has 6 levels of mesh refinement with a refinement ratio of
$2 : 1$, and has a grid spacing of $dx \sim 0.024M$ on the finest level.
The medium and high resolutions are $1.5$ and $2\times$ as high, respectively. The medium
resolution is the default resolution for the results presented in this work.
The transient growth at $\sim 50M$ is due to the slow ramp on of the scalar field.
Similarly, in the right panel of Fig.~\ref{fig:cnst_axi}, we show the integrated
constraint violation for the equal-mass non-spinning head-on collision initially
separated by $75M$ and with $\lambda = 0.1 m^2$. 
Here the lowest resolution has 7 levels of mesh refinement and a 
grid spacing of $dx \sim 0.008M$ on the finest mesh refinement. 
The medium and high resolution
have, respectively, $4/3$ and $2\times$ the resolution of lowest resolution.

\begin{figure*}
     \includegraphics[width=\columnwidth,draft=false]{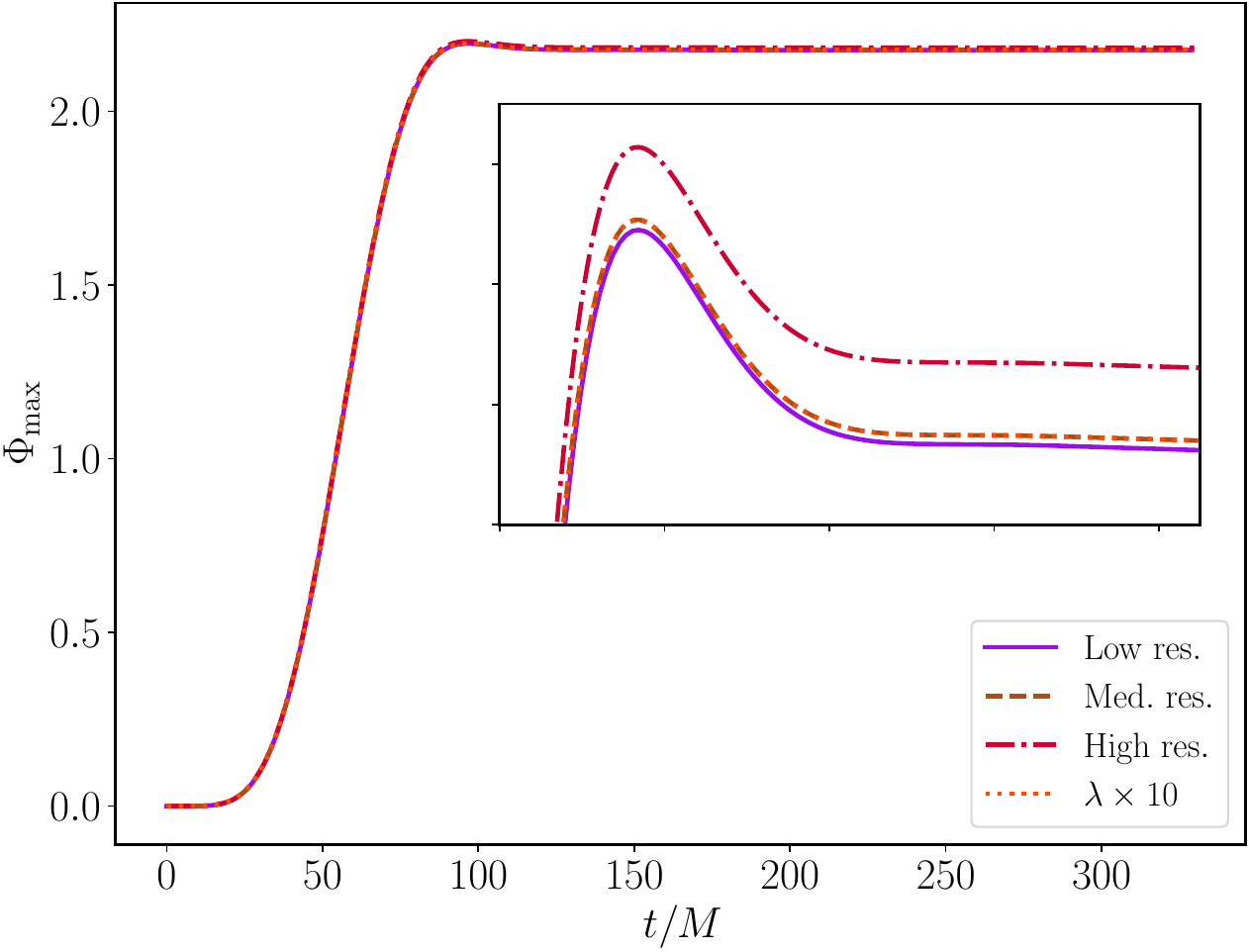}
     \includegraphics[width=\columnwidth,draft=false]{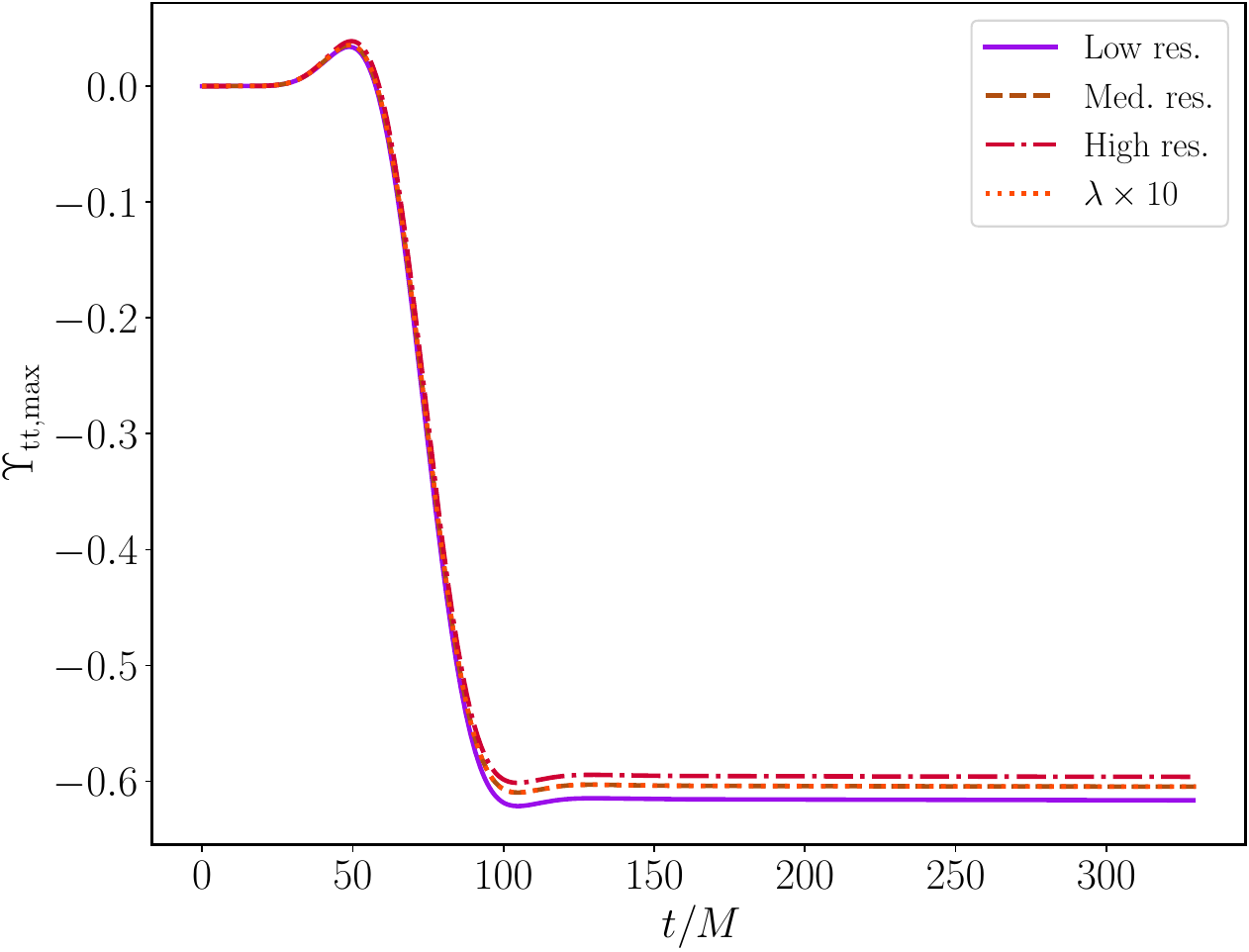}
	\caption{Maximum value of scalar field and second-order correction to the $tt$ 
	component of the metric for the scalarization of single black hole computed for two
	couplings at the
	default resolution (Med. res.) and after scaling out the dependence on $\lambda$.
	We also show the corresponding values for the low and high resolution runs.
\label{fig:scaling_test_bh}
}
\end{figure*}

\begin{figure*}
     \includegraphics[width=\columnwidth,draft=false]{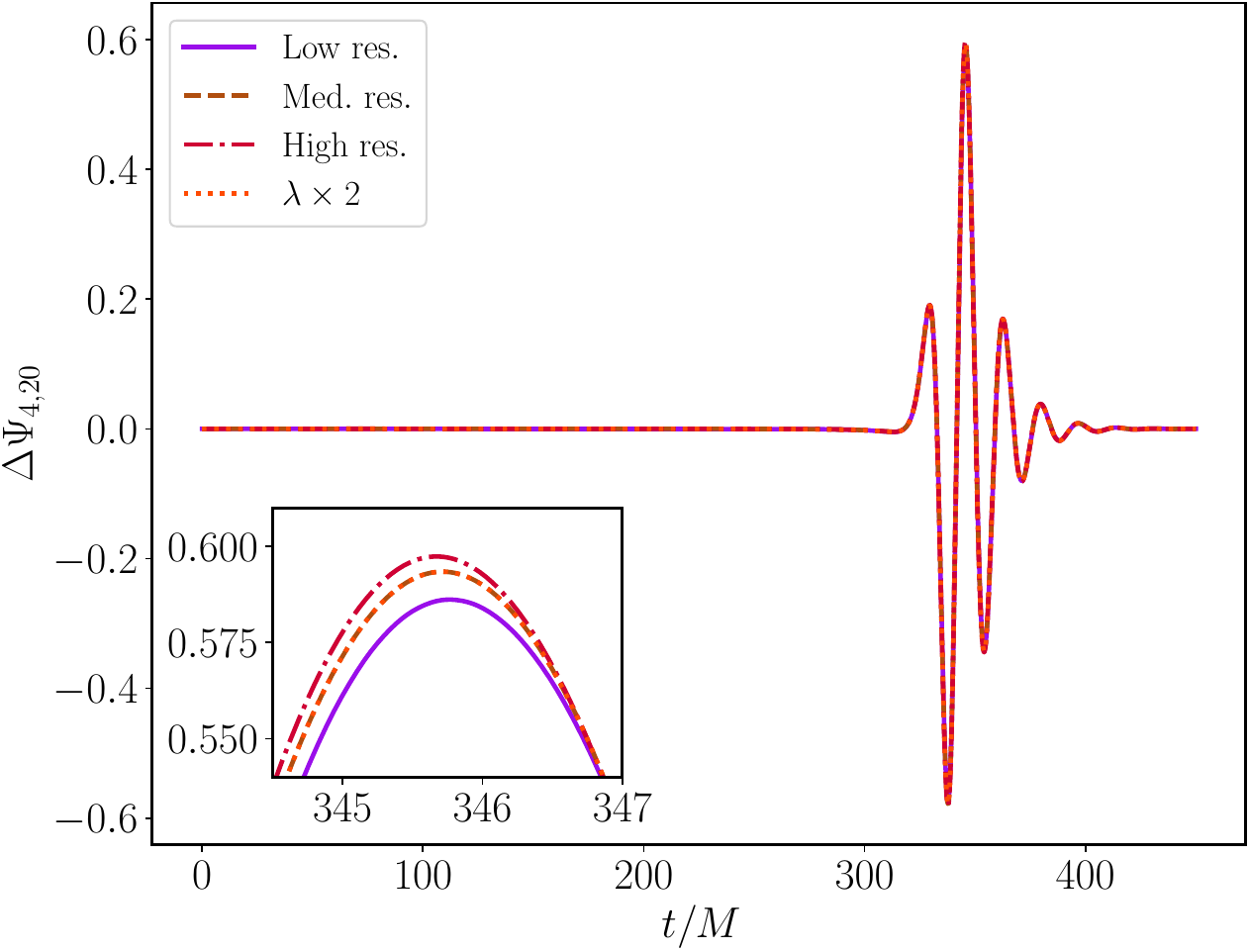}
	\includegraphics[width=\columnwidth,draft=false]{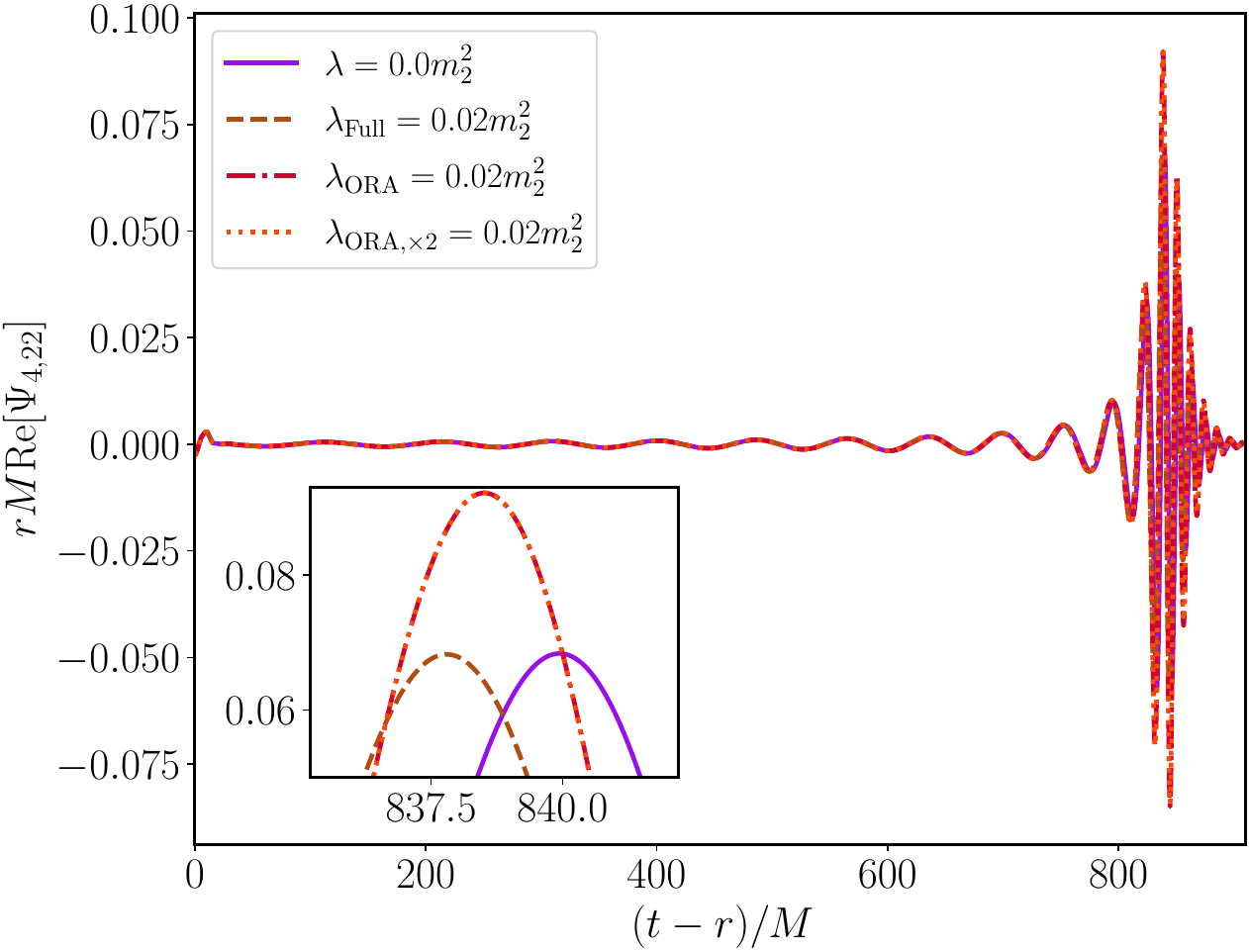}
	\caption{Left: Second-order correction in gravitational waveform from 
	the head-on collisions of equal-mass, non-spinning black holes
	for two different
	couplings and three different resolutions after scaling out the quadratic
	dependence on $\lambda$. We also show the correction at medium resolution
	when using a coupling of twice the canonical value and find that the
	difference is much smaller than the truncation error.
	Right: Gravitational waveform for GW150914-like 
	binary in GR and EsGB assuming a coupling of $\lambda=0.02m_2^2$. We
	compare the full solution to the order-by-order solution computed for two
	couplings and assuming a quadratic dependence with the coupling. The fact that
	the two order-by-order solutions overlap provides evidence we correctly
	implemented the order-by-order approach.
\label{fig:psi4_scaling_tests}
}
\end{figure*}

In Fig.~\ref{fig:scaling_test_bh}, we test our numerical implementation of the
order-by-order approach in our MGH code by comparing the maximum value of the scalar
field and the correction to the $tt$ component of the metric for two different couplings,
after scaling out the dependence on coupling using
the results of Sec.~\ref{sec:scaling_solns}. We find that the difference between the scaled
out quantities computed with two different coupling values is much smaller than the
truncation error of solutions. This provides evidence that we correctly implemented
the order-by-order approach. Similarly,
in Fig.~\ref{fig:psi4_scaling_tests}, we show that in the order-by-order approach the gravitational waveform scales
quadratically with the coupling both for the head-on collision and quasi-circular
inspiral of equal-mass non-spinning black holes.

%============================================================================

\section{Fixing equations and constraints}
\label{app:fixingconstraints}
Here we argue that constraints are preserved by the fixed equations. First,
recall that constraint damping in the generalized harmonic formulation is achieved
by ensuring the effective evolution of the coordinate conditions is damped away (see Ref.~\cite{Gundlach_2005}).
Such a behavior is enforced by adding a suitable contribution to Einstein equations that: (i) is
proportional to the constraints, and (ii) is such that, upon taking the divergence and using the Bianchi
identity, the constraints are seen to obey a homogeneous damped wave equation. As a result,
violations of the constraints are driven towards zero. 

In the case of beyond GR theories, extra contributions appear in the equations of motion
in addition to the Einstein tensor. An analogous constraint damping strategy can be adopted in such 
cases (for an explicit discussion in the case of EsGB, see Ref.~\cite{East:2020hgw}), where one
uses explicitly that the extra contributions should have zero divergence.

When fixing the equations, such an identity is not explicitly enforced. 
The resulting induced evolution equations for the constraint violation contains an additional source from
the divergence of such tensor which must be shown to suitably approach zero
in the adopted fixing-the-equations approach.
We next argue this is the case.

The modified gravity theory in a suitable formulation is governed by
\begin{align}
	G^{ab}
	+
	S^{ab}
	+ 
	M^{ab}
	&=
	Q^{ab},
\end{align}
where the beyond GR contributions are split into $S^{ab}$---containing contributions
with second order derivatives---and
$M^{ab}$---which contains the remaining terms, and $Q^{ab}$ is the constraint function that vanishes when the
constraints are satisfied (for example, $F^{ab}$ in
Eq.~\eqref{eq:eom_esgb_tensor_fix}). Furthermore, the divergence of $Q^{ab}$ gives
rise to a damped evolution equation for the constraints, to converge to the left
hand side, which is zero as $\nabla_a S^{ab} = - \nabla_a M^{ab}$. 
That is
\begin{align}
	\label{eq:constraint_bianchi}
\nabla_a \left( G^{ab} + S^{ab} + M^{ab} \right)=0=\nabla_{a} Q^{ab},
\end{align}
so the constraints are conserved.
The fixed equations replace $S^{ab}$ by ${\Pi^{(g)}}_{ab}$\footnote{In the main text, we label such an object as ${\Pi^{(g)}}_{ab}$. Here,
we will drop the label ${}^{(g)}$ for convenience.}
 and provide an equation of
motion for this tensor to approach $S^{ab}$:
\begin{align}\label{fixconstraint}
	G^{ab}
	+
	\Pi^{ab}
	+
	M^{ab}
	&=
	Q^{ab}.
\end{align}
Notice that $\nabla_a M^{ab}= {\Pi^{\phi}} \nabla_b \phi$ by definition and, due
to the fixed equation introduced for its dynamics, it approaches $-2 \lambda {\cal G} \nabla_b \phi$ .
We next argue that the fixing equation enforces
that $\nabla_a \Pi^{ab}$ approaches $\nabla_a S^{ab}$ in a controlled manner.

Starting with the equation of motion for $\Pi_{ab}$:
\begin{align}
\nabla_{c} \nabla^{c} \Pi^{ab}
&=
\frac{1}{\sigma}
n^{c} \nabla_{c} \Pi^{ab}
+
\frac{\kappa}{\sigma}
\left(
\Pi^{ab}
-
S^{ab}
\right),
\end{align}
where, for simplicity, we adopted a slightly different form of the equation than the one used
explicitly in this work to aid in the presentation. Next, 
we take the covariant derivative
\begin{align}
	\nabla_{a} \nabla_{c} \nabla^{c} \Pi^{ab}
&-
\frac{1}{\sigma}
\nabla_{a}
\left( 
n^{c} \nabla_{c} \Pi^{ab}
\right)\nonumber \\
&-
\frac{\kappa}{\sigma}
\left(
	\nabla_{a}
	\Pi^{ab}
	-
	\nabla_{a}
	S^{ab}
 \right)
=0.
\end{align}

Using the fact that $R_{ab} \simeq \mathcal{O}\left(\lambda\right)$, the first term, 
to order $\mathcal{O}\left(\lambda^{2}\right)$, gives
\begin{align}
	\nabla_{a} \nabla_{c} \nabla^{c} \Pi^{ab}
&=
	\nabla_{c} \nabla_{a} \nabla^{c} \Pi^{ab}
-
R_{ace}{}^b \nabla^{c} \Pi^{ae} 
\nonumber \\
&=
\Box \left( \nabla_{a} \Pi^{ab}
 \right) -
R_{ace}{}^b \nabla^{c} \Pi^{ae}
-
\nabla^{c}
\left(
R_{ace}{}^{b} \Pi^{ae}
\right).
\end{align}
Proceeding analogously, to the same order, the second term gives
\begin{align}
	\nabla_{a}
\left(
n^{c} \nabla_{c} \Pi^{ab}
\right)
&=
\left( 
\nabla_{a} n^{c}
\right) 
\nabla_{c} \Pi^{ab}
+
n^{c} \nabla_{a} \nabla_{c} \Pi^{ab} \nonumber\\
&=
\left( 
\nabla_{a} n^{c}
\right) 
\nabla_{c} \Pi^{ab}
+
n^{c} \nabla_{c} \nabla_{a} \Pi^{ab}
-
n^{c} R_{ace}{}^{b} \Pi^{ae}.
\end{align}

Combining it all together we have
\begin{align}\label{eqnconstdrive}
0
	=&
\Box 
\left( 
\nabla_{a} \Pi^{ab}
 \right)
-
\frac{1}{\sigma}
n^{c} \nabla_{c} \left( \nabla_{a} \Pi^{ab}
 \right)
\nonumber \\
-&
\frac{\kappa}{\sigma}
\left(
\nabla_{a} \Pi^{ab}
-
\nabla_{a} S^{ab}
-
\frac{1}{\kappa}
n^{c} R_{ace}{}^{b} \Pi^{ae}
\right)
\nonumber \\
-&
R_{ace}{}^b \nabla^{c} \Pi^{ae}
-
\nabla^{c}
\left(
R_{ace}{}^{b} \Pi^{ae}
\right)
\nonumber \\
-&
\frac{1}{\sigma}
\left( 
\nabla_{a} n^{c}
\right) 
\nabla_{c} \Pi^{ab}.
\end{align}

The first two lines drive $\nabla_{a} \Pi^{ab}$ to
$\nabla_{a} S^{ab} + (1/\kappa)\, n^c R_{ace}^b \Pi^{ae}$. The
first term is precisely what cancels the contribution from $\nabla_a M^{ab}$,
and the second one is not only subleading with respect to curvature, but also
can be made smaller by a judicious choice of $\kappa$.
To examine the role of the other contributions, we can use the cyclic
symmetries of the Riemann tensor and $\Pi^{ab} = \Pi^{(ab)}$.
The third line simplifies, because
\begin{align}
\Pi^{ae}
\nabla^{c}
R_{ace}{}^{b} 
&=
-
\Pi^{ae}
\nabla^{c}
\left(
R_{cea}{}^{b}
+
R_{eac}{}^{b}
\right)\\
&=
-
\Pi^{ae}
\nabla^{c}
R_{cea}{}^{b}\\
&=
-
\Pi^{ae}
g^{bd}
\nabla_c
R^{c}{}_{ead}\\
&=
\Pi^{ae}
g^{bd}
\left( 
\nabla_a
R^{c}{}_{edc}
+
\nabla_d
R^{c}{}_{eca}
\right) \sim \mathcal{O}\left(\lambda^{2}\right).
\end{align}
So we then have 
\begin{align}
R_{ace}{}^b \nabla^{c} \Pi^{ae}
&=
\nabla^{c}
\left(
R_{ace}{}^{b} \Pi^{ae}
\right)
+
\mathcal{O}\left(\lambda^{2}\right).
\end{align}
The third line and the last term from the second line give
\begin{align}
	& 2 \nabla^{c}
F_{c}{}^{b}
-
\frac{1}{\sigma}
n^{c} F_{c}{}^{b},\\
&F_{c}{}^{b} := R_{ace}{}^{b} \Pi^{ae}.
\end{align}
This contribution is not only proportional to curvature terms, but also 
has an extra order of derivatives, which should be subleading in the valid EFT regime. 
Thus, $\nabla_a \Pi^{ab}$ converges to $\simeq
\nabla_a S^{ab} + (1/\kappa)\, n^c R_{ace}^b \Pi^{ae}$.
The last term can be reduced by the choice of $\kappa$. 
As a consequence, the divergence of Eq.~\eqref{fixconstraint} gives rise
to controlled dynamics for the constraints, approximating the behavior from the original system,
as expected.

\begin{figure*}
	\includegraphics[width=0.99\columnwidth,draft=false]{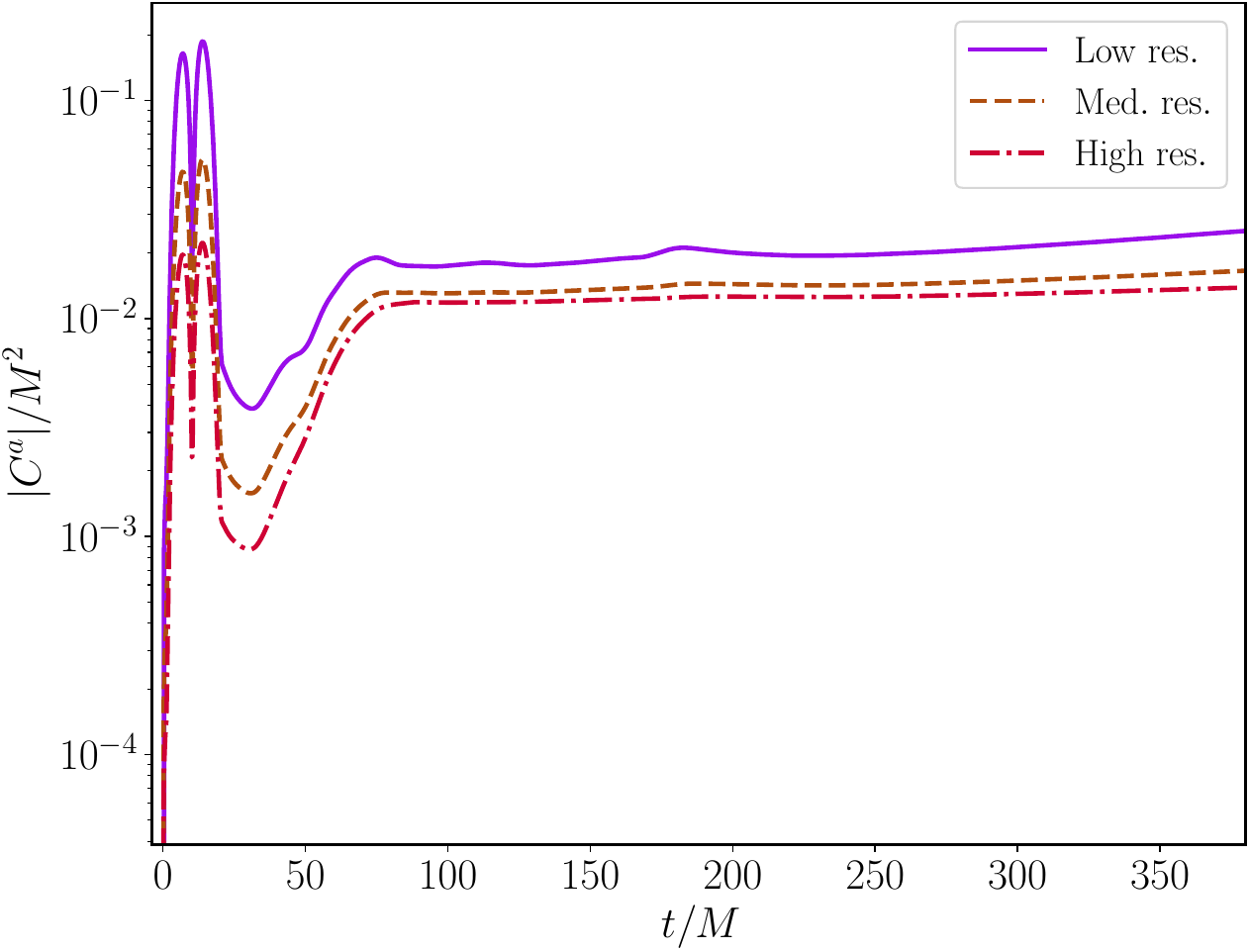}
     \includegraphics[width=0.99\columnwidth,draft=false]{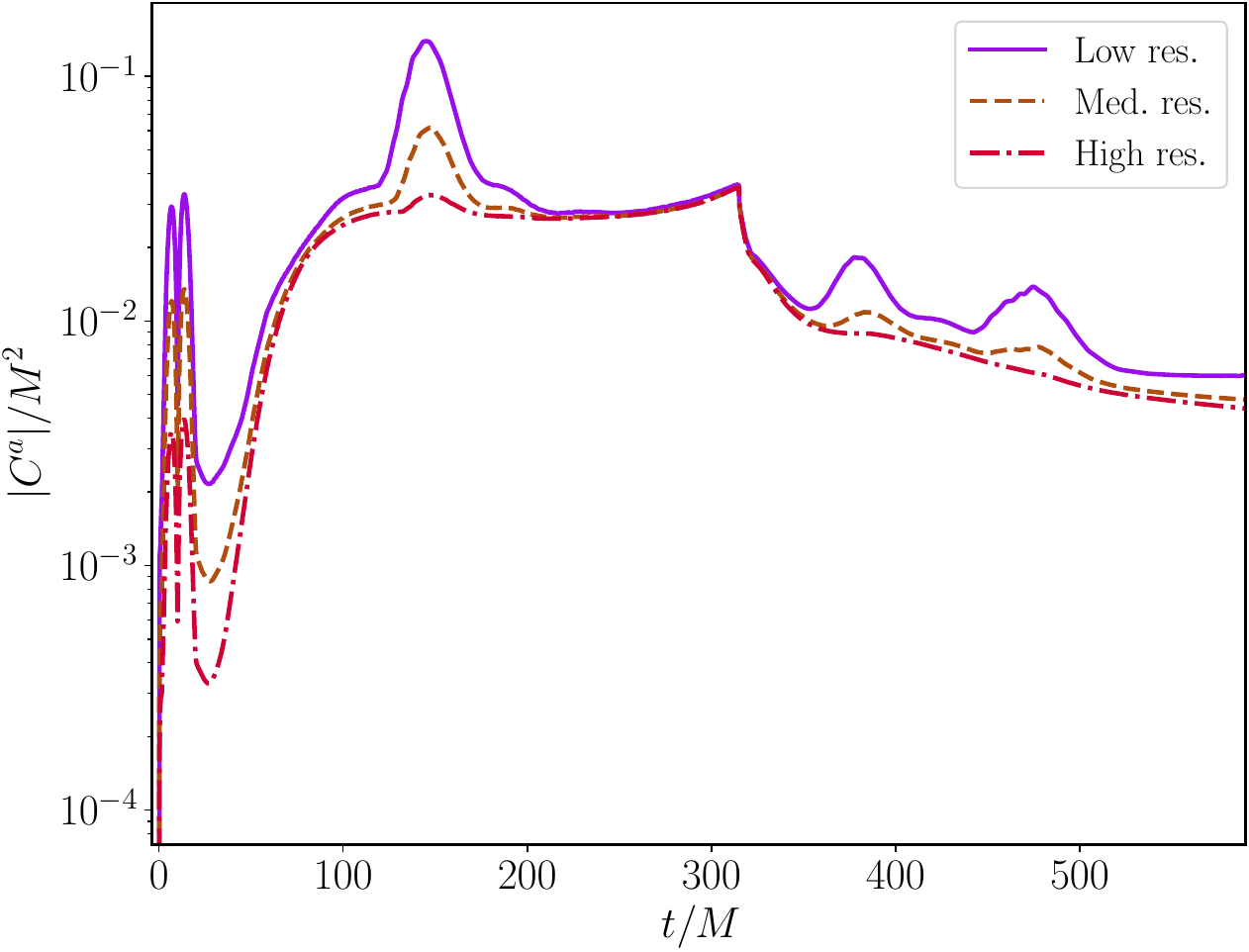}
	\caption{Integrated norm of the constraint violation $C_a$ as a function of
	coordinate time (in units of the total mass) for the scalarization of a single
	black hole spacetime with a coupling value of $\lambda = 0.04 M^2$ (left)
	and for an equal-mass head-on collision with $\lambda=0.04m^2$ (right) 
	at three resolutions 
	when solving the fixed system of equations. The damping parameters for the single
	black hole are
    $\{\sigma,\kappa\}=\{0.2M,0.5/M\}$ 
    so that $T=1.8M$, 
	and similarly for the head-on collision, where $M$ is now mass of each individual
	black hole.
	The medium and high resolutions have $1.5$ and $2 \times$ the resolution of the
	low resolution simulation.
\label{fig:cnst_axi_fix}
}
\end{figure*}

\begin{figure*}
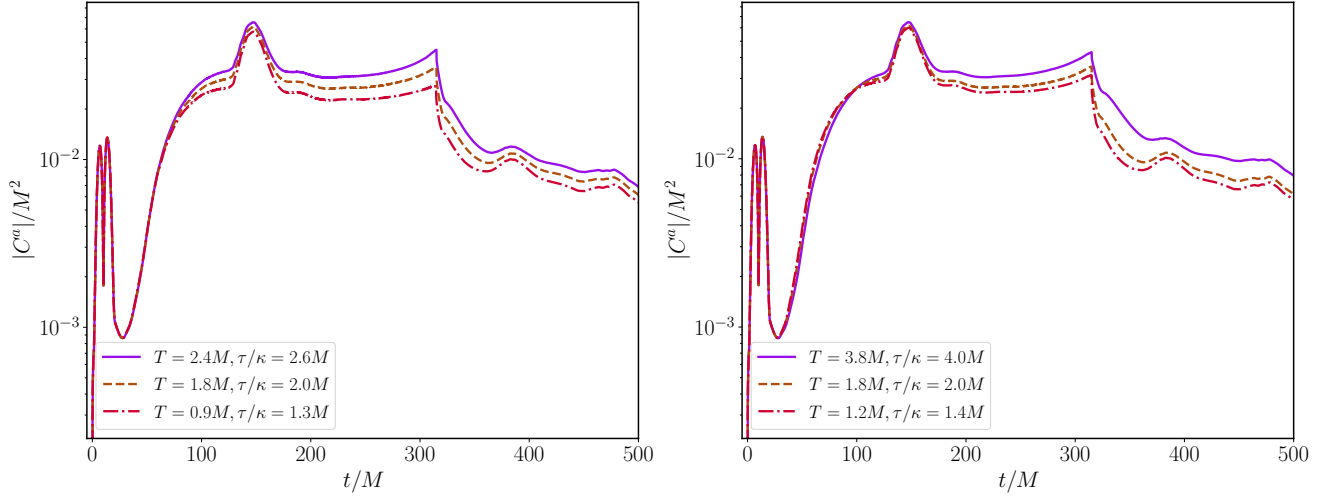

	\includegraphics[width=0.99\columnwidth,draft=false]{{cnst_bhbh_axi_tau}.pdf}
	\includegraphics[width=0.99\columnwidth,draft=false]{{cnst_bhbh_axi_kappa}.pdf}
	\caption{Integrated norm of the constraint violation $C_a$ as a function of
	coordinate time (in units of the total mass) 
    for the same equal-mass head-on collision with $\lambda=0.04m^2$ shown in the right panel of Fig.~\ref{fig:cnst_axi_fix}
	when varying $t_1$ but keeping $t_2$ fixed (left),
	or varying $t_2$ but keeping $t_1$ fixed (right). 
    As expected, the constraints decrease with increasing $\kappa$.
\label{fig:cnst_tau_kappa}
}
\end{figure*}

In our studies, we have confirmed this behavior and, unless the timescale $T$
employed is much larger than the scale of the black holes considered, the
constraints remain under control and the solutions show a convergent
behavior throughout the evolutions. 
We show the behavior of the constraint violation with increasing numerical resolution in Fig.~\ref{fig:cnst_axi_fix},
and when increasing $\kappa$ in Fig.~\ref{fig:cnst_tau_kappa}.
The constraint violation is found to decrease with one or the other, depending on which contribution
dominates at a given time.

%=============================================================================

\section{Secular errors in the order-by-order approach: a toy model}
\label{app:anharm_osc}
As a toy model to illustrate how secular errors appear in the order-by-order approach,
let us consider the anharmonic oscillator
\begin{equation}
    \label{eqn:aho}
    \frac{d^2x}{dt^2}+\left(\frac{2 \pi}{T_0} \right)^2\left(x+\epsilon x^3\right)=0 , 
\end{equation}
where we will consider $\epsilon$ to be a smaller parameter, and choose initial conditions
with 
\begin{equation}
x(t=0)=1,\ \frac{dx}{dt}(t=0)=0. 
\end{equation}    
For $\epsilon>0$, the system will oscillate with 
shorter period compared to $\epsilon=0$ (where the period is $T_0$), but will have the same amplitude of oscillation.
This is similar to binary black hole mergers in EsGB, where we find non-zero coupling
causes the binary to inspiral faster compared to GR, and hence the frequency of the GW
signal increases, but we find that the amplitude of the GW signal 
at fixed frequency is not significantly affected. 

\begin{figure*}
    \includegraphics[width=\columnwidth,draft=false]{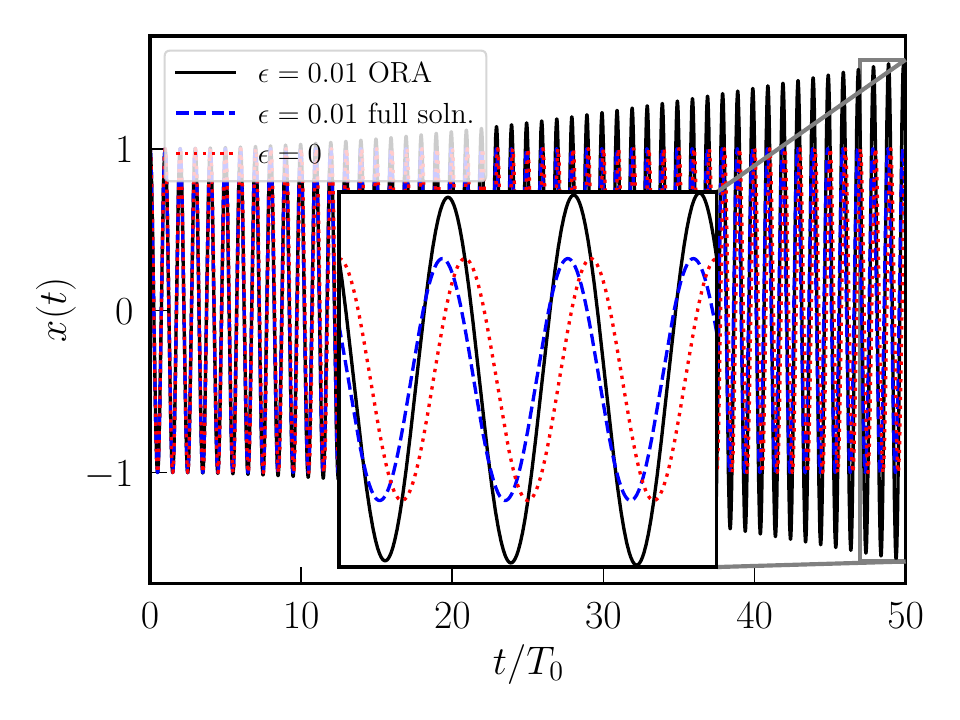}
    \includegraphics[width=\columnwidth,draft=false]{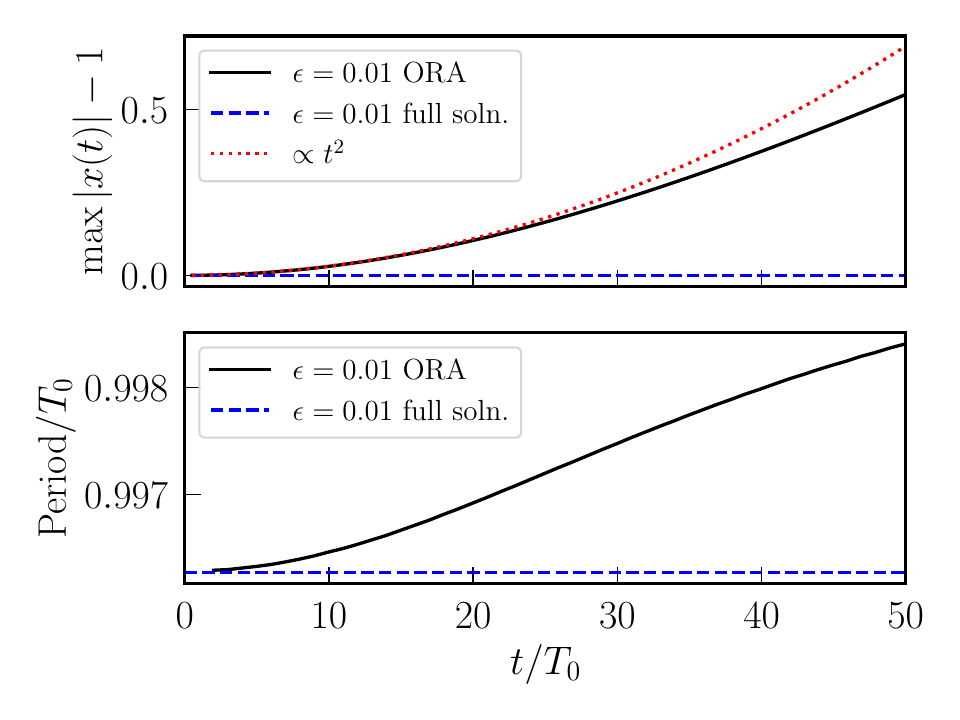}
	\caption{
        A comparison of the solution to the anharmonic oscillator system
        [Eq.~\eqref{eqn:aho_ora} with $\epsilon=0.01$] using the ORA approach
        (solid black curves) to the full solution (dashed blues curves).  Left:
        The time evolution of the anharmonic oscillator compared to the
        harmonic oscillator (dotted red curve). The inset shows that after 50
        oscillations, the anharmonic oscillator is noticeably out-of-phase
        compared to the harmonic ($\epsilon=0$) oscillator, and the ORA approximation has a spurious
        increase in amplitude.  Right: The increase in amplitude (top) and
        period (bottom) of oscillator in the ORA solution. In the full
        solution, these quantities are constant. 
\label{fig:anharm}
}
\end{figure*}

Solving Eq.~\ref{eqn:aho} to linear order in $\epsilon$ in the order-by-order approach
with the specified initial conditions, we
obtain 
\begin{align}
    \label{eqn:aho_ora}
    x(t) = &\cos( \omega_0 t) \\ &+ \frac{\epsilon}{32}\left[\cos(3\omega_0t) - \cos(\omega_0 t) - 12 \omega_0 t \sin(\omega_0 t) \right] 
\nonumber 
\end{align}
where $\omega_0 = 2\pi/T_0$. From this, we can immediately see that this solution
will exhibit secular growth on timescales of the order of $T_0/\epsilon$.  This
is illustrated in Fig.~\ref{fig:anharm}, where we compare the order-by-order
solution to the full solution taking $\epsilon=0.01$. With this value, the
anharmonic oscillator becomes out of phase with the $\epsilon=0$ case by $\sim
1$ radian after $50T_0$. As shown in Fig.~\ref{fig:anharm}, the order-by-order
approach captures the decrease in the period of oscillation initially, but over
time, exhibits a secular increase in the period towards its unperturbed value. 

This approximation also introduces an increase in the amplitude of oscillation,
which at early times grows like $t^2$. By the time the effect of the $\epsilon$
term has led to a noticeable dephasing, the spurious increase in amplitude and
period have also become significant.  This simple model captures the features
that we also find when considering binary black hole mergers in EsGB. In
particular, whenever the modified gravity contribution causes a noticeable
dephasing in the GWs, the ORA approach also gives a significant
spurious increase in the amplitude of the GW signal.

%=============================================================================
\section{Fixing a potentially elliptic second order model}
\label{app:fixingmodel}

Consider the following equation
\begin{equation}
	\partial_{t}^2\phi = \partial_{x}^2\phi + a \partial_{x}^2\phi \, ,
\end{equation}
for some value of $a \in \mathbb{R}$. If $a<-1$, this defines an elliptic second order equation for $\phi$; otherwise,
the problem is hyperbolic. Unless the condition $a>-1$ is satisfied, an attempt treat the equation as normal
hyperbolic evolution equation would fail. Consider instead the following fixed version,
\begin{eqnarray}
\partial_{t}^2\phi &=& \partial_{x}^2\phi + a \Pi \ , \nonumber \\
\sigma \partial_{t}^2\Pi &=& \sigma \partial_{x}^2\Pi - \partial_{t}\Pi - \kappa (\Pi - \partial_{x}^2\phi) \ ,
\end{eqnarray}
which forces the variable $\Pi$ to approach $\partial_{x}^2\phi$ on a timescale $\simeq \kappa^{-1}$, and recover the behavior of
the original equation for long wavelengths, while also forcing the system to be hyperbolic. 
A straightforward Fourier-mode analysis, i.e., assuming $\{\phi,\Pi\}=\{\phi,\Pi\}_0 e^{st} e^{ikx}$, reveals
the fixed system has, for low wavenumbers, four distinct solutions behaving as
\begin{eqnarray}
s_{\pm} &=& \pm i (\sqrt{1+a}) k + \frac{a}{2 \kappa} k^2  \ , \\ \nonumber
s_{1,2} &=& -\frac{1}{2\sigma} \pm \frac{\sqrt{1-4\kappa\sigma}}{2\sigma} 
+ \left(\pm\frac{2(1-a)\kappa \sigma+a}{2 \kappa \sqrt{1-4\kappa\sigma}}-\frac{a}{2 \kappa}  \right) k^2 \nonumber \,.
\end{eqnarray}
Thus, it has two modes coinciding with the behavior expected in the original problem, and two extra ones which are
damped away.
For high wavenumbers, instead, one has two solutions with dual roots,
\begin{eqnarray}
s=\pm i k \, ,
\end{eqnarray}
which corresponds to plane wave solutions propagating with speed unity.
This model highlights some features that are particularly relevant to EsGB gravity: 
If the hyperbolic condition is violated, the fixed system 
controls the high frequency behavior of the solution, but still retains the low frequency dispersion
relation expected for the hyperbolic ($s=\pm \sqrt{1+a} i k$) or elliptic $s=\pm \sqrt{|1+a|} k$) cases. 

\end{document}